\documentclass[preprint,3p]{elsarticle}

\usepackage{aas_macros}
\usepackage{lineno,hyperref}
\hypersetup{
  colorlinks,
  citecolor=Blue,
  linkcolor=Red,
  urlcolor=Blue}

\usepackage{cleveref}

\journal{Icarus}

\biboptions{authoryear}

\renewcommand{\deg}{^\circ}

\newcommand{\arcsec}{^{\prime\prime}}

\newcommand{\au}{\,\mathrm{au}}
\newcommand{\km}{\,\mathrm{km}}
\newcommand{\meter}{\,\mathrm{m}}
\newcommand{\cm}{\,\mathrm{cm}}
\newcommand{\mm}{\,\mathrm{mm}}
\newcommand{\um}{\,\mu \mathrm{m}}

\newcommand{\Myr}{\,\mathrm{Myr}}
\newcommand{\kyr}{\,\mathrm{kyr}}
\newcommand{\yr}{\,\mathrm{yr}}
\newcommand{\days}{\,\mathrm{d}}
\newcommand{\hours}{\,\mathrm{hours}}
\newcommand{\hour}{\,\mathrm{h}}
\newcommand{\minute}{\,\mathrm{min}}
\newcommand{\second}{\,\mathrm{s}}
\newcommand{\magnitude}{\,\mathrm{mag}}
\newcommand{\K}{\,\mathrm{K}}
\newcommand{\J}{\,\mathrm{J}}

\newcommand{\kg}{\,\mathrm{kg}}

\newcommand{\W}{\,\mathrm{W}}

\begin{document}

\begin{frontmatter}

\title{Dynamical Evolution and Thermal History of Asteroids (3200) Phaethon and (155140) 2005~UD}

\author{Eric MacLennan}
\address{Department of Physics, PO Box 64, 00014 University of Helsinki, Finland}

\author{Athanasia Toliou}
\address{Asteroid Engineering Laboratory, Onboard Space Systems, Lule\r{a} University of Technology, Box 848, SE-98128 Kiruna, Sweden}

\author{Mikael Granvik}
\address{Department of Physics, PO Box 64, 00014 University of Helsinki, Finland}
\address{Asteroid Engineering Laboratory, Onboard Space Systems, Lule\r{a} University of Technology, Box 848, SE-98128 Kiruna, Sweden}





\begin{abstract}
The near-Earth objects (NEOs) (3200) Phaethon and (155140) 2005~UD are thought to share a common origin, with the former exhibiting dust activity at perihelion that is thought to directly supply the Geminid meteor stream. Both of these objects currently have very small perihelion distances ($0.140\au$ and $0.163\au$ for Phaethon and 2005~UD, respectively), which results in them having perihelion temperatures around $1000\K$. A comparison between NEO population models to discovery statistics suggests that low-perihelion objects are destroyed over time by a, possibly temperature-dependent, mechanism that is efficient at heliocentric distances less than $0.3\au$. By implication, the current activity from Phaethon is linked to the destruction mechanism of NEOs close to the Sun.

We model the past thermal characteristics of Phaethon and 2005~UD using a combination of a thermophysical model (TPM) and orbital integrations of each object. Temperature characteristics such as maximum daily temperature, maximum thermal gradient, and temperature at different depths are extracted from the model, which is run for a predefined set of semi-major axis and eccentricity values. Next, dynamical integrations of orbital clones of Phaethon and 2005~UD are used to estimate the past orbital elements of each object. These dynamical results are then combined with the temperature characteristics to model the past evolution of thermal characteristics such as maximum (and minimum) surface temperature and thermal gradient. The orbital histories of Phaethon and 2005~UD are characterized by cyclic changes in $e$, resulting in perihelia values periodically shifting between present-day values and $0.3\au$. Currently, Phaethon is experiencing relatively large degrees of heating when compared to the recent $20,000\yr$. We find that the subsurface temperatures are too large over this timescale for water ice to be stable, unless actively supplied somehow. The near-surface thermal gradients strongly suggest that thermal fracturing may be very effective at breaking down and ejecting dust particles. Observations by the DESTINY+ flyby mission will provide important constraints on the mechanics of dust-loss from Phaethon and, potentially, reveal signs of activity from 2005~UD.

In addition to simulating the recent dynamical evolution of these objects, we use orbital integrations that start from the Main Belt to assess their early dynamical evolution (origin and delivery mechanism). We find that dwarf planet (2)~Pallas is unlikely to be the parent body for Phaethon and 2005 UD, and it is more likely that the source is in the inner part of the asteroid belt in the families of, e.g., (329) Svea or (142) Polana. 
\end{abstract}


\end{frontmatter}


\section{Introduction}
 
Observational monitoring of (3200)~Phaethon (perihelion distance $q \approx 0.140\au$), the largest asteroid with $q < 0.3\au$, reveals consistent and repeated activity during perihelion passage in the form of an extended dust tail \citep{Jewitt_etal2013,Hui&Li2016}. This consistent and repetitious activity from Phaethon, along with its orbital similarity to Geminid meteors, is strong evidence that it supplies dust to the Geminid meteor stream \citep{Williams&Wu1993,Jenniskens2006}. As such, is the first asteroid discovered to be associated with a meteor stream \citep{Gustafson1989}. The orbital similarity of the smaller (155140)~2005~UD ($q \approx 0.163\au$) to both Phaethon suggests that it too is genetically related and may also contribute mass to the Geminid stream. The present characteristics of these two asteroids, briefly reviewed below, and their past thermal characteristics thus provide important constraints and insight regarding the destruction of asteroids at small heliocentric distances.

The top-shaped Phaethon has an effective diameter estimated to be in the range of 5--$6\km$ with a rotation period of $\approx3.6 \hours$ \citep{Taylor_etal2019,Hanus_etal2018}. Its spectral classification as a B-type \citep{Licandro_etal2007} combined with knowledge from numerical simulations \citep{deLeon_etal2010,Todorovic18} suggest its parent body to be main-belt dwarf planet (2)~Pallas. The geometric albedo derived from thermophysical modeling, $p_V \approx 0.13$ \citep{Hanus_etal2016,Masiero_etal2019}, is slightly higher than most B-types but consistent with B-types comprising the Pallas collisional family \citep{AliLagoa_etal13,AliLagoa_etal16}. Modeling of Phaethon's thermal inertia ($600 \pm 200 \J \meter^{-2} \K^{-1} \second^{-1/2}$) gives a characteristic grain size of 1--$2\cm$ \citep{Hanus_etal2016,Hanus_etal2018}. \citet{Masiero_etal2019} estimated the near-infrared to visible reflectance ratio to be $p_\mathrm{IR}/p_V = 0.8 \pm 0.06$, which is consistent with the Pallas collisional family \citep{AliLagoa_etal13}. Using measurements of the semi-major axis drift rate, \citet{Hanus_etal2018} calculated a bulk density of $1670 \pm 470 \kg \meter^{-3}$ by modeling the thermal recoil forces off the surface (i.e., the Yarkovsky effect). Anomalous brightening and evidence for a dust tail were observed in NASA STEREO images during Phaethon's 2009, 2012, and 2016 perihelion passages \citep{Jewitt&Li2010,Li&Jewitt2013,Jewitt_etal2013,Hui&Li2016}. Current mass estimates for the dust tail ($\sim 3 \times 10^5 \kg$) over the narrow $\sim2\days$ window of activity at perihelion suggest mass-loss rates ($\sim 0.1-3 \kg \second^{-1}$) that are insufficient to supply the mass of the Geminid stream \citep{Jewitt_etal2013,Hui&Li2016}. This perihelion mass-loss estimate---along with the non-detection of activity elsewhere in the orbit \citep{Jewitt_etal2019}---may suggest that most of stream mass was produced in one or more catastrophic event(s).

The Apollo asteroid 2005~UD (hereafter, UD) has been inferred to be associated with Phaethon on similarity of their orbital parameters \citep{Ohtsuka_etal2006} and colors \citep{Jewitt&Hsieh_2006}. However, the common origin claim has been disputed on dynamical grounds by \citet{Ryabova_etal2019}. Observations of UD's lightcurve show a rotation period of $\sim 5.2 \hour$ \citep{Jewitt&Hsieh_2006,Devogle_etal20} and an equivalent prolate ellipsoid axis ratio of $a/b = 1.45$. \citet{Masiero_etal2019} used NEOWISE data to more precisely estimate the size ($D \approx 1.2 \km$), geometric albedo ($p_V \approx$ 0.14), and infrared reflectance ($p_\mathrm{IR}/p_V \approx 1.41$). Recent work by \citet{Devogle_etal20} indicates that UD's polarimetric phase curve is consistent with Phaethon's. The geometric albedo of UD \citep[$p_V = 0.10 \pm 0.02$;][]{Devogle_etal20} lies between the mean albedos of other B-types and of Pallas family members \citep{AliLagoa_etal16}. A thermal inertia of $300^{+120}_{-110} \J \meter^{-2} \K^{-1} \second^{-1/2}$ was derived using NEOWISE data which gives a characteristic regolith grain size of 0.9--$10\mm$ \citep{Devogle_etal20}.

Current knowledge of the observed near-Earth asteroid (NEA) population indicates a deficiency of objects with perihelion distance $q<0.3\au$, compared to population prediction models \citep{2016Natur.530..303G}. This observation-model mismatch among low-$q$ objects suggests that they are actively destroyed at small heliocentric distances by one or more mechanisms unaccounted-for in the model. Because low-$q$ asteroids are subject to intense solar radiation, their surfaces reach temperatures upwards of $1000\K$---the largest for any solid surface in the Solar System. This fact suggests that a {\it temperature-dependent} process, in particular, destroys asteroids at small heliocentric distances and on timescales shorter than $\sim1 \Myr$---the dynamical lifetime of NEAs.

The exact cause of Phaethon's activity is unknown, but it is likely that its mass-loss is related to the depletion of low-$q$ asteroids in general. \citet{Jewitt12} details many possible activity-causing mechanisms for active asteroids including: radiation pressure, electrostatic repulsion, sublimation, thermal fracturing, dehydration, rotational instability, and meteoroid impacts. Of these mechanisms, \citet{Jewitt12} favored one of the temperature-dependent processes (i.e., sublimation and thermal fracturing) for Phaethon because of the nature of activity observed during perihelion passage---which we investigate in this work. The proposed JAXA DESTINY+\footnote{\href{https://destiny.isas.jaxa.jp/science/}{https://destiny.isas.jaxa.jp/science/}} mission is scheduled to fly by Phaethon in the upcoming decade \citep{Arai_etal2018,Arai_etal2019}, with a possible extension to UD. In-situ observations of dust activity from Phaethon (and possibly UD) will serve as crucial indicators as to the nature of activity and ultimate destruction of these bodies.

The discovery and monitoring of main-belt comets \citep[MBCs;][]{Hsieh&Jewitt06}---a subset of active asteroids distinguished by their periodic activity that occurs around perihelion \citep{Hsieh_etal10,Hsieh_etal11,Hsieh_etal18a,Hsieh_etal18b}---suggests that buried volatiles (likely water ice) exist on main-belt asteroids (MBAs) and trigger dust activity when sufficiently heated. A similar line of reasoning can be made for Phaethon, as its activity has only been observed during perihelion passage. The clear factor that separates Phaethon from MBCs is the drastically higher maximum surface temperatures and shorter duration of activity. If volatile sublimation is responsible for Phaethon's activity and/or the destruction of low-$q$ NEAs, it must be shown that it is possible for volatiles to exist somewhere in the body and that the activity is triggered by the rising perihelion temperatures. Directly comparing Phaethon, UD and low-$q$ NEAs to MBCs is difficult because of the drastically different thermal environment and history, which we aim to investigate in this work.

Cyclic heating and cooling of a rock will cause heterogeneous internal thermal expansion and a corresponding stress field \citep{Molaro_etal15,Hazeli_etal18}. If the local stress exceeds the material strength then material failure will occur, ultimately resulting in the development of one or more cracks \citep{Molaro&Byrne12,Molaro_etal15}. Experimental heating of meteorite samples show that crack growth, and material breakdown, can occur on rates consistent with hypothesized regolith generation timescales \citep{Delbo_etal14}. This experimental evidence and modeling work \citep{Molaro_etal17,Hazeli_etal18} together suggest that this process could be more efficient than impacts at breaking down asteroid regolith \citep{Basilevsky_etal15}, particularly for asteroids that come close to the Sun and that have fast rotation periods \citep{Molaro&Byrne12,Molaro_etal15}. Thus, thermal fracturing may contribute to the destruction of low-$q$ asteroids over large timescales \citep{Jewitt12}. Thermophysical models (TPMs) are powerful tools that can be used to accurately estimate thermal gradients and the efficiency of thermal fracturing during perihelion passage.

Motivated by these previous findings, our aims in this paper are thus: 1) investigate potential temperature-dependent mechanisms causing Phaethon's activity, 2) hypothesize as to the temperature-dependent destruction of low-$q$ asteroids, and 3) speculate on the surface properties of Phaethon and UD ahead of the DESTINY+ mission. To do so, we quantitatively characterize the current and past temperature characteristics of Phaethon and UD by combining thermophysical and dynamical techniques. The temperature-dependent mechanisms considered in this work---sublimation, thermal fracturing, and dehydration---act differently according to the temperature characteristics. Sublimation of volatiles and material dehydration will become relevant after a threshold temperature is reached, and the processes of thermal fracturing is expected to be more efficient when the thermal gradient increases. We thus model the maximum temperatures and thermal gradients for these two objects. Our results are then used to postulate on the broader context of probable temperature-dependent processes affecting this subset of objects.

\section{Theory and methods}

\subsection{Dynamical Modeling}

\subsubsection{Orbit Computation}

Our dynamical analysis relies on accurate initial orbit determination. Samples of orbital elements for the present epoch for Phaethon and UD are obtained in a two-step process by utilizing the OpenOrb software \citep{2009M&PS...44.1853G}. We begin by computing the best-fit orbital elements in the least-squares sense, as well as the covariance matrix that describes the uncertainty hyperellipsoid for the orbital elements. We make a conservative assumption for the astrometric uncertainty by using $\sigma_\mathrm{RA}=\sigma_\mathrm{Dec}=0.5\arcsec$ for all astrometry available. Then we sample the resulting uncertainty hyperellipsoid to generate orbits that reproduce the available astrometry. For the sampling we model a Gaussian distribution centered on the least-squares solution with $\sigma_j=2\sigma_{\mathrm{nominal},j}$ where $j$ refers to one of the six orbital elements ($a,e,i,\Omega,\omega,M_0$). This approach is used in order to focus the orbit sample on the best-fitting orbits while ensuring that a significant fraction of the uncertainty region is covered. Note that we do not automatically accept each sample orbit, but require that each one also agrees with the available astrometry given our conservative estimate for the astrometric uncertainty. In addition, we also require that $\Delta\chi^2 = \chi^2 - \chi_\mathrm{LS}^2 < 20.1$, where $\chi^2$ corresponds to the orbit in question and $\chi_\mathrm{LS}^2$ corresponds to the best-fit least-squares orbit. The chosen upper limit corresponds to the $3\sigma$ limit for a six-dimensional problem. The dynamical model includes gravitational perturbations by the eight planets, Pluto, the Moon \citep[DE430;][]{2014IPNPR.196C...1F}, and the 25 largest asteroids \citep[BC430;][]{Bae17}. We account for the relativistic perihelion shift, but do not account for non-gravitational perturbations. We discard observations as outliers if the R.A. and/or Dec residuals are $>2\arcsec$. Next, we will feed the computed orbits (later referred to as clones) to an orbit integrator and propagate them backwards in time.

\subsubsection{N-body Orbital Simulation}

Our integrator of choice was the SWIFT\_RMVS integrator, which can efficiently handle a close encounter between orbital clones (having zero mass) and the massive objects in our simulation \citep{1994Icar..108...18L}. Since there is no straightforward way to perform backward integrations with the SWIFT package, we adopted negative values for all time parameters and modified the relevant checks in the main loop of the integrator accordingly.

We chose a small time step equal to $10^{-3}\yr$, which is an appropriate value for integrating orbits that reach small heliocentric distances. Close encounters are defined as having a distance of less than 3.5 Hill radii from a massive object and a clone. During a close encounter SWIFT uses an adaptive time step that is shorter than the nominal value. We stopped the integration after $1\Myr$, at which point the divergence of the orbits has made it impossible to make a reasonably accurate assessment of the evolution of $q$. In addition, we integrated 100 random orbits for $5\Myr$ to observe the effect of close encounters on the dynamical history of the clones. We extracted the orbital elements every $100\yr$ from the simulations.

We also investigated the importance of general relativity in the orbital evolution of these NEOs, due to the low $q$ values. We included post-Newtonian effects in a Bulirsch-Stoer integrator, incorporating the commonly used relation for relativistic corrections derived by \citet{Newhall1983}. Next, we integrated the nominal orbits of Phaethon and UD for $0.5\Myr$ both with and without relativistic corrections. We found that the effect of close encounters in the secular evolution of both objects is much more significant than the small change in the precession of the argument of perihelion ($\omega$) induced by the relativistic effects. Hence, the results that follow are computed without relativistic corrections.

\subsection{Thermophysical Modeling}

The TPM we use to compute temperatures for Phaethon and UD during an entire orbital revolution is modified from that presented in \citet{MacLennan&Emery2019}, which is used to calculate diurnal surface temperatures. These modifications, described in \autoref{subsec:tpmdescr}, account for the circumstances associated with the longer heating timescale, time-varying insolation (incoming solar radiation), and changing size of the solar disk throughout the orbital revolution of the objects. We refer to the TPM after these modifications as the orbTPM and to the unmodified version of \citet{MacLennan&Emery2019} as the diTPM. The orbTPM is run for an entire orbital period instead of for individual positions along the orbit. This feature of the orbTPM accounts for the rapid changes in insolation around each object's perihelion passage and for the deeper thermal wave propagation due to insolation changes throughout the orbit. In \autoref{TPMimpl} we investigate the significance and importance of accounting for these phenomena, as well as the possible implications for the interpretation thermal infrared emission of NEAs on high-$e$ orbits.

\subsubsection{Model Description}\label{subsec:tpmdescr}

The orbTPM numerically calculates surface and sub-surface temperatures, $T$, on a rotating spherical object revolving around the Sun. The solution to the one-dimensional heat diffusion (Fourier's Law) equation,
\begin{equation}\label{FourierEq}
    \frac{\partial T(x, t)}{\partial t} = \frac{k}{\rho c_s}\frac{\partial^2 T(x, t)}{\partial x^2}
\end{equation}
is computed using the surface insolation as the only energy input. \autoref{FourierEq} is evaluated at discrete depth ($x$) and time ($t$) interval by using the Crank-Nicolson finite-difference method \citep{Press_etal07}. The effective thermal conductivity ($k$), regolith density ($\rho$), and heat capacity ($c_s$) are assumed to be constant with $x$ and $t$ and together represent the {\it bulk} properties of the regolith, as opposed to the properties of individual rock grains. The square-root of the product of these three thermophysical properties defines the thermal inertia, $\Gamma = \sqrt{k \rho c_s}$, which is a measure of a material's resistance to temperature change. The energy balance for a surface facet with a solar incidence angle of $z_i$ and a bolometric Bond albedo of $A_\mathrm{b}$,
\begin{equation}\label{eq:energy}
    \frac{S_\odot(1-A_\mathrm{b})}{r^2_\textrm{au}} \cos(z_i) = \varepsilon \sigma T^4_\textrm{eq} - k \frac{dT}{dx}\bigg|_\mathit{surf},
\end{equation}
is used as the upper boundary condition for solving \autoref{FourierEq}. The left hand side of \autoref{eq:energy} is the absorbed insolation where the solar flux at $1\au$, $S_\odot$, is taken to be $1367 \W \meter^{-2}$ and is scaled using the inverse square of the heliocentric distance given in au ($r_\textrm{au}$). The right hand side is the energy radiated assuming a blackbody with emissivity $\varepsilon$, with Stefan-Boltzmann constant ($\sigma$), and thermal energy conducted into the subsurface. The lower boundary condition is taken to be $\frac{dT}{dx} = 0$, implying zero heat transfer, at a sufficiently large depth---as mentioned below.

Additionally, we partly account for the Sun being a disk, by altering the insolation near the local sunrise and sunset of a surface facet. This is done, crudely, by using a simple function to modulate the input energy when the complement of the solar incidence angle is less than half of angular diameter of the Sun ($\delta_\odot$): 
\begin{equation}
E_\mathrm{input} \propto \cos(\delta_\odot/2)\cos(z_i) + \sin(\delta_\odot/2)\sqrt{1 - \cos(z_i)^2}.
\end{equation}
This scheme does not {\it explicitly} approximate the Sun as a disk, but does approximate a sunset or sunrise well enough for our application to objects with rotation periods much shorter than their orbital periods. More importantly, the approximation produces reasonably accurate facet temperatures in the polar regions, which is relevant to estimating seasonal changes in temperatures.

Surface temperatures are calculated for a set of discrete latitude values as the object rotates about its spin axis and orbits the Sun. This means that we do not explicitly calculate temperatures for a complete set of latitude and longitude values at each point in time. Thus, each time step corresponds to a single longitude strip on the object. Because rotation periods for asteroids are significantly shorter than the orbital periods, we are able to linearly interpolate across time steps in order to estimate temperatures at other longitude.

The depth variable, $x$, is discretized into increments small enough such that the heat transport from one to the other results in model stability. We set the total number of depth steps to be 300, in order to model heat transport at large depths---which is many factors greater than the diurnal insolation variation. The thermal skin depth, $l_s$, gives the length over which the temperature variation changes by a factor of $e = 2.718\ldots$ over a specified timescale, $\tau$:
\begin{equation}\label{eq:skindepth}
    l_s = \sqrt{\frac{k}{\rho c_s}\frac{\tau}{2\pi}}.
\end{equation}
Typically, TPMs that are constructed to interpret observations collected over a diurnal timescale ($\tau \sim10\hours$) must model subsurface heat transport to a sufficient depth for which there is no thermal gradient (and thus no heat transport). Because we wish to model temperatures for an entire orbit ($\tau \sim10^5 \hours$), we must account for subsurface heat transport associated with insolation changes throughout an orbital period. Additionally, for the sake of increasing the computation efficiency, we increase the spacing between subsequent depth steps as the depth increases such that $\Delta x_i = \Delta x_0 \times 1.02^{i}$. The topmost depth step is taken to be $l_s/5$ and the next depth step is 2\% larger, and so on. This does not sacrifice model accuracy because the changes in temperature as a function of depth decrease exponentially. To satisfy the lower boundary condition, the temperature difference between the last two depth points is set to zero.

The size of discrete time steps is held constant throughout the orbit of the object and set according to the number of time steps per rotation. The number of time steps per rotation is chosen to be large enough such that for large diurnal temperature changes (i.e., at perihelion) the temperature change between subsequent time steps is small enough to ensure numerical stability. For the most extreme cases the diurnal temperature range can exceed 400 K, for which we find that 1800 time steps per rotation is sufficient. After each orbital period we adjust the temperature profile (i.e., the temperature across all depth steps) by a particular value---calculated from the conservation of energy---in order to achieve faster convergence. The model was run for 20 orbital periods with constant orbital elements, which allowed for temperature convergence (the temperature difference at a given orbital position is $<1\K$ from the previous orbital revolution). A final check for model convergence is performed by ensuring that the orbit-averaged surface temperature for a given latitude is $<1\K$ different from the temperature at the largest depth (a consequence of energy conservation).

\subsubsection{TPM Implications and Caveats}\label{TPMimpl}
As a sanity check, we compare the surface temperatures from the orbTPM implementation with surface temperatures computed from the diTPM at a given heliocentric distance. For this demonstration, we model the emitted thermal flux from a set of asteroids with varying $e$ and with $a = 1.27\au$, $A_\mathrm{b}$ = 0.05, $\Gamma = 450 \J \meter^{-2}\K^\textrm{-1} \second^{-1/2}$, $P_\mathrm{rot} = 4 \hours$, and zero spin obliquity using both models. We find differences in the maximum and minimum surface temperatures throughout the orbit, as well as the diurnal temperature range at small heliocentric distances. In all cases, the difference becomes larger with increasing $e$.

\begin{figure}[h!]
  \centering
  \begin{tabular}[b]{c}
	\includegraphics[clip,trim=0.4cm 1.5cm 0.4cm 1.2cm,width=0.5\linewidth]{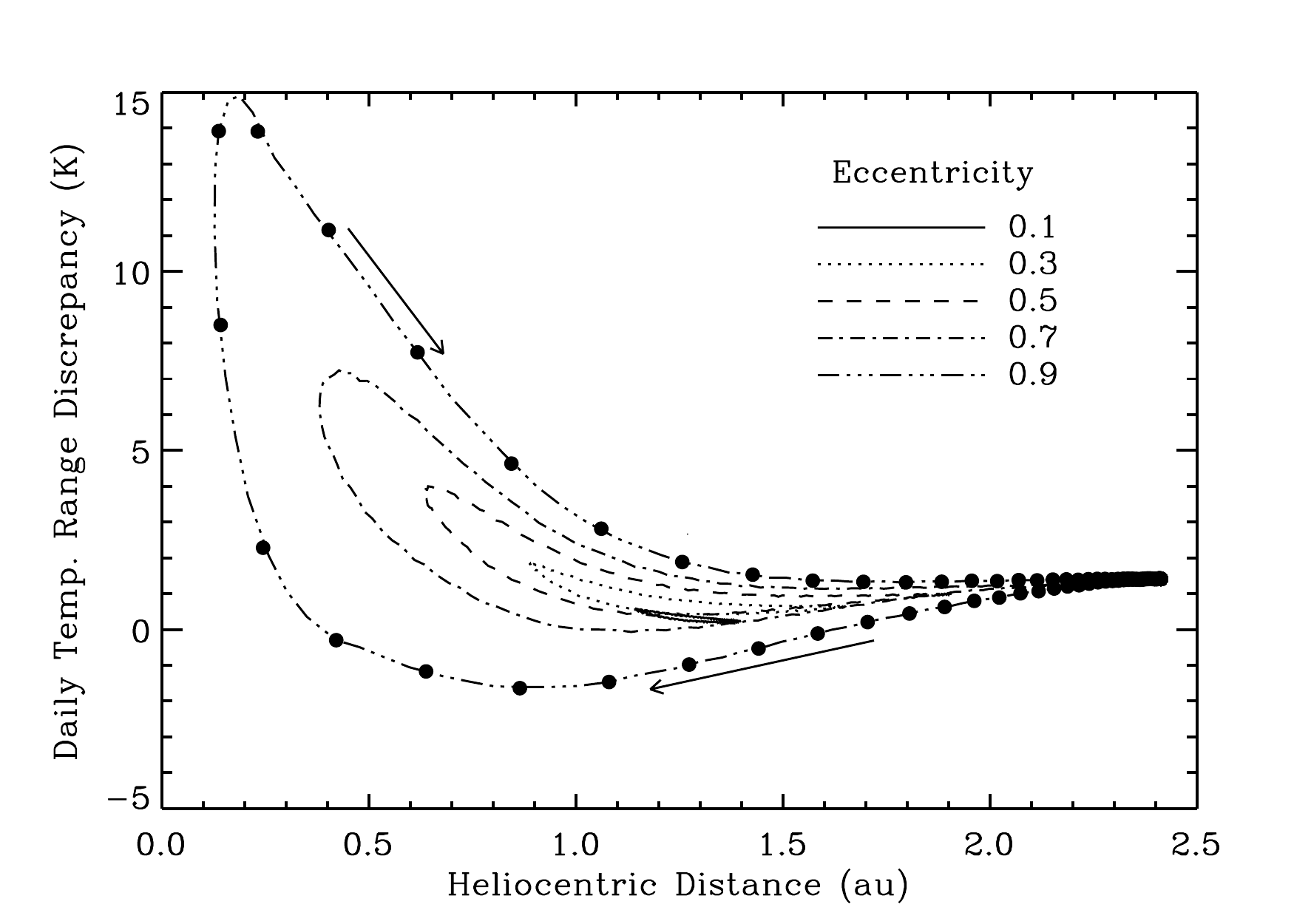} \\
	\includegraphics[clip,trim=0.4cm 1.5cm 0.4cm 1.2cm,width=0.5\linewidth]{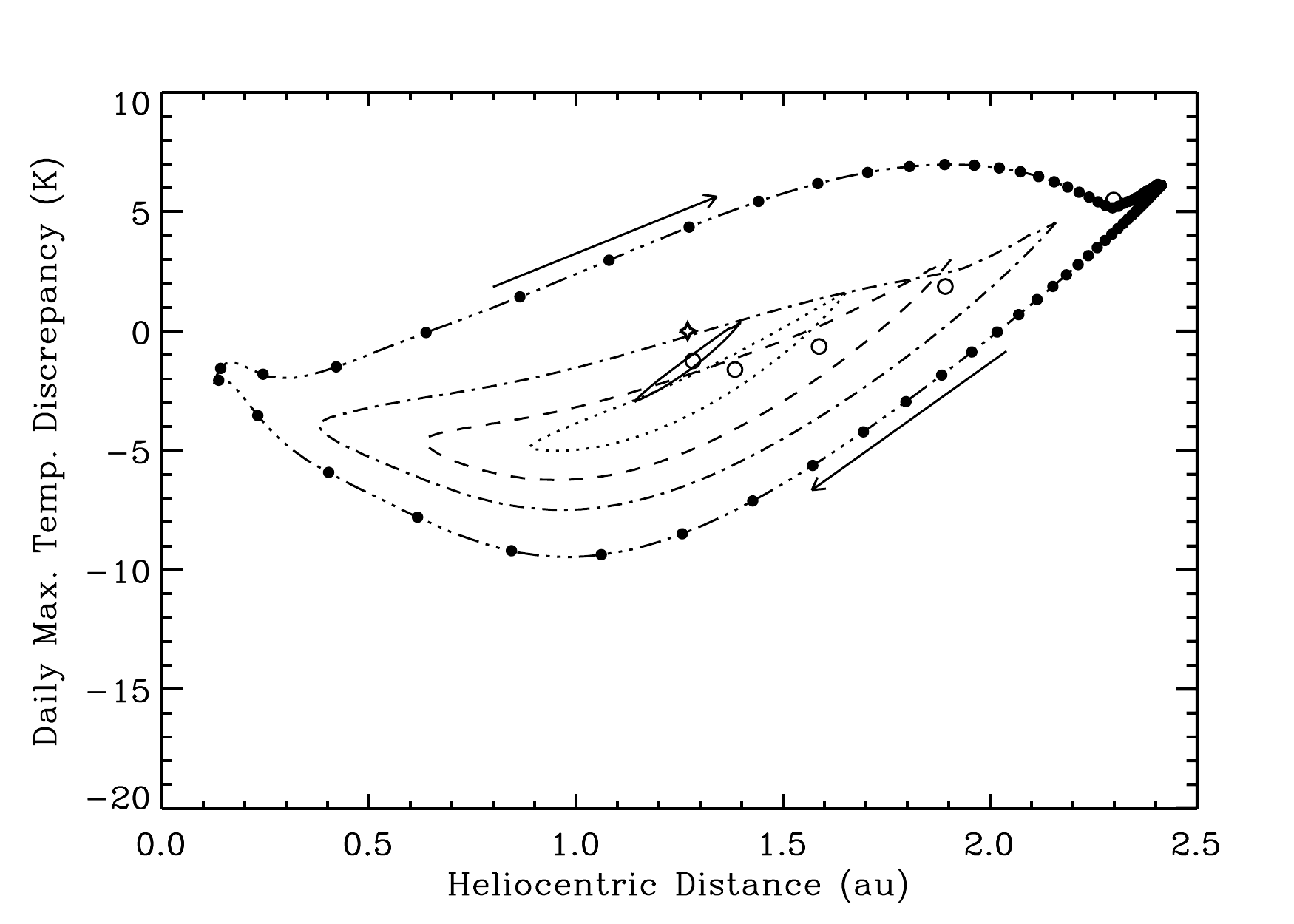}\\
	\includegraphics[clip,trim=0.4cm 0.2cm 0.4cm 1.1cm,width=0.5\linewidth]{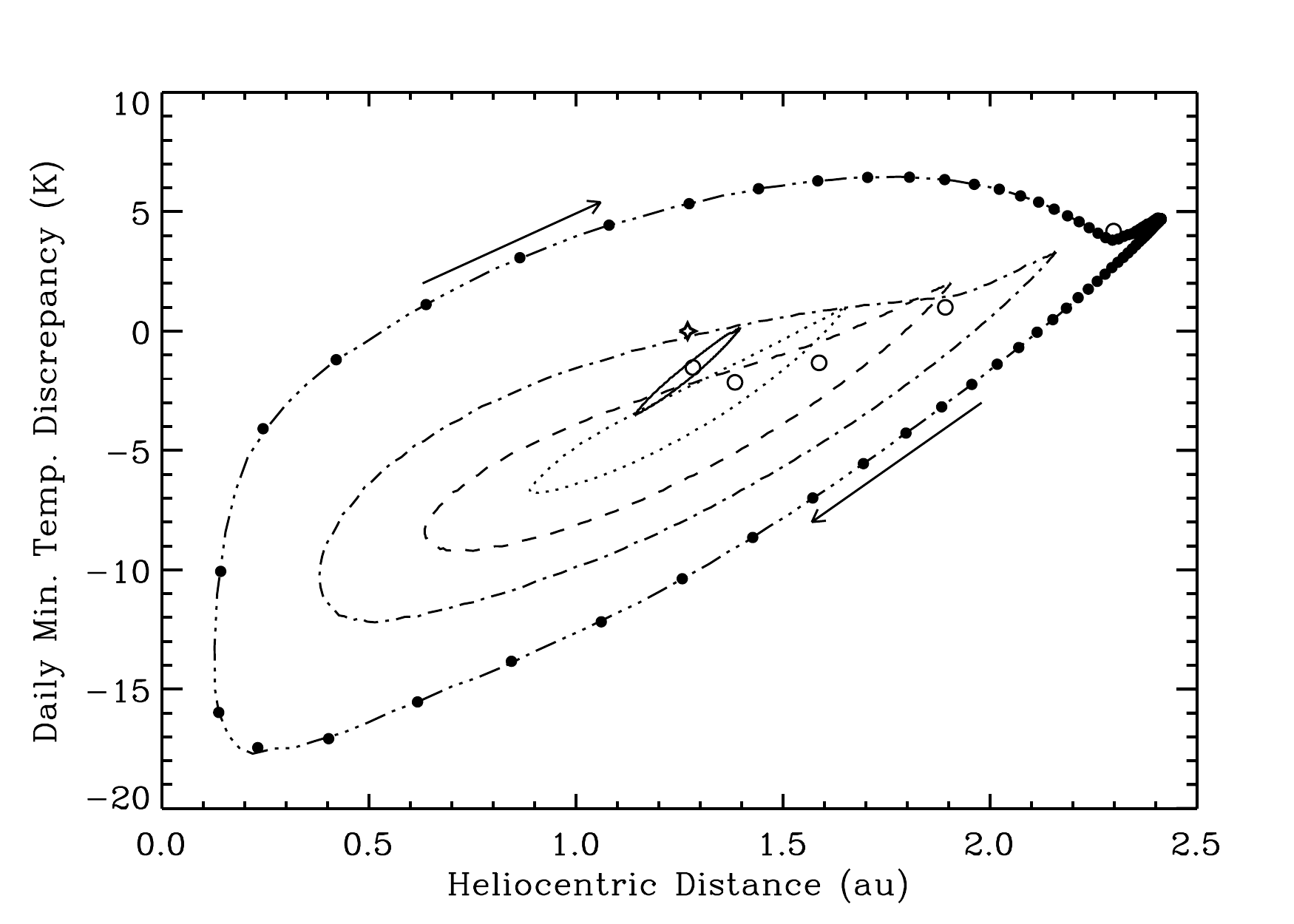} \\
  \end{tabular}
  \caption{The difference between temperatures calculated using diTPM and orbTPM. Dots indicate time steps separated by $\sim2 \days$. Open circles show the median values of temperature difference and heliocentric distance. A positive value for the difference indicates a larger orbTPM temperature.}\label{fig:benchtemp}
\end{figure}

The difference in diurnal temperature range, which is mainly influenced by thermal inertia and rotation period of the object, increases from nearly zero at $r>1.5\au$ to nearly $15\K$ at the perihelion point for $e = 0.9$ (top panel of \autoref{fig:benchtemp}). The temperature differences are likely due to the seasonal changes in thermal energy that is accounted for in the orbTPM but not in the diTPM; specifically, the time lag difference between the propagation of the diurnal and orbital thermal wave and how much total insolation is absorbed in one orbit. The difference is greater after perihelion passage and smallest before. Maximum and minimum temperature differences (middle and bottom panels of \autoref{fig:benchtemp}) show a trend opposite to that of the diurnal temperature range---temperatures calculated using the orbTPM are cooler for small heliocentric distances and warmer further away from the Sun. The difference in maximum and minimum temperatures is greater (furthest from zero) before perihelion passage and for objects with larger $e$. This disagreement between the orbTPM and diTPM, particularly in the maximum temperature, imply that asteroid diameters and albedos calculated over diurnal timescales may be wrong by a significant fraction---particularly for high-$e$ NEAs observed at large heliocentric distances when the relative temperature difference is higher. In this work we are not concerned with these temperature differences because we use the orbTPM, and we leave such investigations for future studies.

The orbTPM has a few built-in assumptions that sacrifice real-world accuracy for computational convenience. First, the model assumes that the thermophysical properties of the regolith do not vary with depth, or across the surface of an asteroid. The former is likely not accurate for planetary regoliths, as Lunar regolith core samples show that deeply buried material is more compacted, which increases the effective thermal inertia \citep{Keihm_etal73}. Yet, compaction of regolith is likely not highly relevant on asteroids with low gravitational environments. Because the daytime net heat flow is downward from the surface and any change in material properties would affect the temperature at greater depths, we claim that result of compacted sub-surface material on daytime (i.e. maximum) surface temperatures and thermal gradients is negligible. Additionally, spacecraft missions to asteroids \citep{Robinson_etal02,Mazrouei_etal14,Sugita_etal19,Lauretta_etal19a} show that their surfaces are non-homogeneous and that regolith grain sizes (and thus, thermal inertia) can vary by varying amounts \citep{Rozitis_etal20}. Generally speaking, surfaces at small heliocentric distances are in equilibrium with the insolation, and are thus controlled by the albedo, but not thermal inertia. Given that Phaethon and UD's surfaces are highly absorbing, any small changes in albedo across the surface will not influence the surface temperatures much (\autoref{eq:energy}). In any case, the orbTPM temperature calculations here represent an average of the temperature characteristics on the surface and the results depend on the (unknown) variation in thermophysical properties of Phaethon and UD's regolith.

Another caveat in our model assumptions is that of a spherical-shaped asteroid with a smooth surface. If we were to model other convex shapes the surface temperatures and gradients would not be affected, although it would alter the proportion of surface area at different latitudes. Since Phaethon and UD have low Bond albedos of just a few percent, we expect that multiple scattering of sunlight due to large-scale shape topography and small-scale roughness is not a large factor. However we expect shadowing and self-heating from topography and roughness to have an measurable influence on the thermal characteristics. Generally speaking, shadowing and self-heating effects would result in more extreme thermal gradients and increase the maximum surface temperatures relative to those estimated in this work. We come back to this point in our analysis of our model results and in our discussion.

\citet{Hanus_etal2016} showed that thermal torques acting on Phaethon will cause a drift in its spin axis. This drift is mostly in the longitude at pericenter \cite[see Figure 8 in][]{Hanus_etal2016}, or in our model the position of the solstice point, with minimal change in the spin obliquity (around $\pm10\deg$). The uncertainty in both the orbit, spin, and thermophysical parameters make it difficult to accurately predict the spin axis orientation at a given moment in the past, but \citet{Hanus_etal2016} demonstrated that the solstice point oscillates with a range of $\sim80\deg$ centered around a value close to the present-day spin orientation. Thus, while the current spin axis direction may be known to high accuracy, it is difficult to estimate its past evolution with similar accuracy.

Finally, it has been theoretically and experimentally shown that thermophysical parameters such as thermal conductivity and heat capacity are dependent on the temperature of a material ($k \propto T^3$; $c_s \propto T$) \citep{Watson64,Macke_etal19}. Furthermore, a temperature-dependent thermal inertia was confirmed for three asteroids with high orbital eccentricity by \cite{Rozitis_etal2018}. In addition, latent heat capacity may be relevant if volatiles exist (albeit deeply buried) beneath the surface of asteroids. At certain temperatures (i.e. when sublimation occurs) the latent heat becomes relevant \citep[e.g.,][]{Schorghofer&Hsieh18} and thermal energy will be sequestered in the subsurface, subsequently lowering the surface temperature. The effect of latent heat is the opposite from a situation of a radioactive heat source that contributes thermal energy and would raise the surface temperature at the surface. While these temperature-dependent factors are certainly relevant for the calculation of temperatures at a specific time and heliocentric distance, we wish to study here the evolution of the temperature characteristics as the orbits evolve through time. In this context, the perihelion distance has a much larger influence on the maximum surface temperature ($T_{max} \propto q^{-2}$) than does the chosen values of thermal inertia or heat capacity or their dependence on temperature.

\section{Results}

\subsection{Dynamical evolution}

\subsubsection{Orbit Solutions}

\begin{figure}[h!]
  \centering
  \begin{tabular}[b]{c}
	\includegraphics[width=0.333\linewidth]{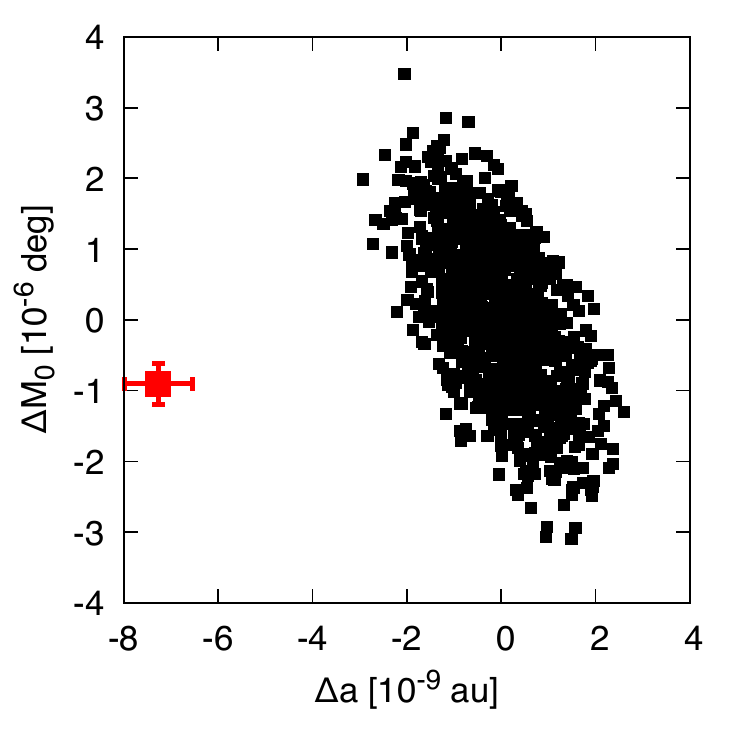}\includegraphics[width=0.333\linewidth]{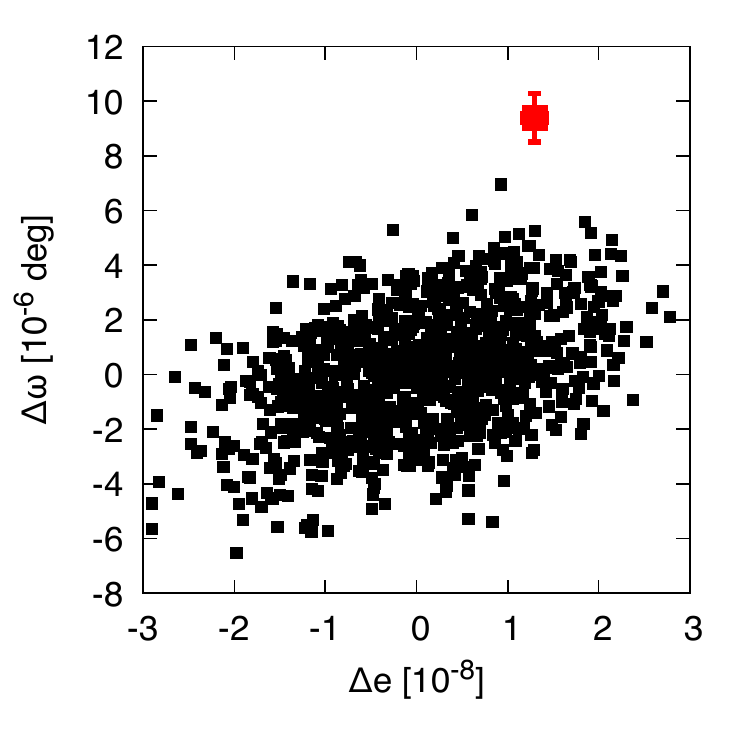}\includegraphics[width=0.333\linewidth]{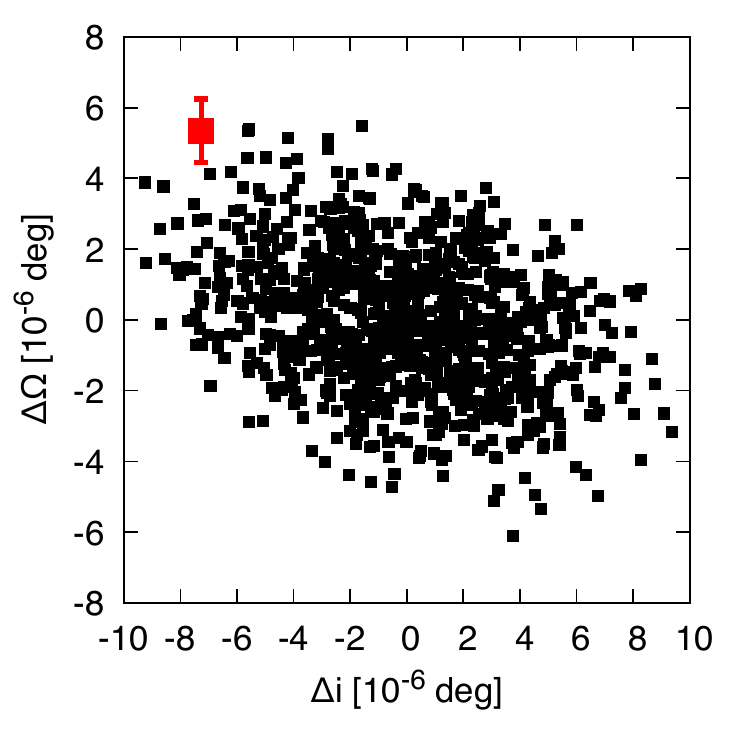} \\
	\includegraphics[width=0.333\linewidth]{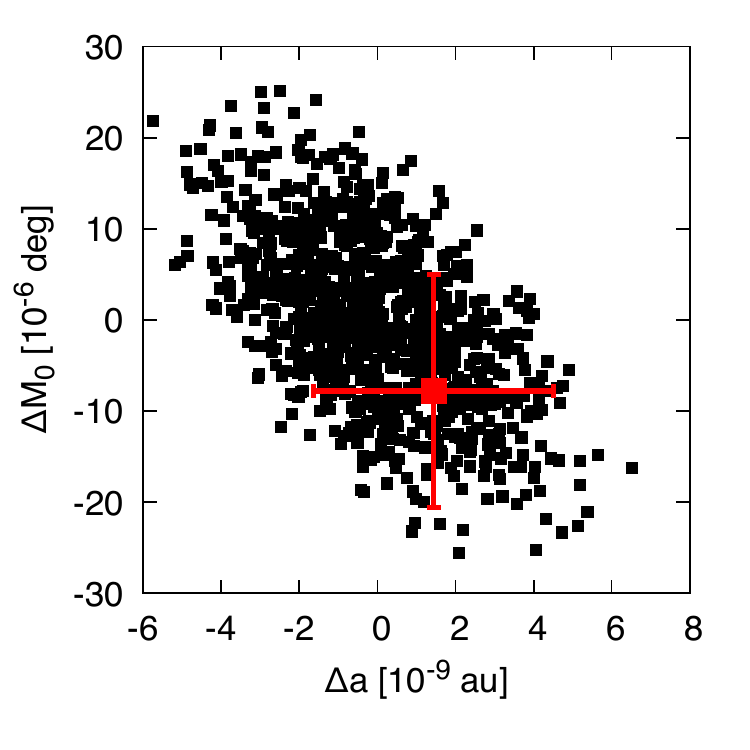}\includegraphics[width=0.333\linewidth]{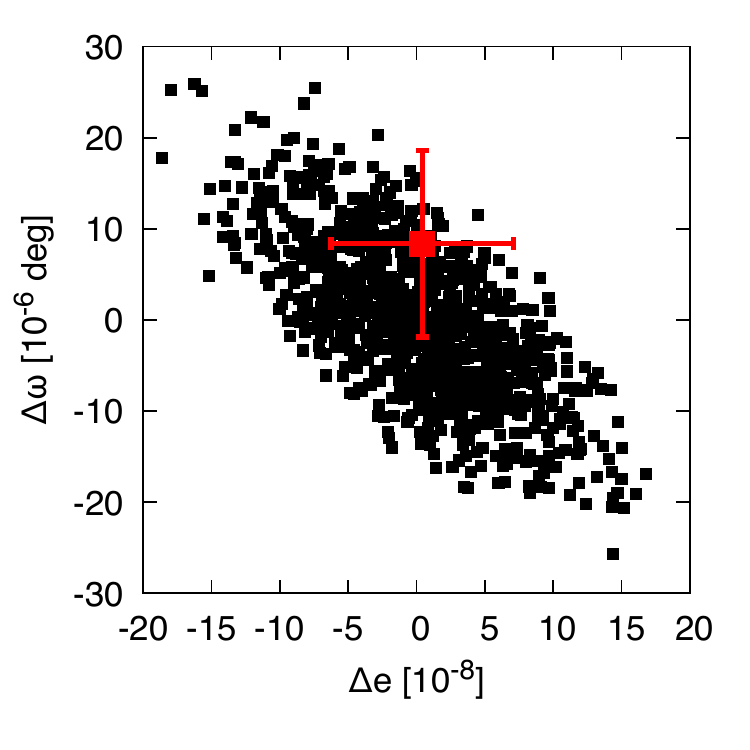}\includegraphics[width=0.333\linewidth]{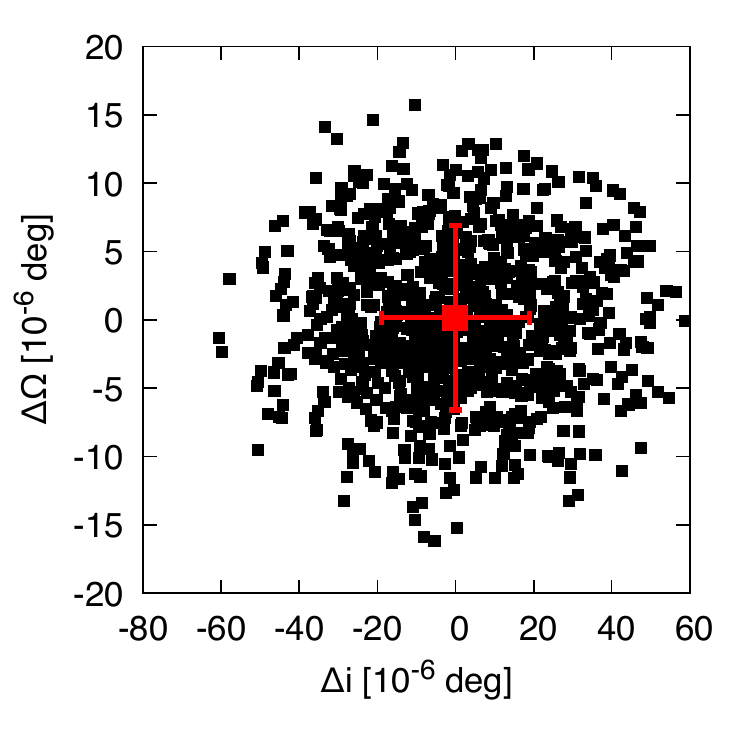} \\
  \end{tabular}
  \caption{Clones for Phaethon (top panels) and UD (bottom panels) shown as black squares. The nominal least-squares solution is in the origin by construction. The JPL solutions (JPL712 for Phaethon and JPL91 for UD) with $1\sigma$ errorbars for the same epoch (MJD 58600.0 TT) in red.}\label{fig:clones}
\end{figure}

The entire publicly available astrometric data set for Phaethon comprises a total of 5545 optical observations spanning from 1983-10-11 to 2020-02-20. We correct the astrometry for zonal errors in reference star catalogues with the model developed by \citet{Far15} and use only these observations for which corrections can be performed. The resulting orbit solutions are based on 4675 optical observations obtained between 1994-11-29 and 2020-02-20 out of which we discarded 48 observations as outliers. The least-squares orbit solution and the derived orbit sample obtained for Phaethon statistically agree with solution \#712 computed and reported by the Jet Propulsion Laboratory (JPL; top panels in \autoref{fig:clones}) to within $\sim3\sigma$. The remaining, fairly minor, differences in absolute terms between the orbit solutions are likely a consequence of different astrometric weighting schemes and because we did not account for non-gravitational effects, nor did we include radar observations in our solution. We decided to omit non-gravitational forces in orbit computation because it is not clear how comet-like non-gravitational forces should be accurately accounted for in long-term orbital (backward) integrations. That is, we do not know when the observed comet-like activity on Phaethon began or how the activity evolved before the object was discovered.

For UD the entire publicly available astrometric data set comprises a total of 697 optical observations spanning from 1982-11-06 to 2019-10-30. After correcting for zonal errors in reference star catalogues we are left with 668 optical observations over the aforementioned time span. The resulting orbit solutions are based on 666 optical observations which means that we discarded two observations as outliers. The least-squares orbit solution and the derived orbit sample obtained for UD are in statistical agreement with solution \#91 computed and reported by JPL using the same input data (bottom panels in \autoref{fig:clones}).

\subsubsection{Late Dynamical Evolution}

\begin{figure}[h!]
  \centering
  {\LARGE Phaethon}\par\medskip
  \begin{tabular}[b]{c}
	\includegraphics[clip,trim=0.15cm 0.1cm 0.4cm 0.3cm,width=0.5\linewidth]{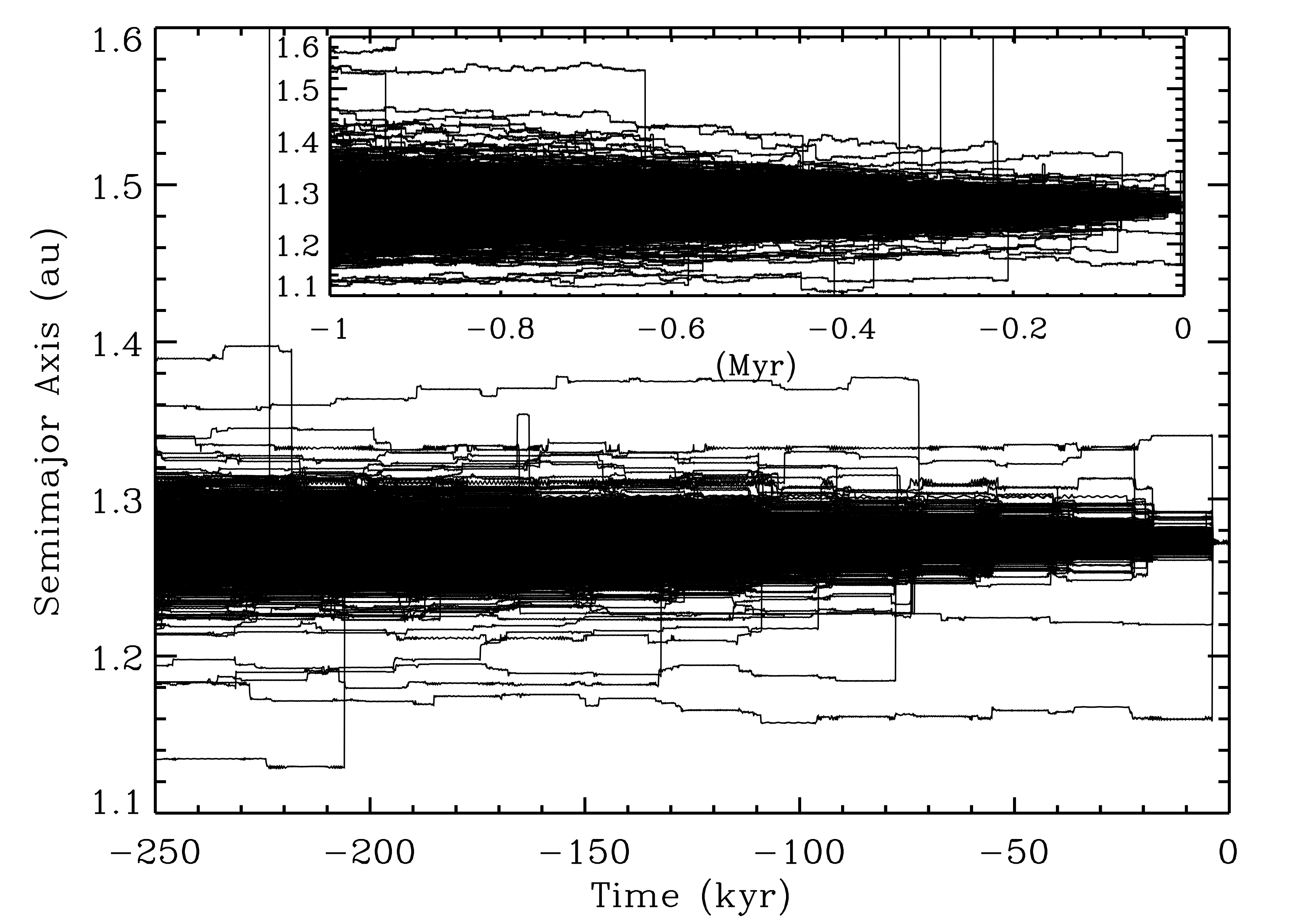}\includegraphics[clip,trim=0.15cm 0.1cm 0.4cm 0.3cm,width=0.5\linewidth]{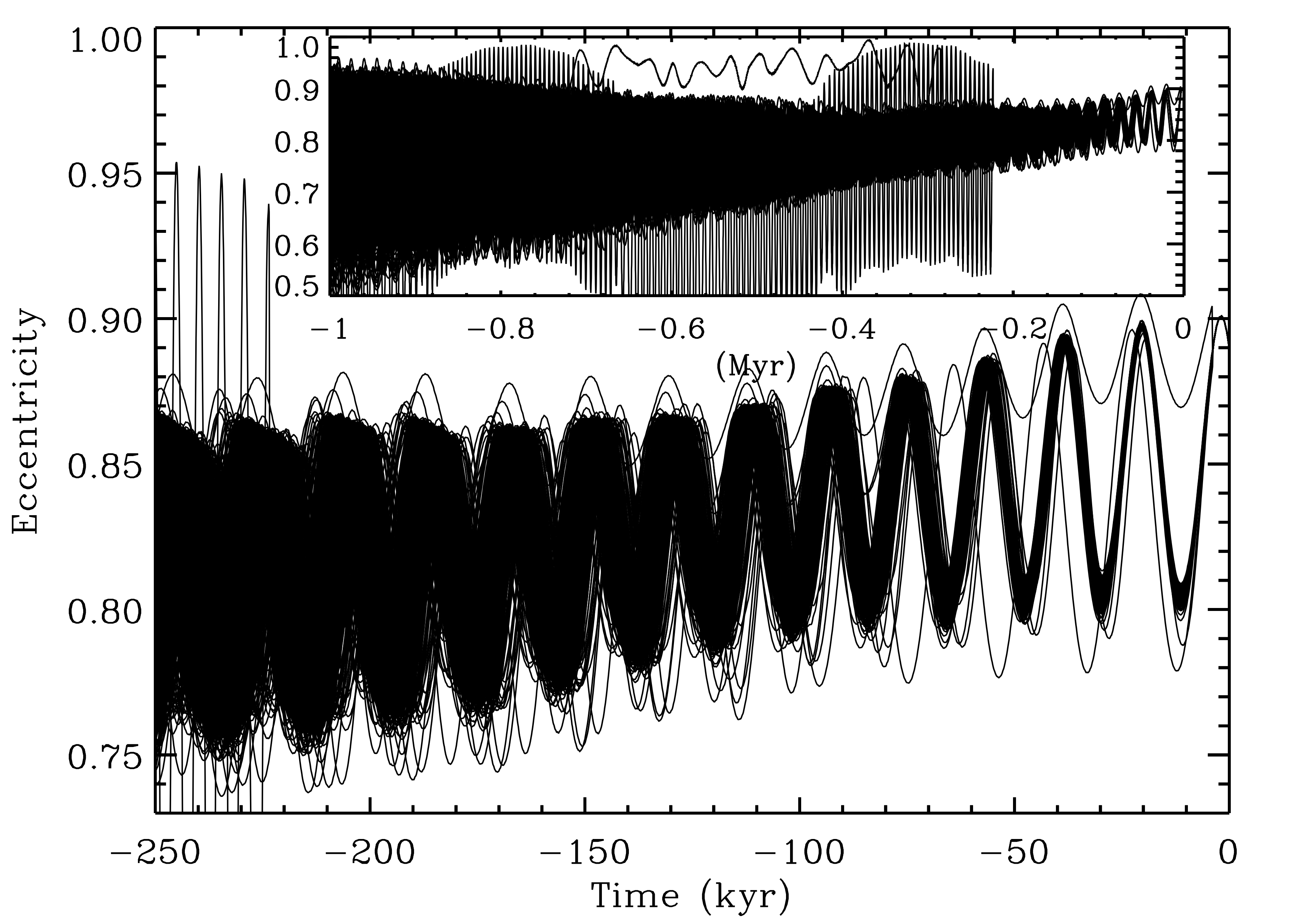} \\
	\includegraphics[clip,trim=0.15cm 0.1cm 0.4cm 0.3cm,width=0.5\linewidth]{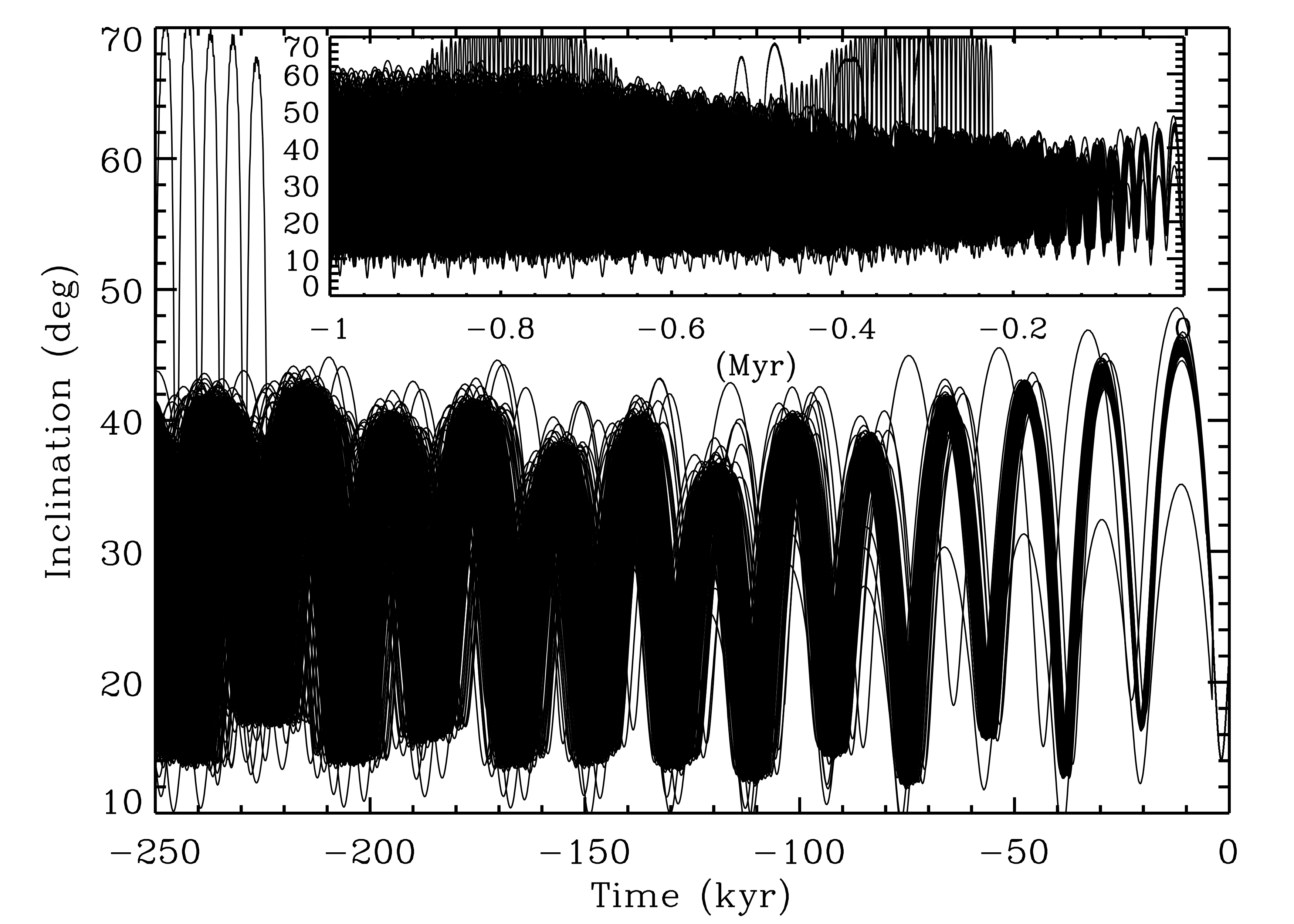}\includegraphics[clip,trim=0.15cm 0.1cm 0.4cm 0.3cm,width=0.5\linewidth]{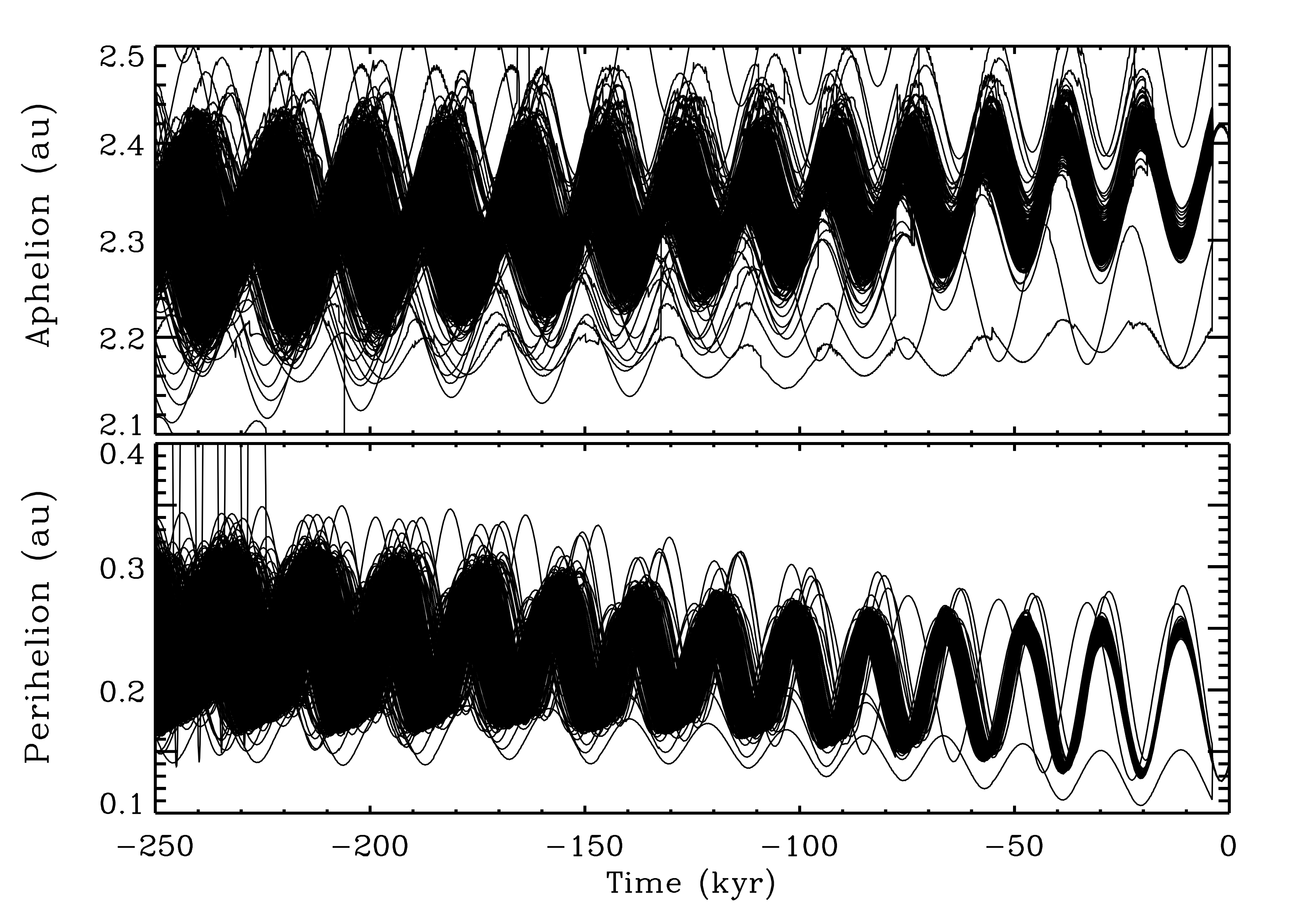}
  \end{tabular}
  \caption{Backward time evolution of the $a$ (top left panel) of 1000 Phaethon-like orbits (nominal Phaethon orbit and 999 additional clones), of $e$ (top right panel), of $i$ (bottom left panel) and of $q$ and $Q$ (bottom right panel). }\label{fig:pha_orbel}
\end{figure}

The late dynamical histories of Phaethon and UD in the inner Solar System are characterized by numerous close encounters with the inner planets. The semi-major axes $a$ of Phaethon clones experience incremental and nearly instantaneous changes during each encounter and diverge from each other as we integrate their orbits further into the past (\autoref{fig:pha_orbel}). This behavior is typical of NEAs in general and can be easily seen in the leftmost panel \autoref{fig:comp}. These jumps in $a$ are, on average, larger in magnitude than the Yarkovsky orbital drift over the same timescale---justifying our decision to omit it from our backwards integration. Because the drift rate measured for Phaethon ($\sim 10^{-4} \au \Myr^{-1}$) is much smaller than the rate of clone divergence over our integrations, including non-gravitational forces such as the Yarkovsky effect would not change the results or interpretation.

It is also apparent that all the clones undergo similar variations in eccentricity $e$ during the last $3-4 \times 10^5 \yr$. The mean value of $e$ oscillates around $\sim 0.85$, a result which is in accordance with \citet{Hanus_etal2016}. Going further back in time, their secular phases begin to mix mainly due to close encounters with the inner planets. When we go even further back in time, the eccentricities of individual clones start diverging, so in practice this means that we cannot meaningfully determine the eccentricity of Phaethon or UD further back in time than $3-4 \times 10^5 \yr$.

As far as the evolution of inclination $i$ is concerned, the secular phases of the clones' orbits also become fully mixed when going further back in time than a few hundreds of thousands of years. If we track the orbits until $t=-5\Myr$, the mean value of $i$ approaches $\sim 33^{\circ}$, which is comparable to the proper inclination of Pallas\footnote{Calculating proper elements for NEOs is not possible using the methods by \citet{Milani1990, Knezevic2000}, because the orbits intersect those of the terrestrial planets. Using the mean inclination as an approximation of the proper inclination is therefore a reasonable alternative.}. However, due to the chaotic nature of Phaethon's orbital evolution this is not enough evidence to establish a statistically-robust connection with Pallas.

\begin{figure}[h!]
  \centering
  \begin{tabular}[b]{c}
	\includegraphics[clip,width=0.5\linewidth]{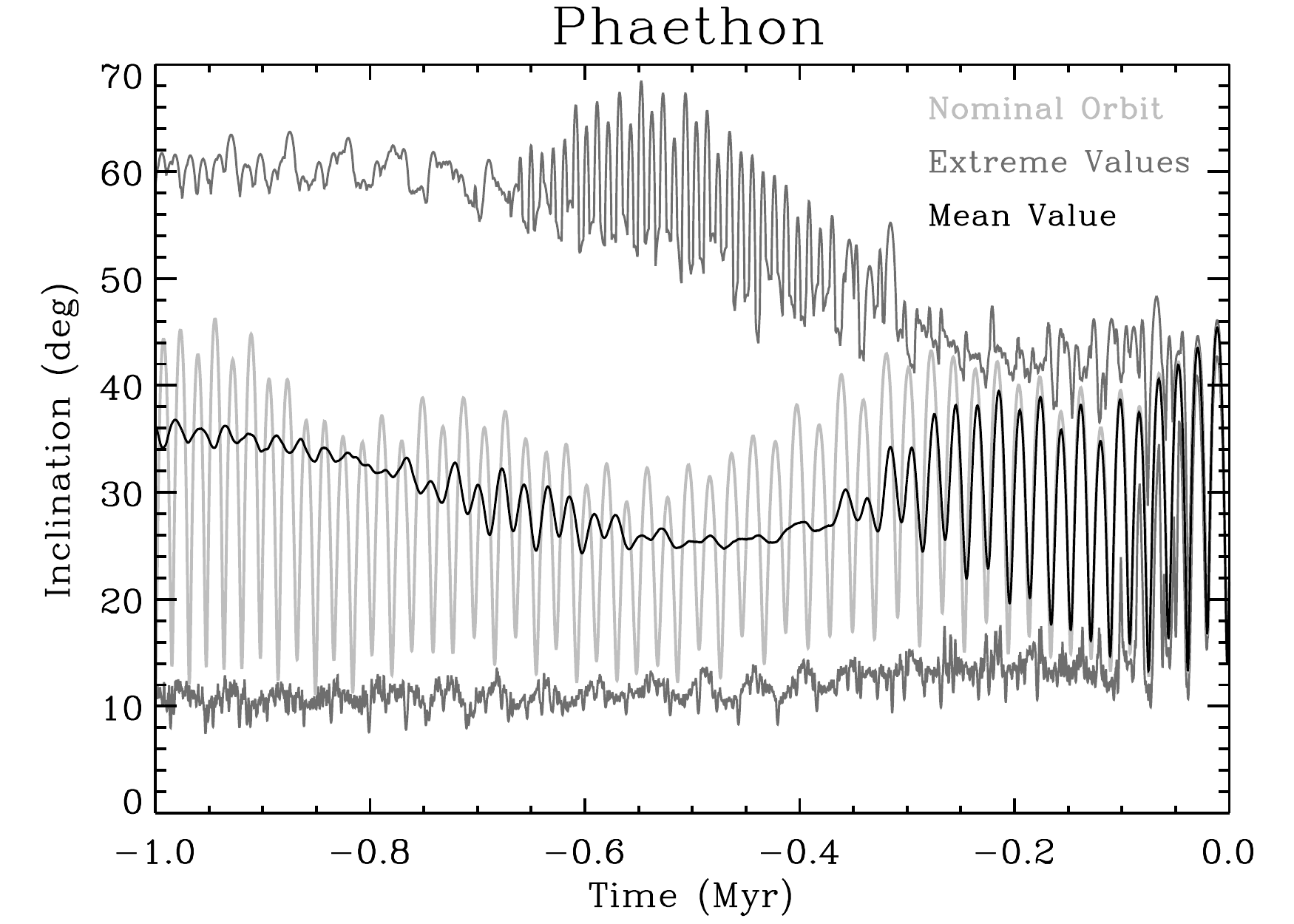}\includegraphics[clip,width=0.5\linewidth]{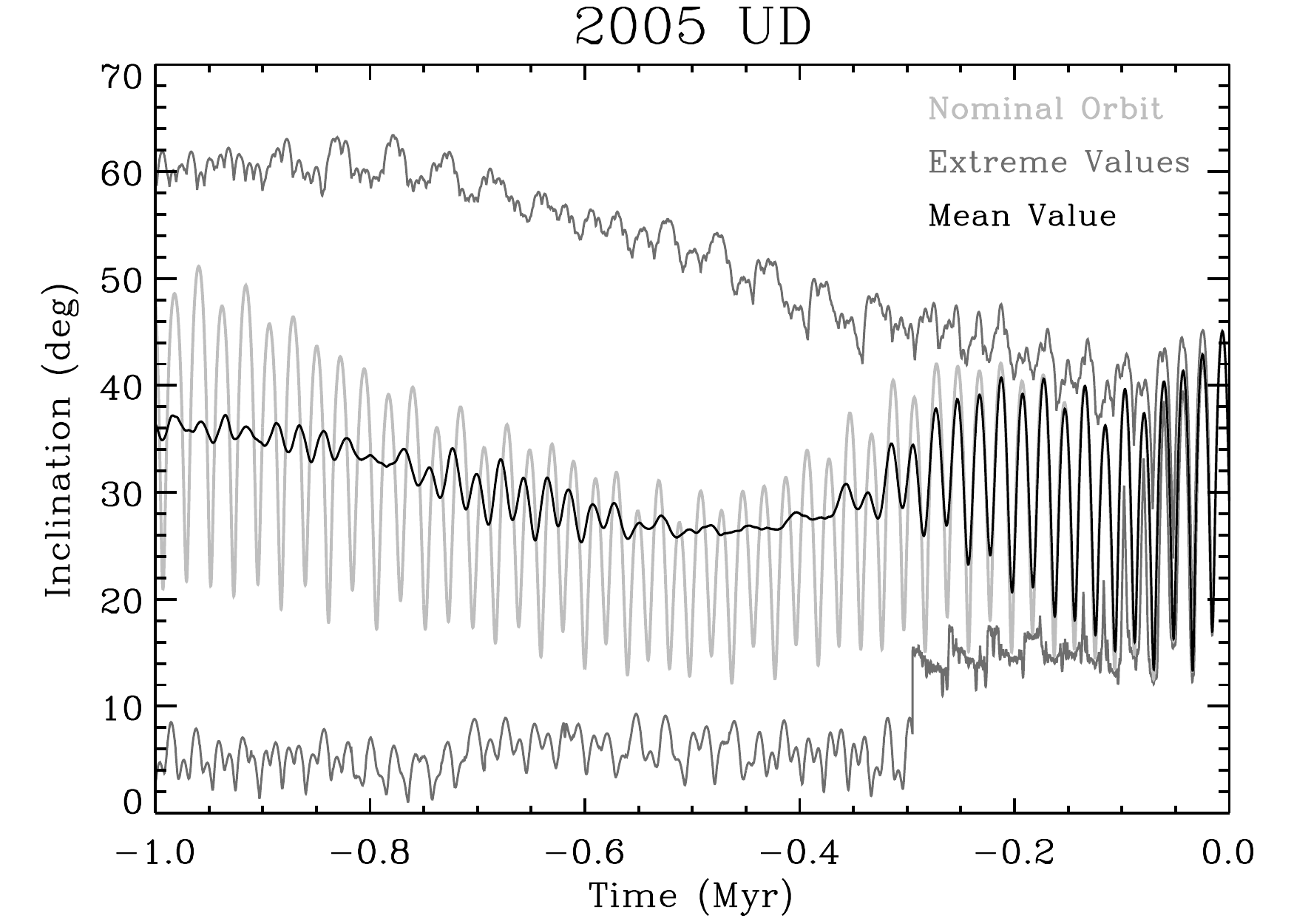}
  \end{tabular}
  \caption{The backward time evolution during $1\Myr$ of the mean value of $i$ (blue) and of the nominal orbit of Phaethon (black). The red lines correspond to the maximum and minimum values that $i$ takes at each specific time. The same for UD in the right panel.}\label{fig:istat}
\end{figure}

\begin{figure}[h!]
  \centering
  {\LARGE 2005 UD}\par\smallskip
  \begin{tabular}[b]{c}
	\includegraphics[clip,trim=0.15cm 0.1cm 0.4cm 0.3cm,width=0.5\linewidth]{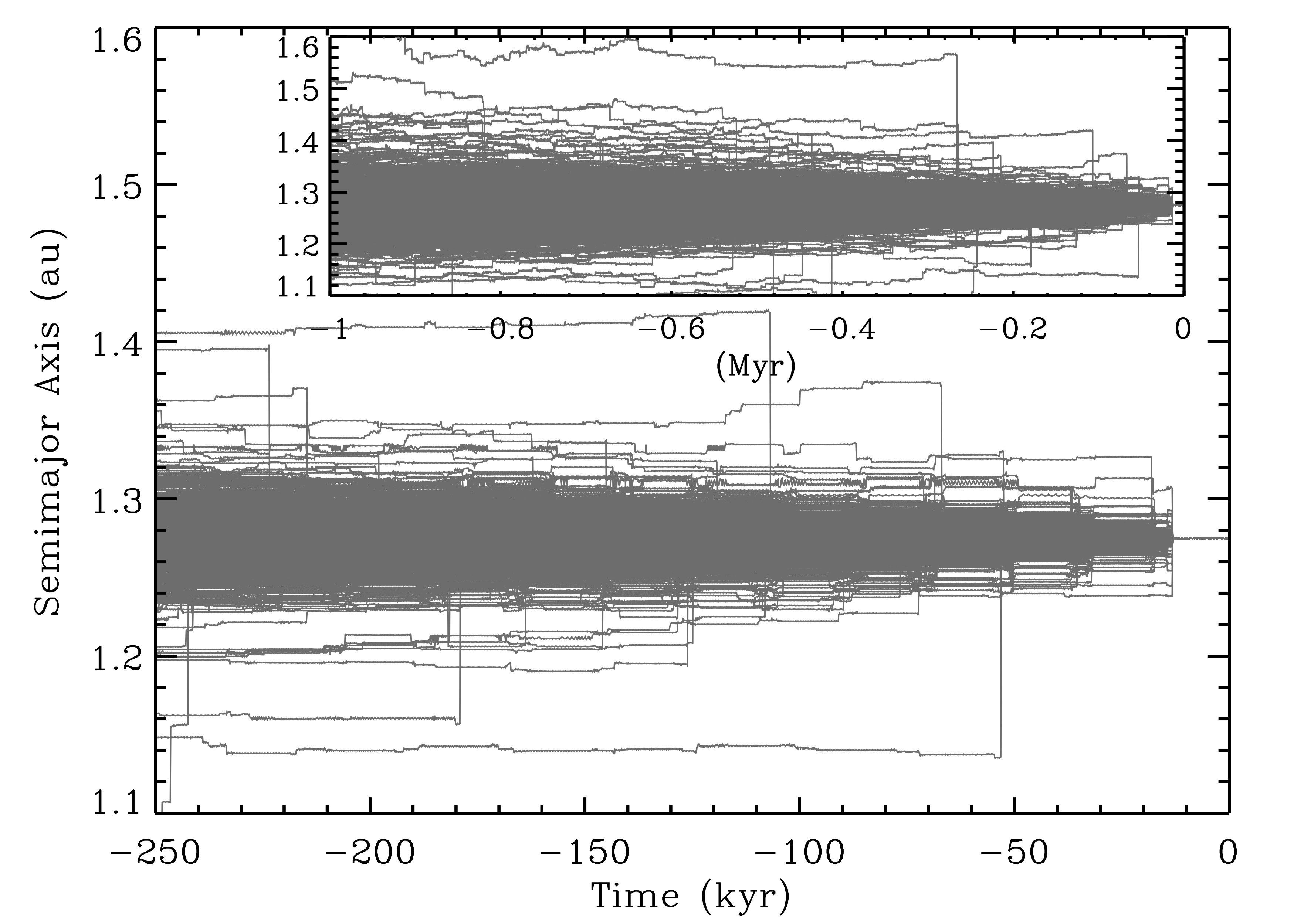}\includegraphics[clip,trim=0.15cm 0.1cm 0.4cm 0.3cm,width=0.5\linewidth]{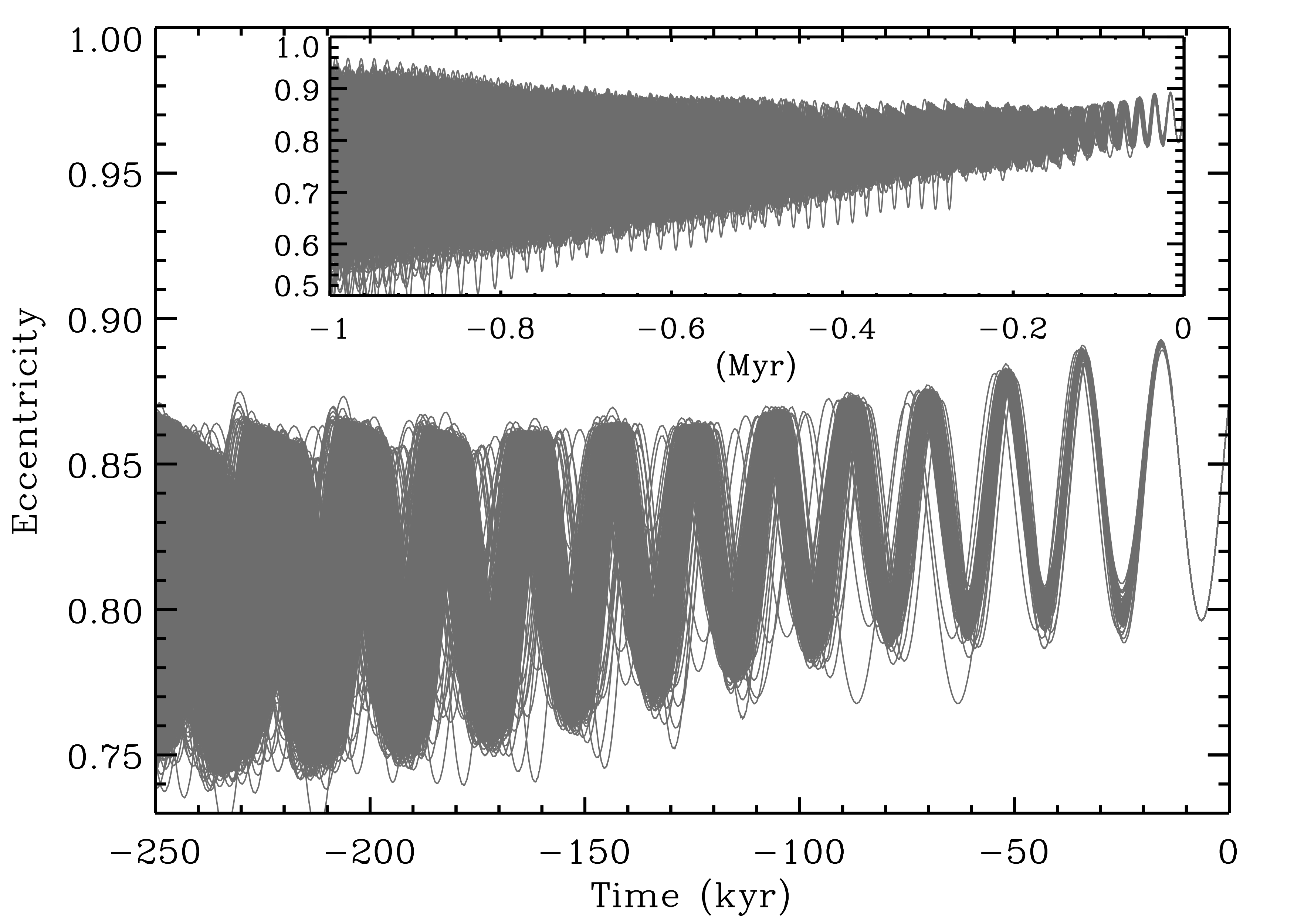} \\
	\includegraphics[clip,trim=0.15cm 0.1cm 0.4cm 0.3cm,width=0.5\linewidth]{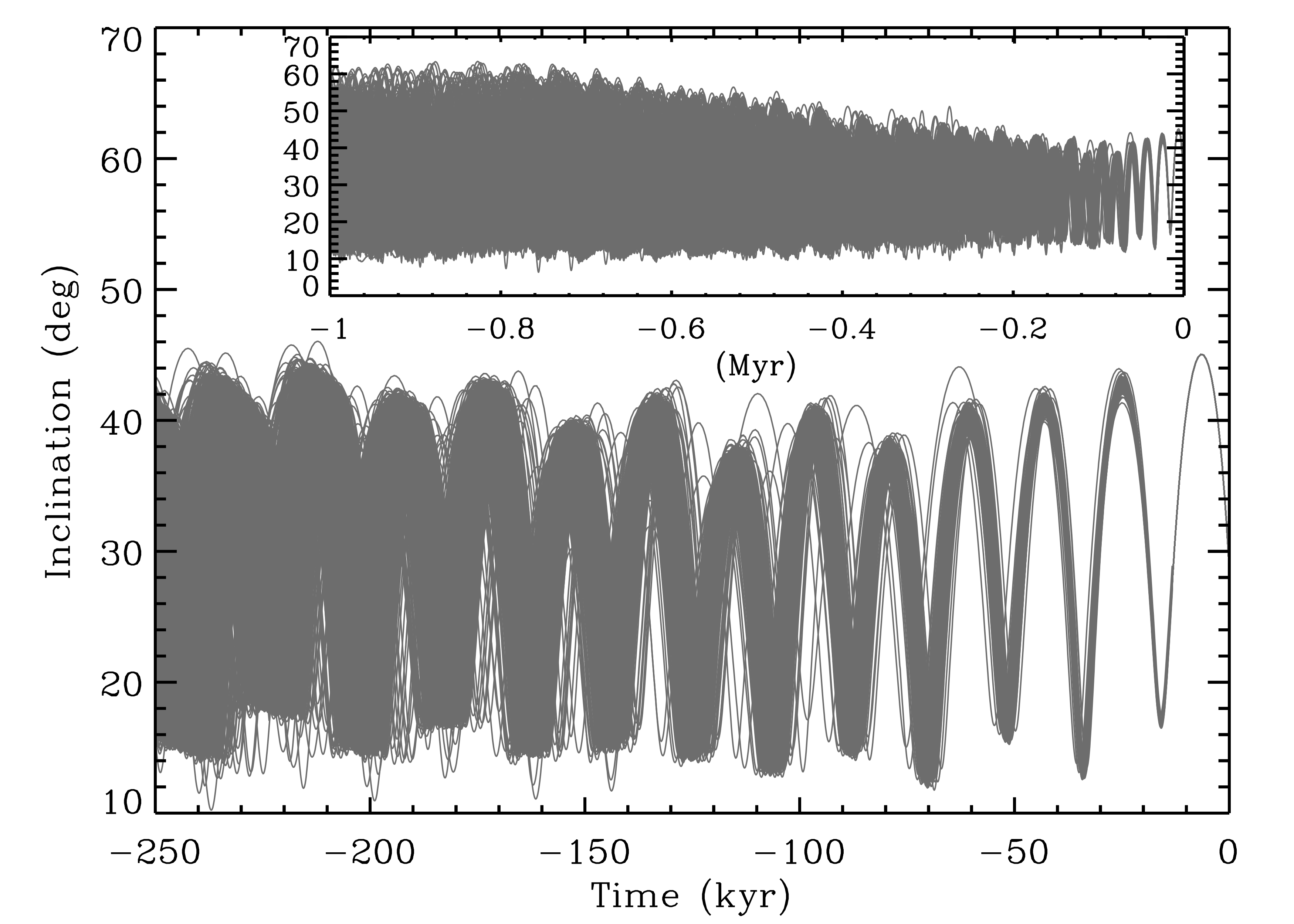}\includegraphics[clip,trim=0.15cm 0.1cm 0.4cm 0.3cm,width=0.5\linewidth]{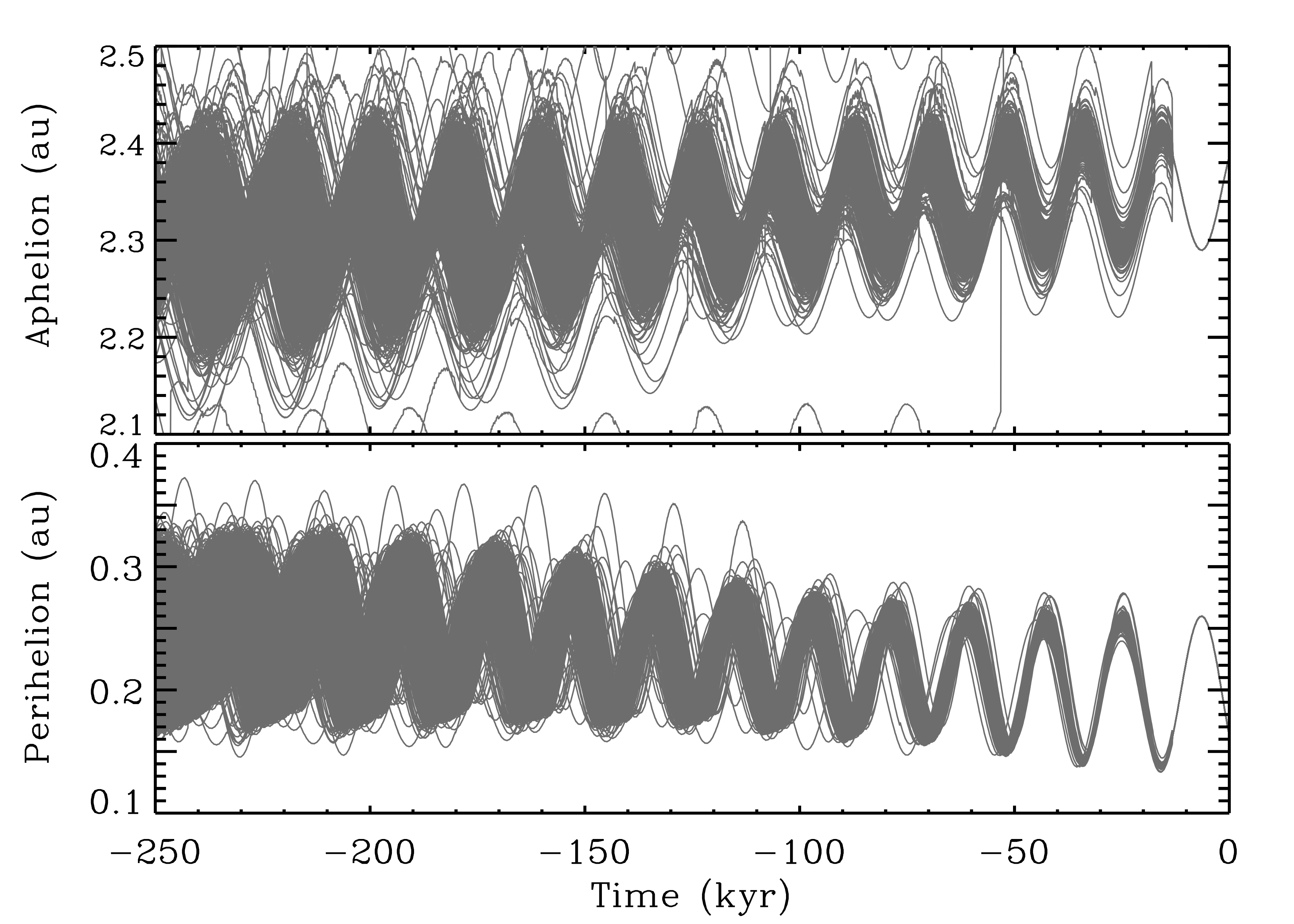}
  \end{tabular}
  \caption{Backward time evolution of $a$ (top left panel) of 1000 UD-like orbits (nominal UD orbit and 999 additional clones), of $e$ (top right panel), of $i$ (bottom left panel) and of $q$ and $Q$ (bottom right panel).}\label{fig:ud_orbel}
\end{figure}

\begin{figure}[h!]
  \centering
  \begin{tabular}[b]{c}
	\includegraphics[clip,width=0.33\linewidth]{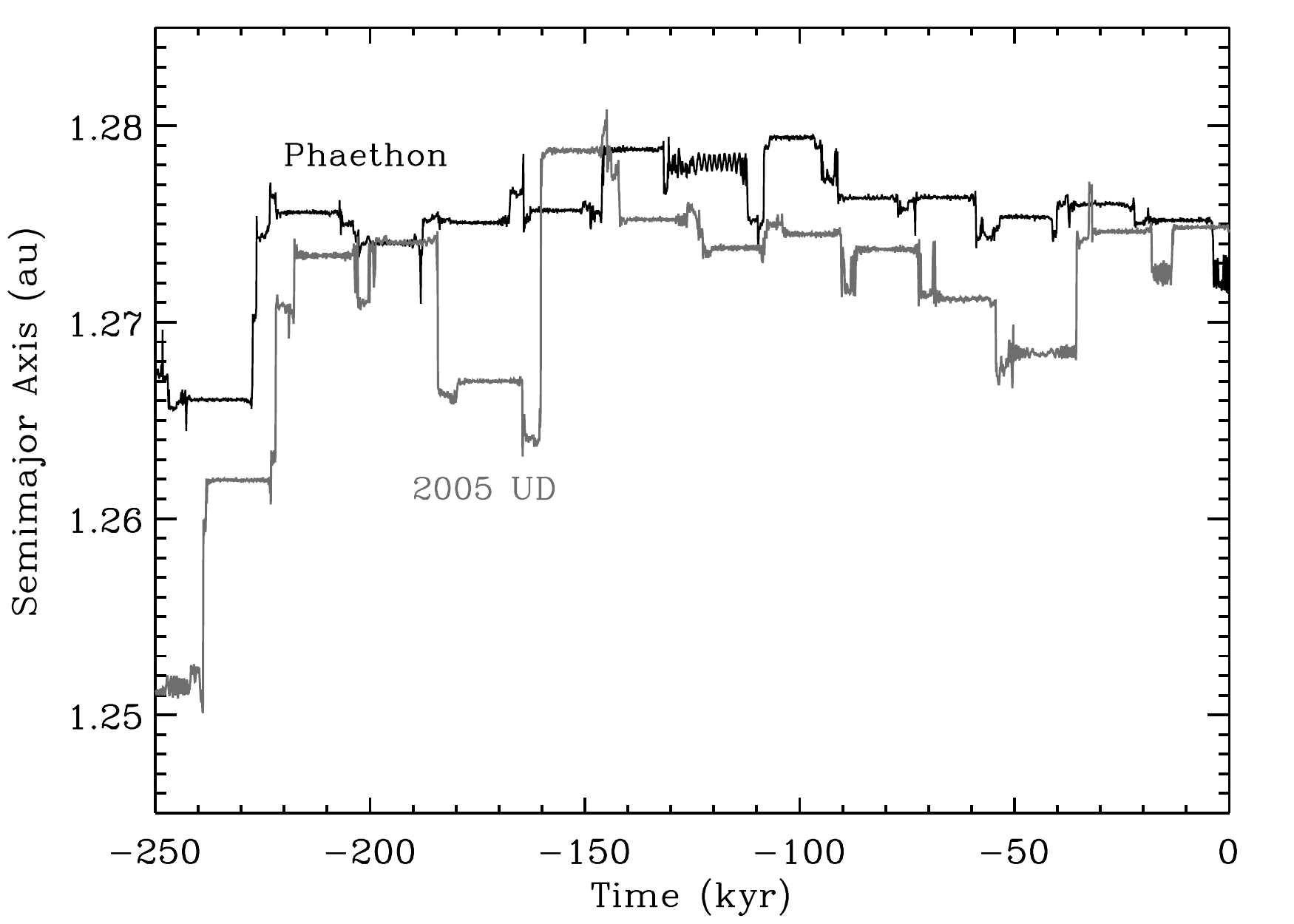}\includegraphics[clip,width=0.33\linewidth]{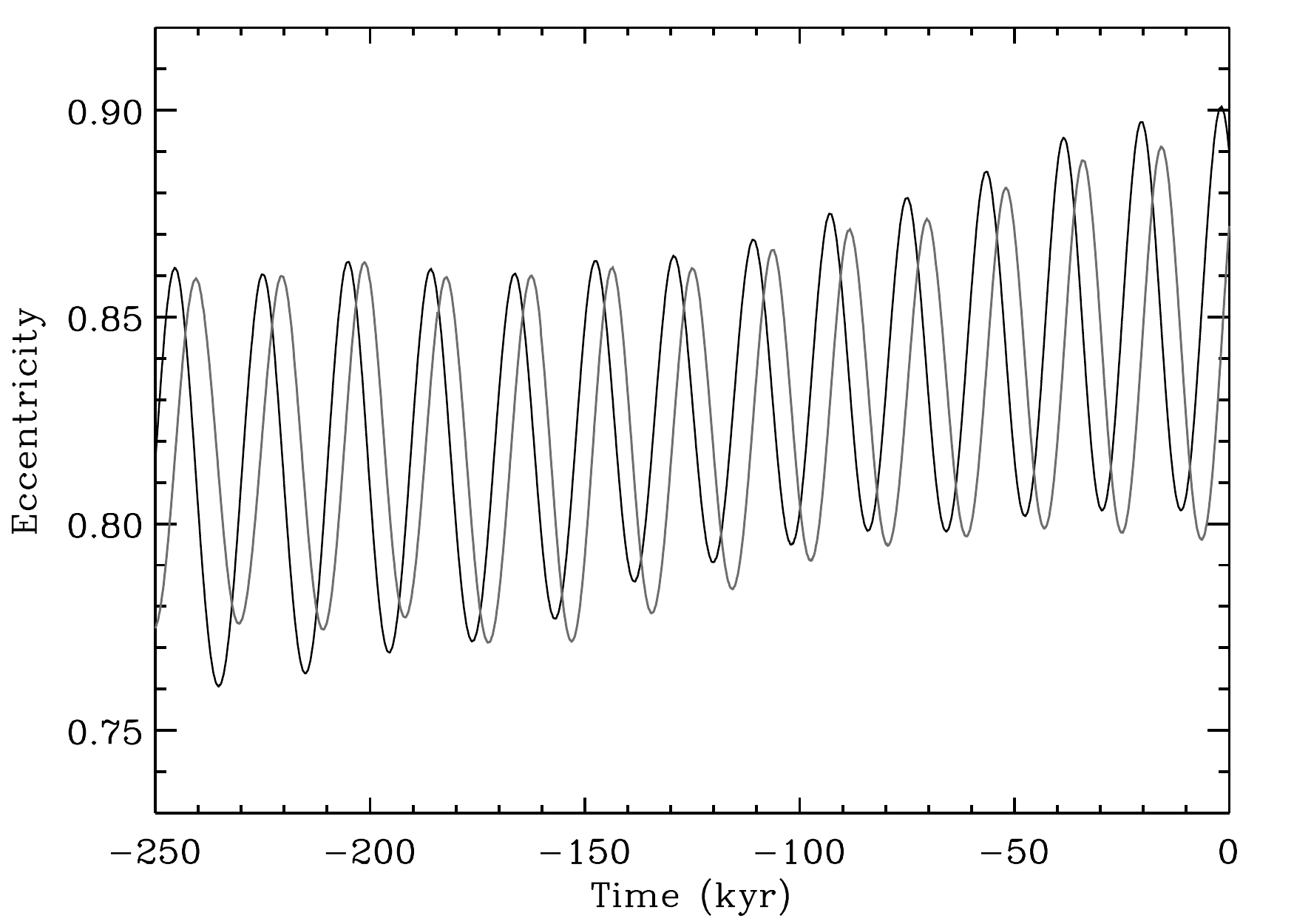}\includegraphics[clip,width=0.33\linewidth]{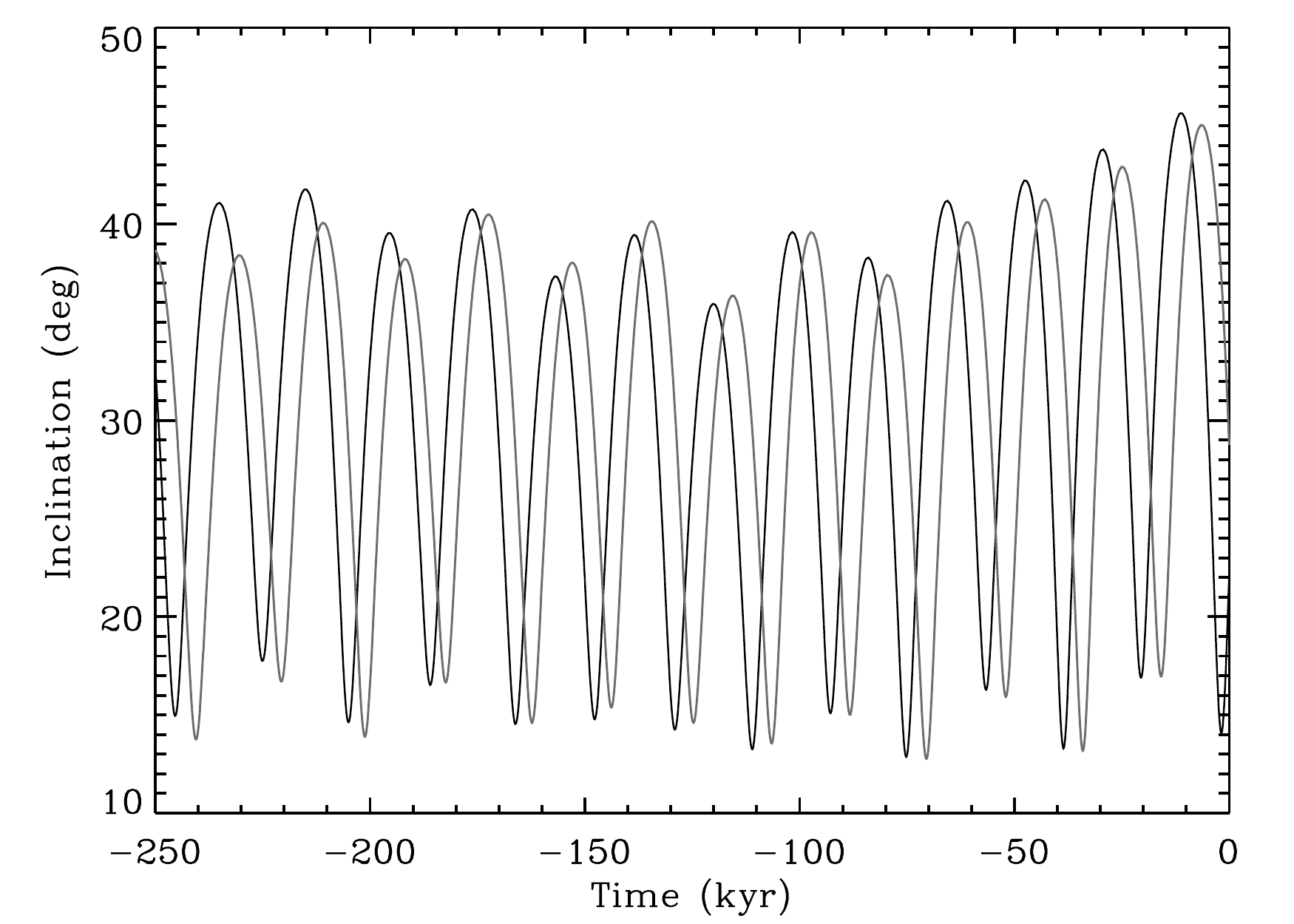}
  \end{tabular}
  \caption{Comparison between the backward time evolution of the $a$ (left), $e$ (middle) and $i$ (right) of the nominal Phaethon and UD orbits.}\label{fig:comp}
\end{figure}

Integrating the orbits of UD and its clones we find a qualitatively similar orbital history as Phaethon (\autoref{fig:ud_orbel}) suggesting a common origin. Currently, UD is at a different secular phase compared to Phaethon (\autoref{fig:comp}). Considering the difference in secular phase, \citet{Hanus_etal2016} propose that these objects must have separated earlier than $100\kyr$ ago. In addition, \citet{Ryabova_etal2019} suggest this fragmentation could not have happened during the last minimum of $q$ since no orbital clone lies within the stream of fragments formed by a breakup of the other body.

\subsubsection{Early Dynamical Evolution and Source Regions}

Direct backward integrations only allow us to obtain meaningful information about Phaethon and UD up to the last few hundred kyr. Hence in order to study the early dynamical evolution---including their likely source regions in the asteroid belt---we take an alternative modeling approach that relies on the integration of orbits forward in time from the asteroid belt into and within the NEO region. In particular, we use the orbital integrations described by \citet{2017A&A...598A..52G}. Their simulation started with an unbiased sample of nearly 80,000 known asteroid orbits in the asteroid belt and includes the Yarkovsky effect. They assumed a typical value for the semi-major axis drift rate of $0.0002\au\Myr^{-1}(D/1\km)^{-1}$, where $D$ is the diameter of the asteroid.

We consider two possibilities for the source of Phaethon and UD: first, that either UD is a fragment of Phaethon which, in turn, is a fragment of Pallas or, second, that both are fragments of Pallas. From the unbiased sample of 80,000 orbits we select those with $|a-2.77\au|<0.05\au$, $|e-0.23|<0.1$, and $|i-34.8\deg|<4\deg$, which cover most of the present-day Pallas family, and thus represent the likely orbital element of Pallas in the past. Focusing on these Pallas-like initial orbits, we find that it takes a $0.1\km$ and $3.0\km$ test asteroid in these Pallas-like orbits $34\pm28\Myr$ and $45\pm29\Myr$, respectively, to reach the NEO region ($q \leq 1.3\au$)---indicating that size does not significantly affect the travel timescale.

We found that, in total, 60 test asteroids with $D=0.1\km$ reach an entrance route to the NEO region, out of which 30 reach the 5:2 mean-motion resonance (MMR) with Jupiter, 16 reach the so-called outer part of the $\nu_6$ secular resonance, and 13 reach the 8:3 MMR with Jupiter. The delivery mechanism remains unspecified for one of the test asteroids. Among the test asteroids with $D=3.0\km$ we find eight that reach the NEO region. Six of these reach the outer part of the $\nu_6$ secular resonance and two reach the 5:2 MMR with Jupiter.

To follow the evolution of the test asteroids to Phaethon-like orbits---$|a-1.271\au|<0.05\au$, $|e-0.890|<0.05$, and $|i-22.260\deg|<1.0\deg$---or to UD-like orbits---$|a-1.274\au|<0.05\au$, $|e-0.872|<0.05$, and $|i-28.675\deg|<1.0\deg$---we use the continuation of the aforementioned orbital integrations in NEO space \citep{2016Natur.530..303G,2018Icar..312..181G}. To account for the destruction at small perihelion distance we consider two different critical perihelion distances, $0.005\au$ (essentially an impact with the Sun) and $0.076\au$ (average super-catastrophic destruction distance). As a result of cloning test asteroids entering the NEO region through the outer part of the $\nu_6$ secular resonance or the 8:3 MMR with Jupiter, a total of 204 test asteroids can be linked back to Pallas-like initial orbits. Not a single one of these 204 test asteroids reach Phaethon-like or UD-like orbits when integrated in the NEO region. In fact, there are only 3--4 test asteroids in each scenario that enter the NEO region through the 5:2 MMR complex which covers the region where the Pallas family resides (Tables \ref{tab:phaethonsource} and \ref{tab:udsource}).

To better understand the source regions for Phaethon and UD we compute the relative likelihoods of the source regions or escape routes from the asteroid belt as defined in \citet{2018Icar..312..181G}. The results suggest that it is unlikely that Phaethon or UD would have resided on a Pallas-like orbit in the asteroid belt, because the likelihood for the 5:2 MMR complex is at most around 5\% (Tables \ref{tab:phaethonsource} and \ref{tab:udsource}). An inner asteroid belt source appears more likely based on the dynamical evidence alone.

The typical lifetimes for test asteroids in the NEO region ($q<1.3\au$) until they reach either a Phaethon-like or UD-like orbit range from about $10\Myr$ to about $80\Myr$ depending on the entrance route or source region (Tables \ref{tab:phaethonsource} and \ref{tab:udsource}). For the $\nu_6$ resonance complex and the Hungaria group, that is, the most likely candidates to explain the origin of both Phaethon and UD, their typical lifetimes range from $20\Myr$ to $80\Myr$. Similar timescales are also found for outer-belt sources.

\begin{table}[h]
    \caption{Source-region or escape-route probabilities for Phaethon according to the model by \citet{2018Icar..312..181G}, and the time ($\Delta t$) from $q=1.3\au$ until $(a,e,i)$ are similar to Phaethon's current orbit. For $\Delta t$ we provide the 10\%, 50\% (that is, median, in bold), and 90\% quantiles, and $N$ is the number of test asteroids used for the analysis. Two different assumptions are used for the critical distance at which asteroids are destroyed close to the Sun: $q=0.005\au$ and $q=0.076\au$ (see text for further details).}
    \label{tab:phaethonsource}
    \centering
    \begin{tabular}{cr|cr|cr}
     & Probability & \multicolumn{4}{c}{Evolution time} \\
        Source/Escape & [\%] & $N_{0.005}$ & $\Delta t_{0.005}$ [Myr] & $N_{0.076}$ & $\Delta t_{0.076}$ [Myr] \\
        \hline
        Hungaria        & $18.5\pm3.7$  & 409 & 13.6/{\bf 77.9}/238  & 258 & 13.3/{\bf 71.0}/195 \\
        $\nu_6$ complex & $64.2\pm10.6$ & 411 & 6.7/{\bf 21.2}/81.5  & 280 & 6.3/{\bf 18.8}/60.8 \\
        3:1J complex    &  $9.5\pm2.6$  &  23 & 6.0/{\bf 25.2}/86.7  &  13 & 6.0/{\bf 10.2}/43.7 \\
        Phocaea         &  $2.5\pm0.9$  &  44 & 17.6/{\bf 62.3}/202  &  21 & 14.9/{\bf 53.9}/188 \\
        5:2J complex    &  $5.4\pm1.5$  &   4 & 17.5/{\bf 19.0}/56.6 &   3 & 19.0/{\bf 38.0}/56.6 \\
        2:1J complex    & -- & -- & -- & -- & -- \\
        JFC             & -- & -- & -- & -- & -- \\
        \hline
    \end{tabular}
\end{table}

\begin{table}[h]
    \caption{Source-region or escape-route probabilities for UD according to the model by \citet{2018Icar..312..181G}, and the time ($\Delta t$) from $q=1.3\au$ until $(a,e,i)$ are similar to UD's current orbit. For $\Delta t$ we provide the 10\%, 50\% (that is, median, in bold), and 90\% quantiles, and $N$ is the number of test asteroids used for the analysis. Two different assumptions are used for the critical distance at which asteroids are destroyed close to the Sun: $q=0.005\au$ and $q=0.076\au$ (see text for further details).}
    \label{tab:udsource}
    \centering
    \begin{tabular}{cr|rc|rc}
     & Probability & \multicolumn{4}{c}{Evolution time} \\
        Source/Escape & [\%] & $N_{0.005}$ & $\Delta t_{0.005}$ [Myr] & $N_{0.076}$ & $\Delta t_{0.076}$ [Myr] \\
        \hline
        Hungaria        & $17.5\pm2.8$ & 427 & 14.9/{\bf 81.6}/246  & 260 & 14.4/{\bf 75.3}/200 \\
        $\nu_6$ complex & $72.4\pm4.2$ & 399 & 7.9/{\bf 24.1}/86.4  & 248 & 6.9/{\bf 21.4}/69.9 \\
        3:1J complex    & $9.9\pm0.9$  &  21 & 6.0/{\bf 25.1}/78.7  &  11 & 6.0/{\bf 10.1}/35.6 \\
        Phocaea         & $0.1\pm0.6$  &  60 & 17.2/{\bf 59.1}/194  &  30 & 8.2/{\bf 39.2}/125 \\
        5:2J complex    & $0.2\pm0.2$  &   3 & 19.1/{\bf 38.1}/42.0 &   3 & 19.1/{\bf 38.1}/42.0 \\
        2:1J complex    & -- & -- & -- & -- & -- \\
        JFC             & -- & -- & -- & -- & -- \\
        \hline
    \end{tabular}
\end{table}

\begin{figure}[h!]
    \centering
    \includegraphics[width=0.45\textwidth]{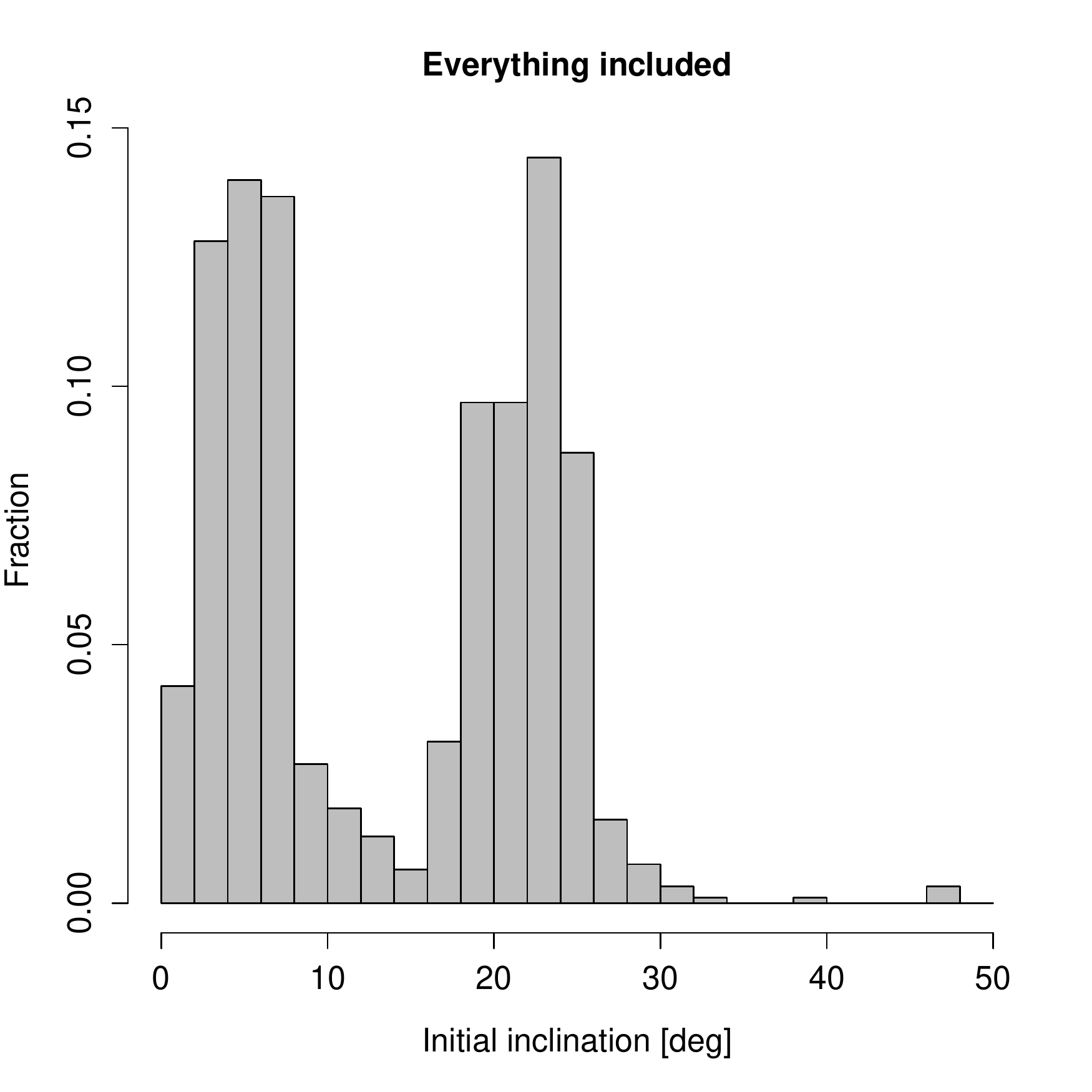}
    \includegraphics[width=0.45\textwidth]{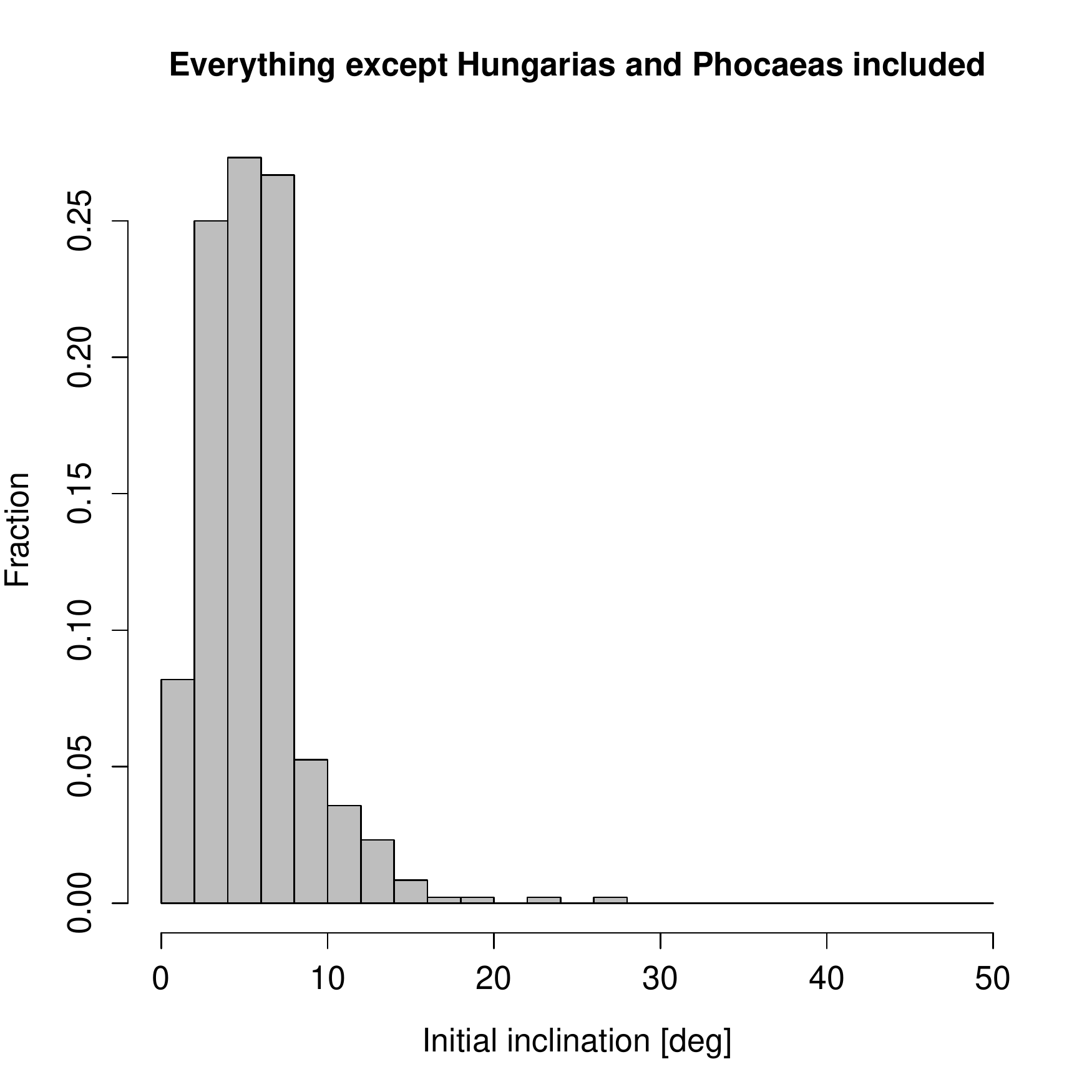}
    \caption{Inclination distribution for the initial orbits in the asteroid belt that eventually evolve to Phaethon-like orbits. The left plot includes test asteroids from all source regions and escape routes and for the right plot we removed test asteroids belonging to the Hungaria and Phocaea groups.}
    \label{fig:initi}
\end{figure}

Considering the possibility that Phaethon and UD originated from the inner Main Belt, we now ask: does the relatively high inclination of Phaethon place a limitation on the inclination of the origin? An examination of the initial inclination distribution of test asteroids that eventually evolve into Phaethon-like orbits shows a range from about $0\deg$ to $50\deg$ (left plot in \autoref{fig:initi}). Excluding the Hungaria and Phocaea groups as sources for Phaethon (right plot in \autoref{fig:initi}), the inclination range reduces to about $0\deg$ to $30\deg$ with more than 90\% of the test asteroids having $i<10\deg$. The small semi-major axes of Phaethon and UD implies that these objects have evolved for a relatively long time in the region of the terrestrial planets and they have thus had plenty of time for close planetary encounters that have changed their inclinations. Hence we posit that focusing on the recent average inclination for Phaethon and UD is not very useful for determining their source region or parent body. We cannot exclusively rule out a high-inclination source, however.

\subsection{Application of orbTPM to Phaethon and UD}

We ran the orbTPM for two sets input parameters that represent the nominal surface properties of Phaethon and UD: $A_\mathrm{b}$, $\Gamma$, $P_\mathrm{rot}$, and spin axis orientation, which is expressed in terms of the obliquity ($\epsilon$) and solstice point ($\lambda_\nu$)---that is, the true anomaly at which northern hemisphere experiences summer solstice (\autoref{tab:TPMinput}). For each object, the orbTPM was run for a predefined set of semi-major axes ($a$) and eccentricities ($e$): values of $a$ range from $1.1\au$ to $1.6\au$ using increments of $0.1\au$, and values of $e$ range from 0.50 to 0.95 with increments of 0.05. Additionally, calculations are also executed using the current nominal orbits of both objects. The effect of the spin obliquity on surface temperatures is apparent when comparing the results, particularly near perihelion and aphelion in the polar regions.

\begin{table}[h]
\caption{Input Thermophysical Parameters}
\label{tab:TPMinput}
\begin{center}
\begin{tabular}[h!]{r|cccccc}

 & $A_\mathrm{b}$ & $\Gamma (\J \meter^{-2}\K^\textrm{-1} \second^{-1/2})$ & $\lambda_\nu$ ($\deg$) & $\epsilon$ ($\deg$) & $P_\mathrm{rot}$ ($\hours$) & References \\
\hline
Phaethon & 0.048 & 600 & {90, 0, \textbf{270}} & \textbf{30} & 3.6 & \citet{Hanus_etal2016,Hanus_etal2018}\\
2005 UD & 0.052 & 300 & 90, {\it 0}, 260 & {\it75} & 5.2 & \citet{Devogle_etal20,Huang_etal20}

\end{tabular}
\end{center}
\end{table}

\begin{figure}[h!]
  \centering
  \begin{tabular}[b]{c}
	\includegraphics[clip,trim=0.4cm 1.35cm 0 0,width=0.5\linewidth]{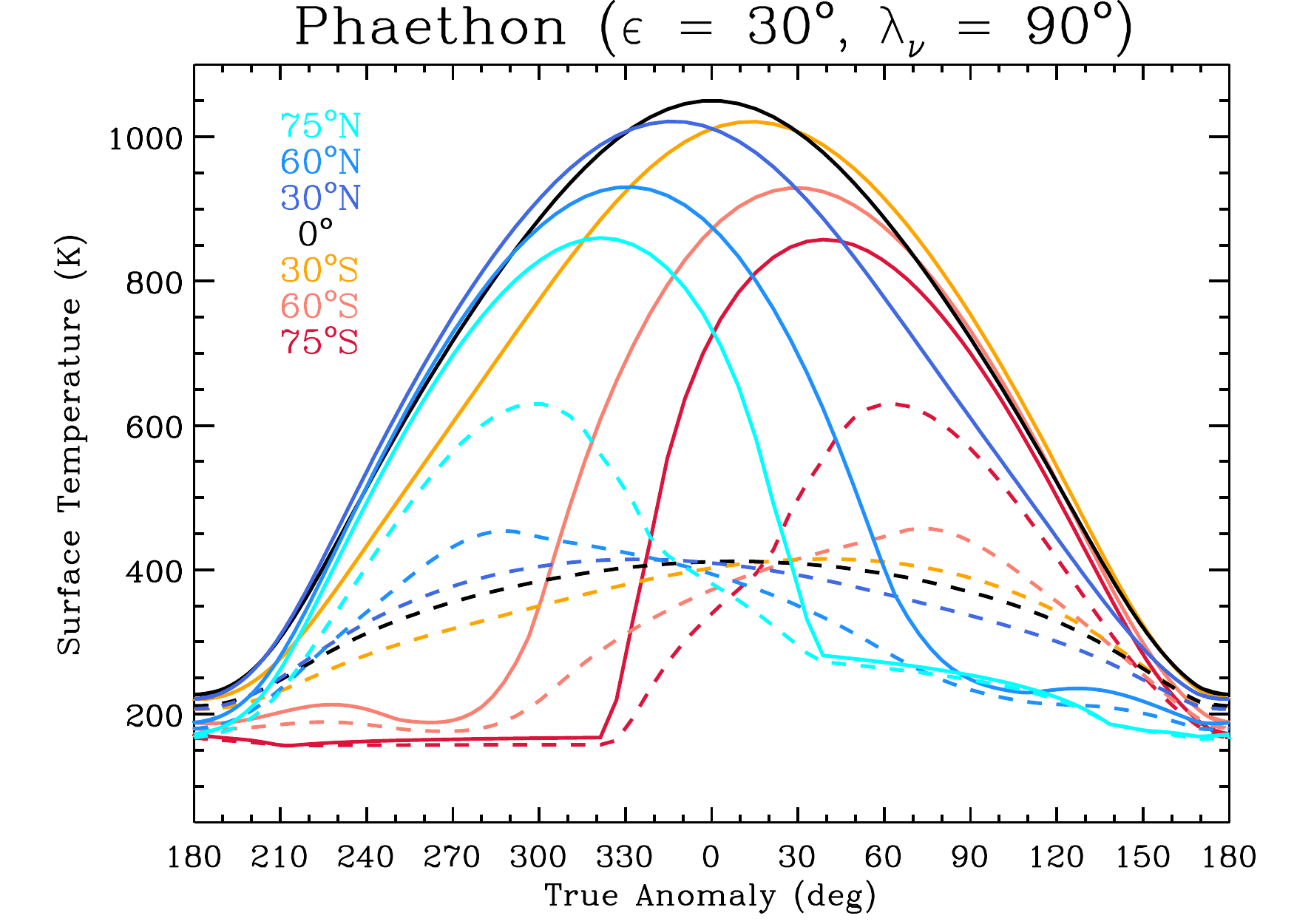}\includegraphics[clip,trim=0.4cm 1.35cm 0 0,width=0.5\linewidth]{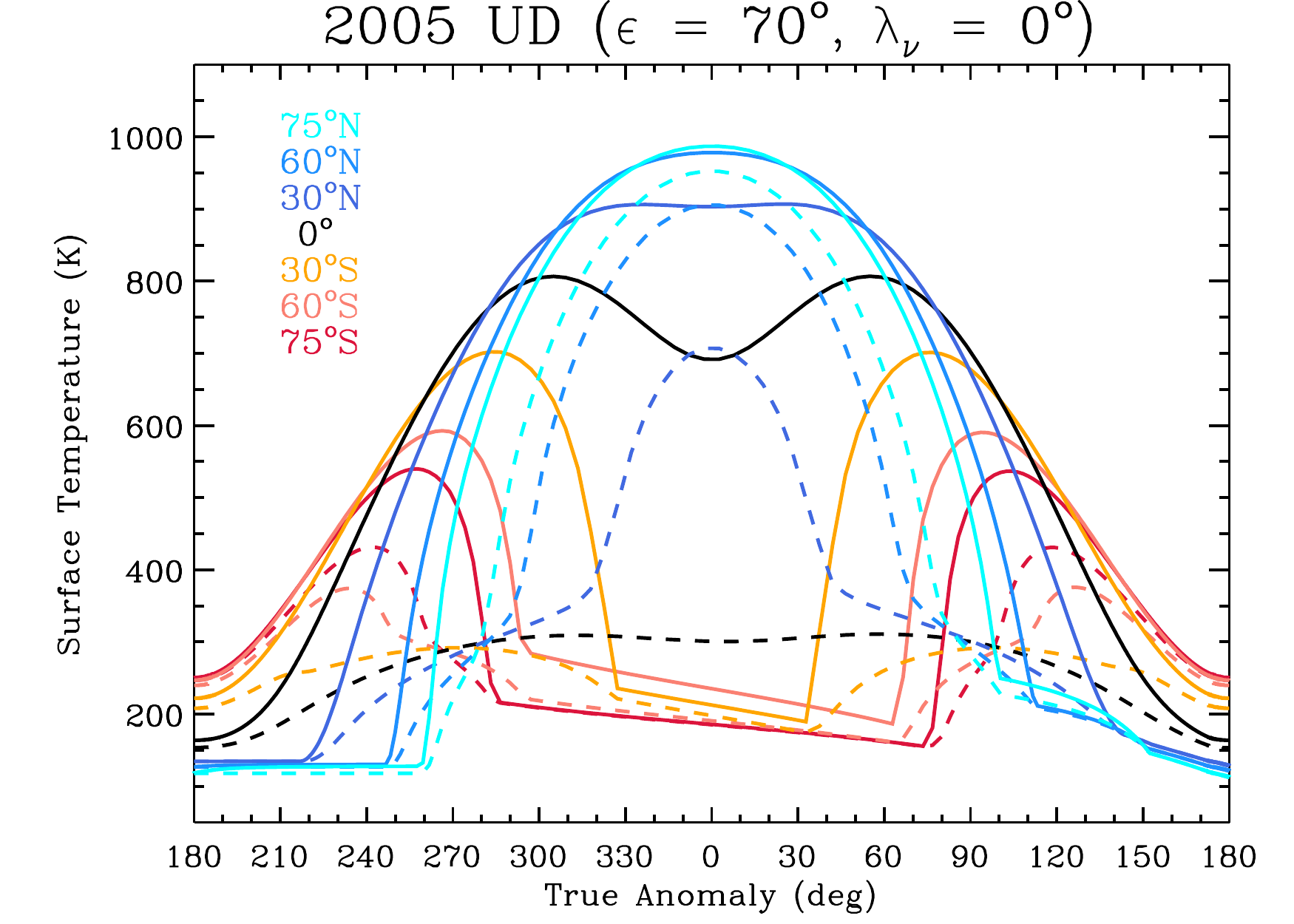} \\
	\includegraphics[clip,trim=0.4cm 1.35cm 0 0.8cm,width=0.5\linewidth]{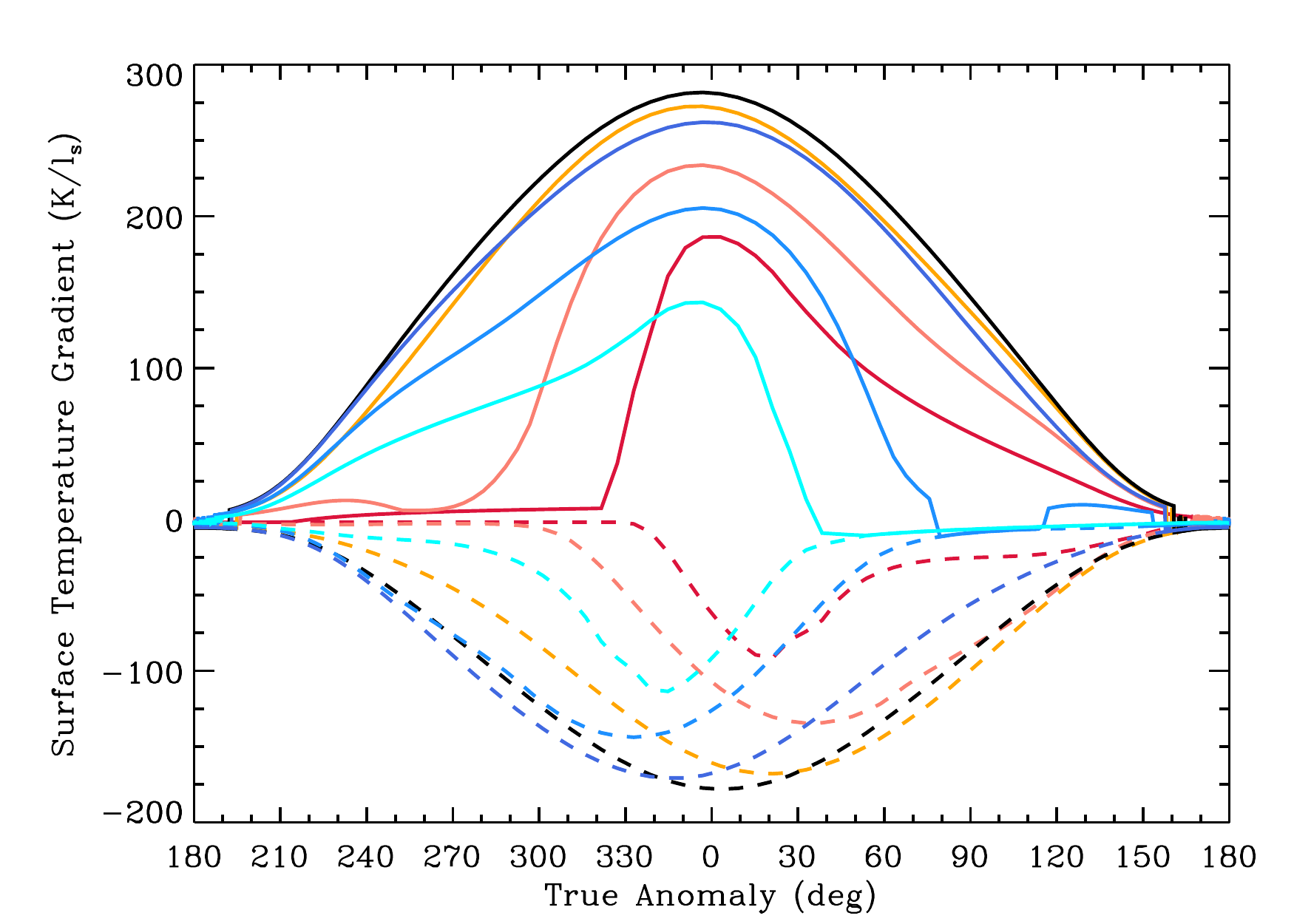}\includegraphics[clip,trim=0.4cm 1.35cm 0 0.8cm,width=0.5\linewidth]{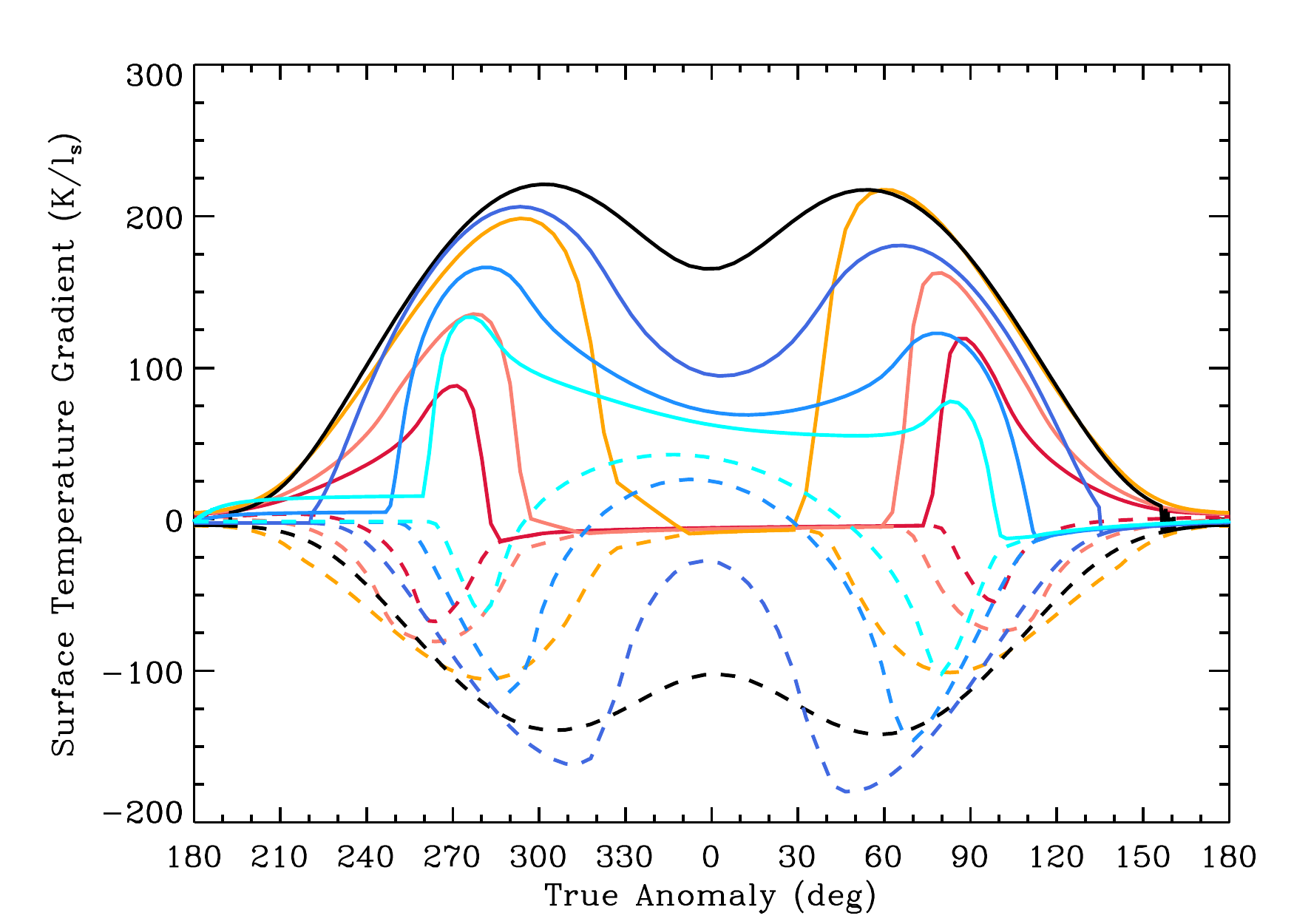} \\
	\includegraphics[clip,trim=0.4cm 0cm 0 0.8cm,width=0.5\linewidth]{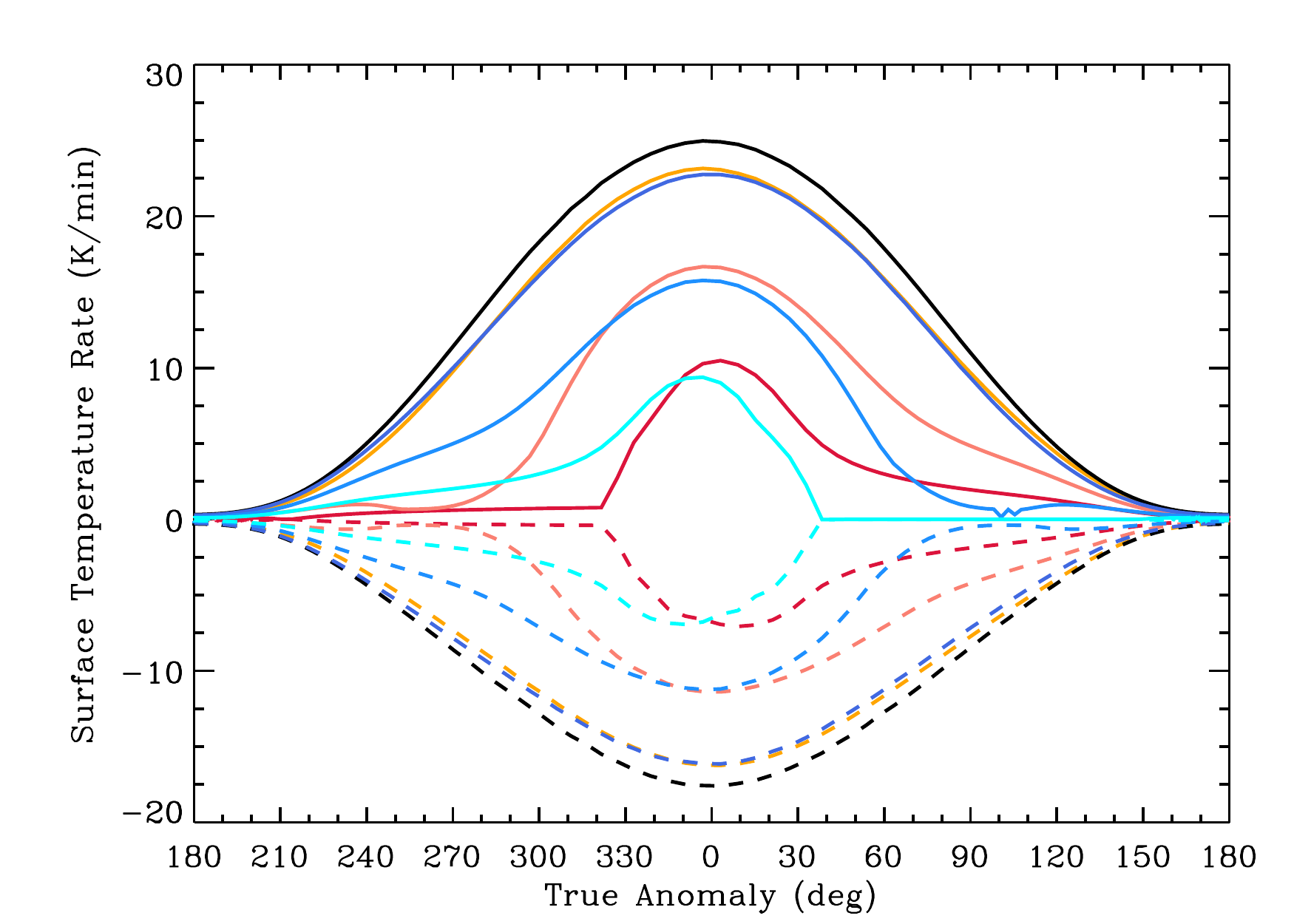}\includegraphics[clip,trim=0.4cm 0cm 0 0.8cm,width=0.5\linewidth]{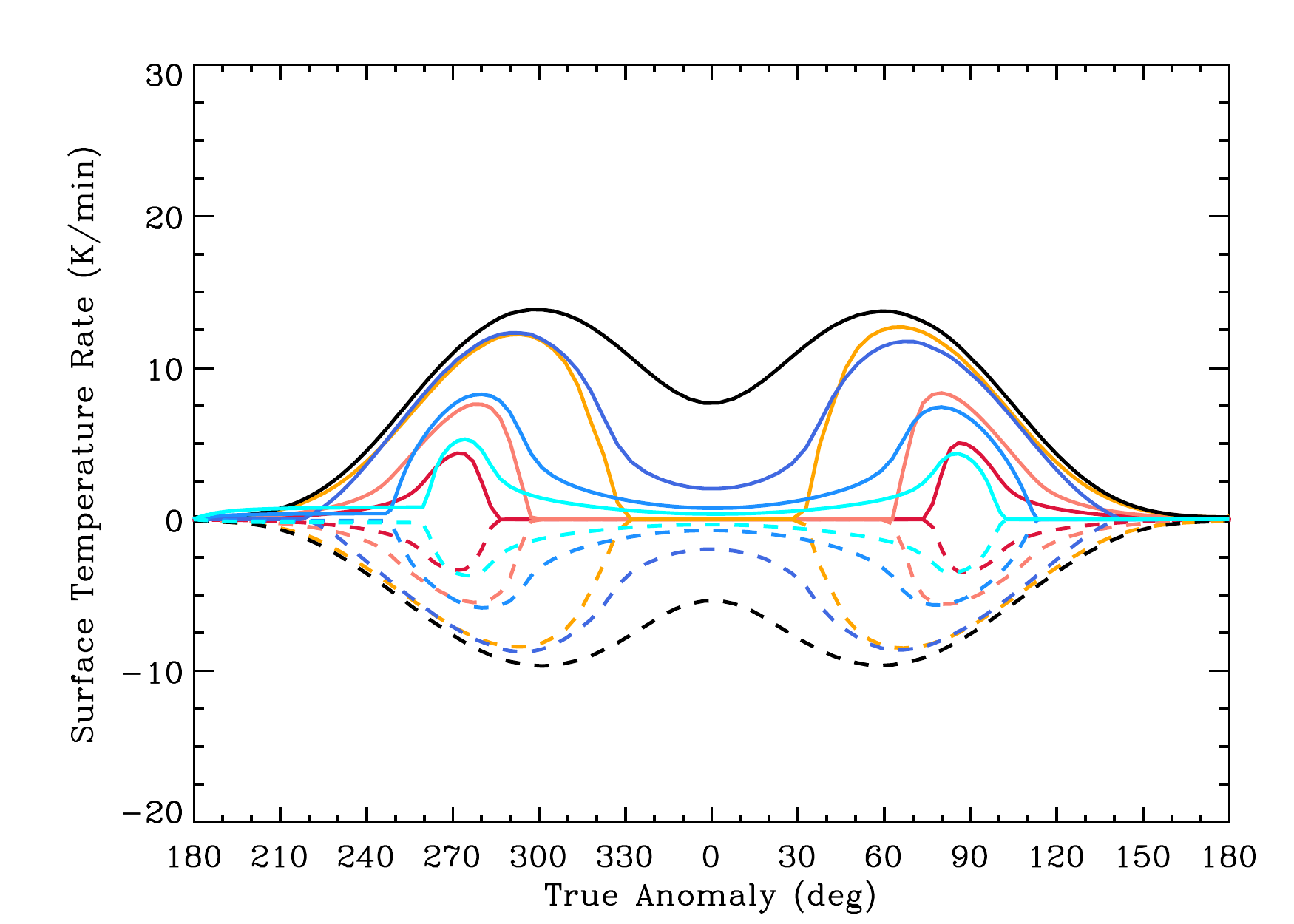}
  \end{tabular}
  \caption{Maximum (solid lines) and minimum (dashed lines) surface temperatures (top panels), thermal gradients (middle panels), and rates of temperature change (bottom panels) as a function of true anomaly at different latitudes for Phaethon (left panels) and UD (right panels).}\label{fig:tpmoutput}
\end{figure}

The calculated temperatures resulting from one run of the orbTPM contains a large amount of information about the temperature characteristics of an object such as surface and sub-surface temperatures across several latitudes and for several thousand points along the orbit. This can be seen in \autoref{fig:tpmoutput}, which depicts the maximum and minimum values of surface temperature, thermal gradient, and temperature rate of change for different latitudes. In order to distill this information and best characterize the temperature characteristics we calculate a few diagnostic parameters for each latitude value. In particular, we select the extrema (maximum and minimum) of: 1) the surface temperature, $T$, 2) the thermal gradient, $dT/dx$, (calculated as the temperature difference between the surface and one diurnal skin depth), and 3) the time rate of change in surface temperature, $dT/dt$. Each of these three parameters is calculated for 1000 points evenly spaced in time throughout the orbit of the object.

\subsubsection{Analysis of Current Temperature Characteristics}\label{sub:tpmanalysis}

The surface properties and current orbital configuration of Phaethon and UD result in both similar and contrasting temperature characteristics. We depict various parameters, as discussed in greater detail below, as a function of true anomaly ($\nu$) in \autoref{fig:tpmoutput}. Since UD's spin axis is not constrained well enough, we analyze the case for which solstice occurs at perihelion (italicized values in \autoref{tab:TPMinput}) in order to contrast with Phaethon's spin pole, and to investigate the temperature dependencies on spin axis orientation. The best estimate for Phaethon's spin axis orientation implies a solstice point of $\lambda_\nu \approx 270\deg$ ($90\deg$ before perihelion) and an obliquity of $\epsilon \approx 30\deg$ (bold values in \autoref{tab:TPMinput}), so we choose this scenario to analyze here.

In general, the maximum surface temperature reached by Phaethon ($\approx1050\K$) is larger than that of UD ($\approx975\K$), due to the lower perihelion distance of the former (\autoref{fig:tpmoutput}). While the perihelion distance strongly influences the value of surface temperatures, the spin-axis orientation (the obliquity, $\epsilon$) primarily influences the {\it surface temperature distribution}. For both objects, the maximum temperatures are reached at perihelion. Maximum global temperatures can be found at equatorial latitudes for Phaethon, due to the location of the sub-solar point at perihelion, and our spin-axis choice for UD puts the maximum temperatures near $75\deg$N. When approaching perihelion a nuanced pattern is observed in which the hottest temperatures happen on either side of perihelion, and a local temperature minimum occurs between these maxima, just before perihelion. Because the northern hemisphere is heated on Phaethon's approach toward the Sun, the nighttime temperatures are higher than the nighttime temperatures on the southern hemisphere, which experiences local sunset at aphelion. This thermal behavior is caused by the interplay between the changes in insolation and illumination angle for each hemisphere.

We calculate the thermal gradient as the temperature change over the topmost thermal skin depth ($l_s$). Using $k = 0.55 \W \meter^{-1} \K^{-1}$, $c_s = 560 \J \kg^{-1} \K^{-1}$, and $\rho = 3110 \kg \meter^{-3}$ \citep{Gundlach&Blum13,Hanus_etal2018} and its rotation period of $3.6\hour$ we obtain $l_s = 2.56\cm$ for Phaethon. The thermal inertia of UD is half that of Phaethon, and with $P_\mathrm{rot} = 5.2\hour$ we obtain $l_s = 1.53\cm$. The thermal gradient throughout the diurnal skin depth shows a similar pattern as the maximum temperature history. Extreme thermal gradients occur either at perihelion (for northern latitudes) or briefly before and after perihelion (for southern latitudes) for both objects. Positive and negative values of the thermal gradient represent gradients occurring during the daytime and nighttime, respectively (\autoref{fig:tpmoutput}). The equator exhibits the maximum global thermal gradient regardless of orbital position or axial tilt, as the diurnal temperature range are maximized there. Less extreme thermal gradients occur at higher latitudes and are, in general, larger during the daytime hours. When regions are cast into permanent shadow, such as the northern pole of Phaethon after perihelion, the thermal gradient quickly approaches zero. Conversely, the southern pole region of Phaethon experiences a significant thermal gradient increase when it is exposed to sunlight just before perihelion.

Looking at the rates of temperature change (bottom panel of \autoref{fig:tpmoutput}), we find that the latitude is a highly influential controlling factor. Broadly speaking, the maximum rates of change are always found at the equatorial latitudes, with the polar regions exhibiting the lowest values. This latitude dependence is independent of spin-axis orientation because the change in insolation over a diurnal cycle is always maximized at the equator. Surface temperature rates of change are also strongly dependent on the rotation period of the object, with higher rates of change correlating with shorter rotation periods. For example, both Phaethon and UD experience a similar diurnal temperature range, but Phaethon's faster rotation rate causes it to experience maximum temperature rates of change that are more than 1.5 times compared to UD's.

\begin{figure}[h!]
  \centering
  \begin{tabular}[b]{c}
	\includegraphics[clip,trim=0.2cm 1.3cm 0 3.1cm,width=.5\linewidth]{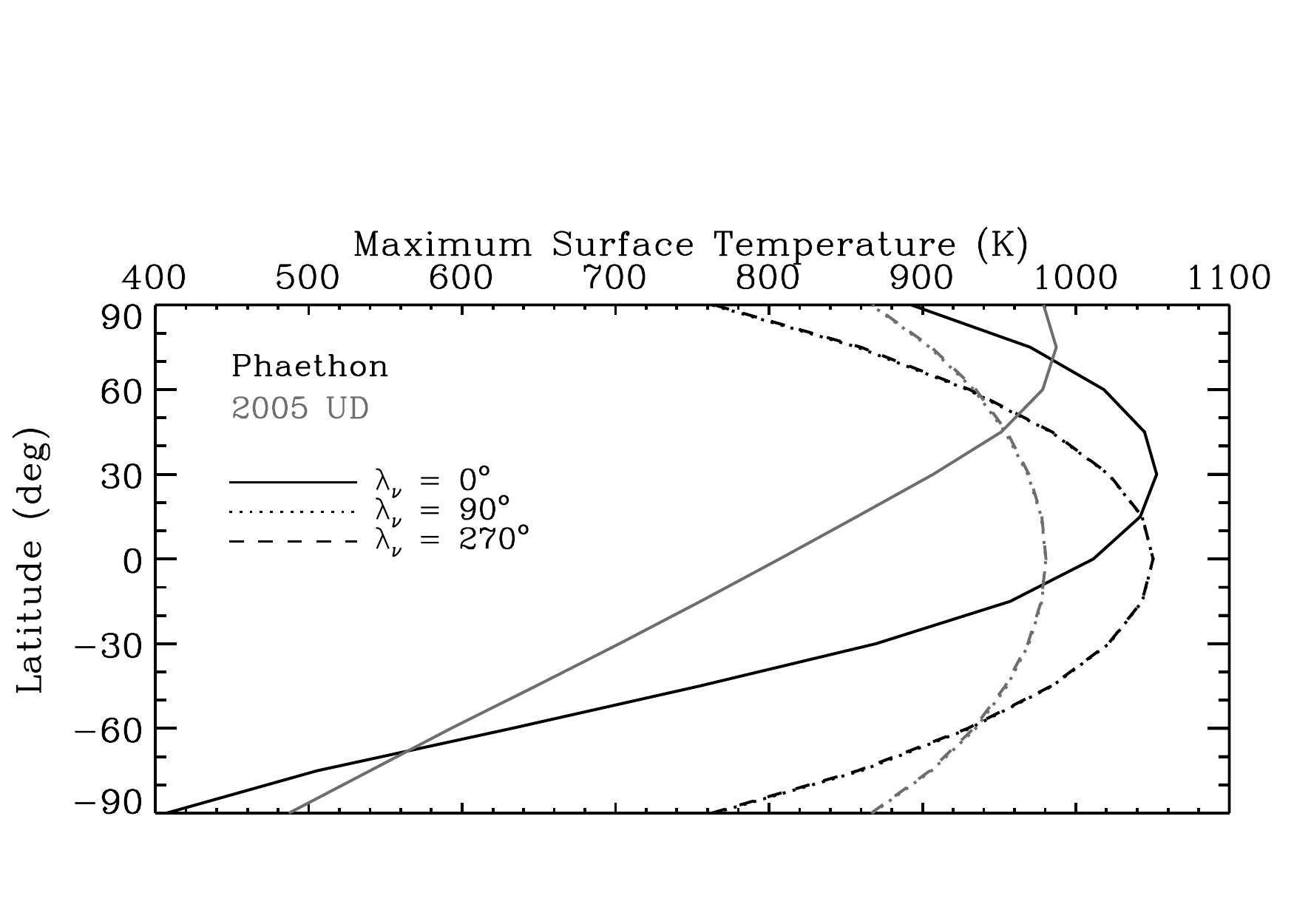}\\ \includegraphics[clip,trim=0.2cm 1.3cm 0 3.1cm,width=.5\linewidth]{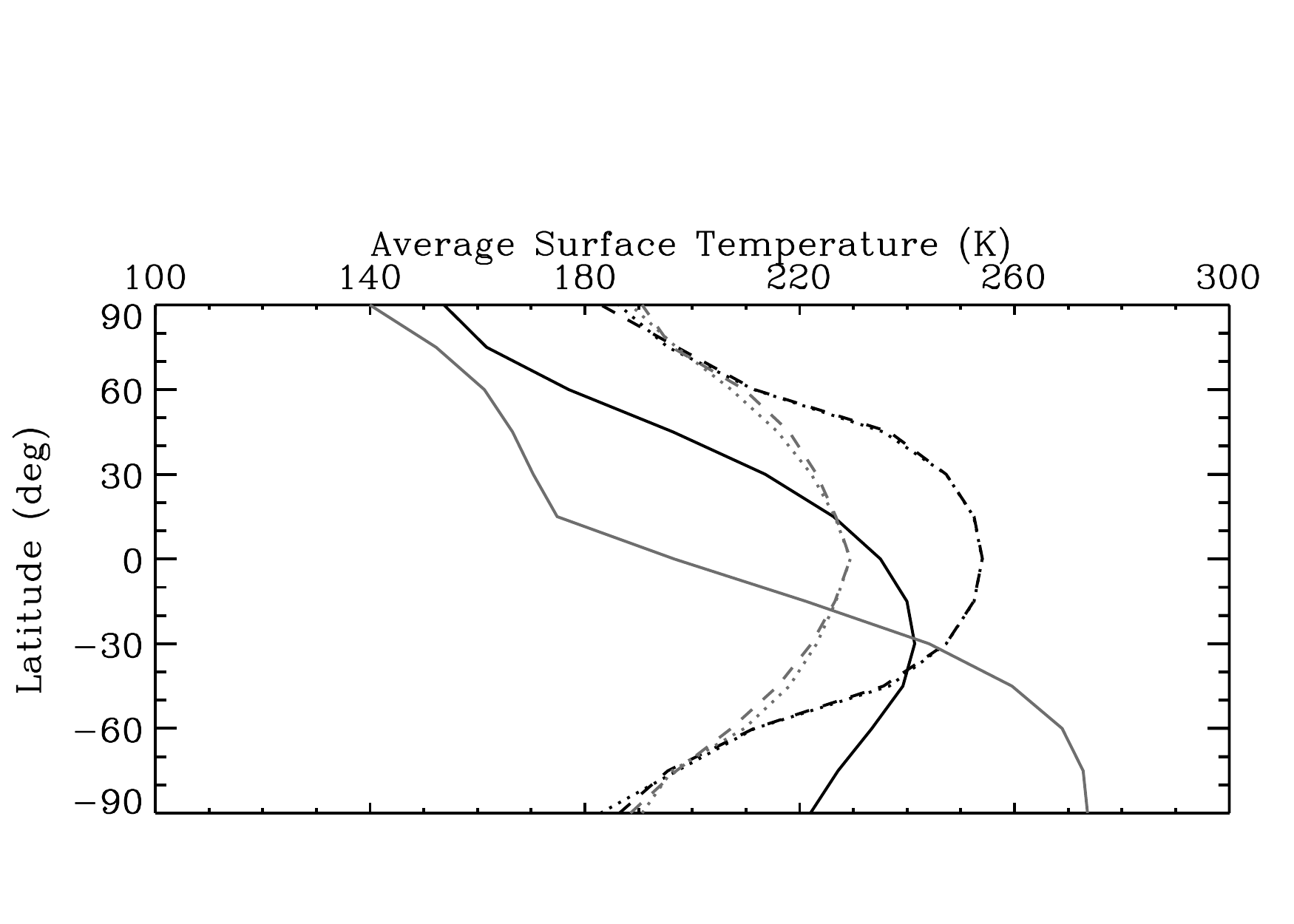}\\ \includegraphics[clip,trim=0.2cm 0cm 0 4.1cm,width=.5\linewidth]{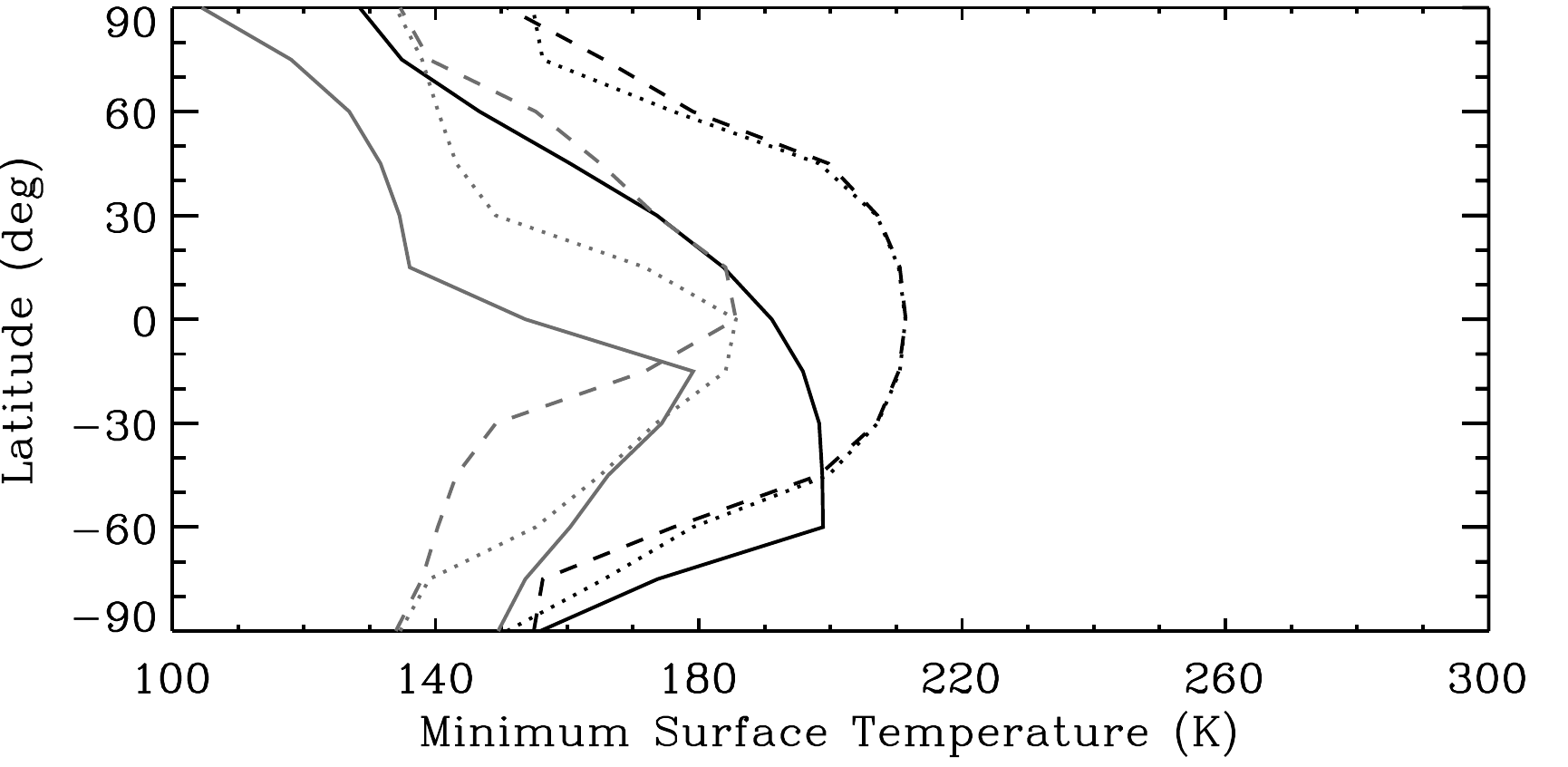}
  \end{tabular}
  \caption{The maximum (top), orbit-averaged (middle), and minimum (bottom) surface temperatures across an entire orbital revolution as a function of latitude. As indicated in the top panel, the solid, dashed, and dotted lines represent solstice points of $0\deg$, $90\deg$, and $270\deg$, respectively.}\label{fig:latorb}
\end{figure}

We have shown that the solstice point will influence the timing of peak perihelion temperatures whereas the obliquity changes where on the surface the peak temperatures are found, particularly in the polar regions. A solstice point at perihelion results in one of the polar regions existing in permanent shadow, which will drastically reduce the total insolation that it receives over an orbit, compared to the rest of the surface and to other spin orientations. This effect can be seen by computing the orbit-averaged temperatures across the surface (\autoref{fig:latorb}). Permanent sunlight at perihelion for the northern polar regions results in higher temperatures at perihelion, and lower orbit-averaged temperatures. Comparing the orbit-averaged temperatures between Phaethon and UD, we can see that the solstice point is more influential than the spin obliquity (middle panel of \autoref{fig:latorb}). The southern polar region shows a drastic drop in the peak temperature, due to permanent shadowing at perihelion, but little change in the minimum temperature. Comparing Phaethon to UD we can see that a larger obliquity has a greater influence on the magnitude of the effects described here.

\begin{figure}[h!]
  \centering
  \begin{tabular}[b]{c}
	\includegraphics[clip,trim=0.4cm 1.3cm 0 1.3,width=.5\linewidth]{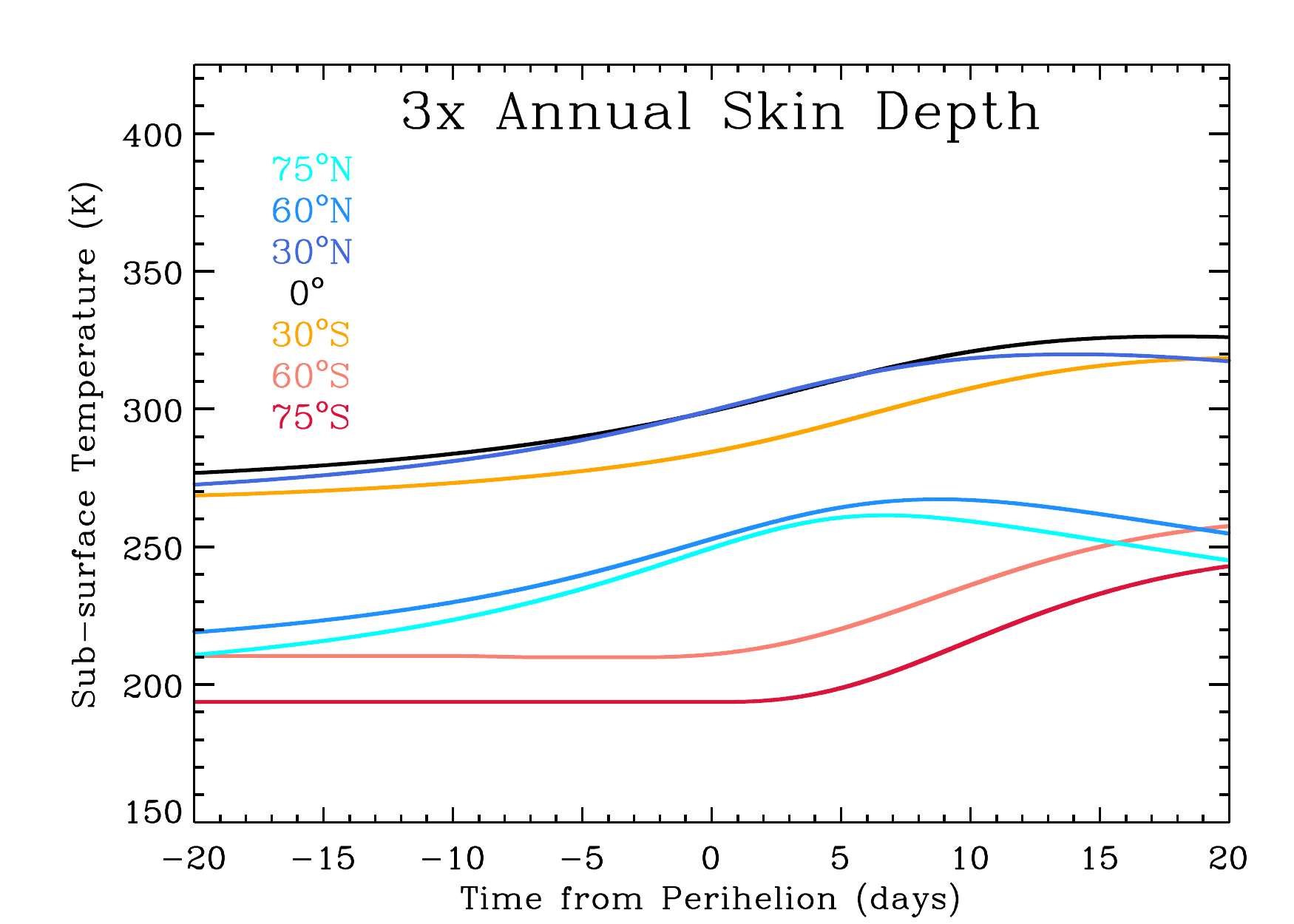}\\ \includegraphics[clip,trim=0.4cm 1.3cm 0 1.3,width=.5\linewidth]{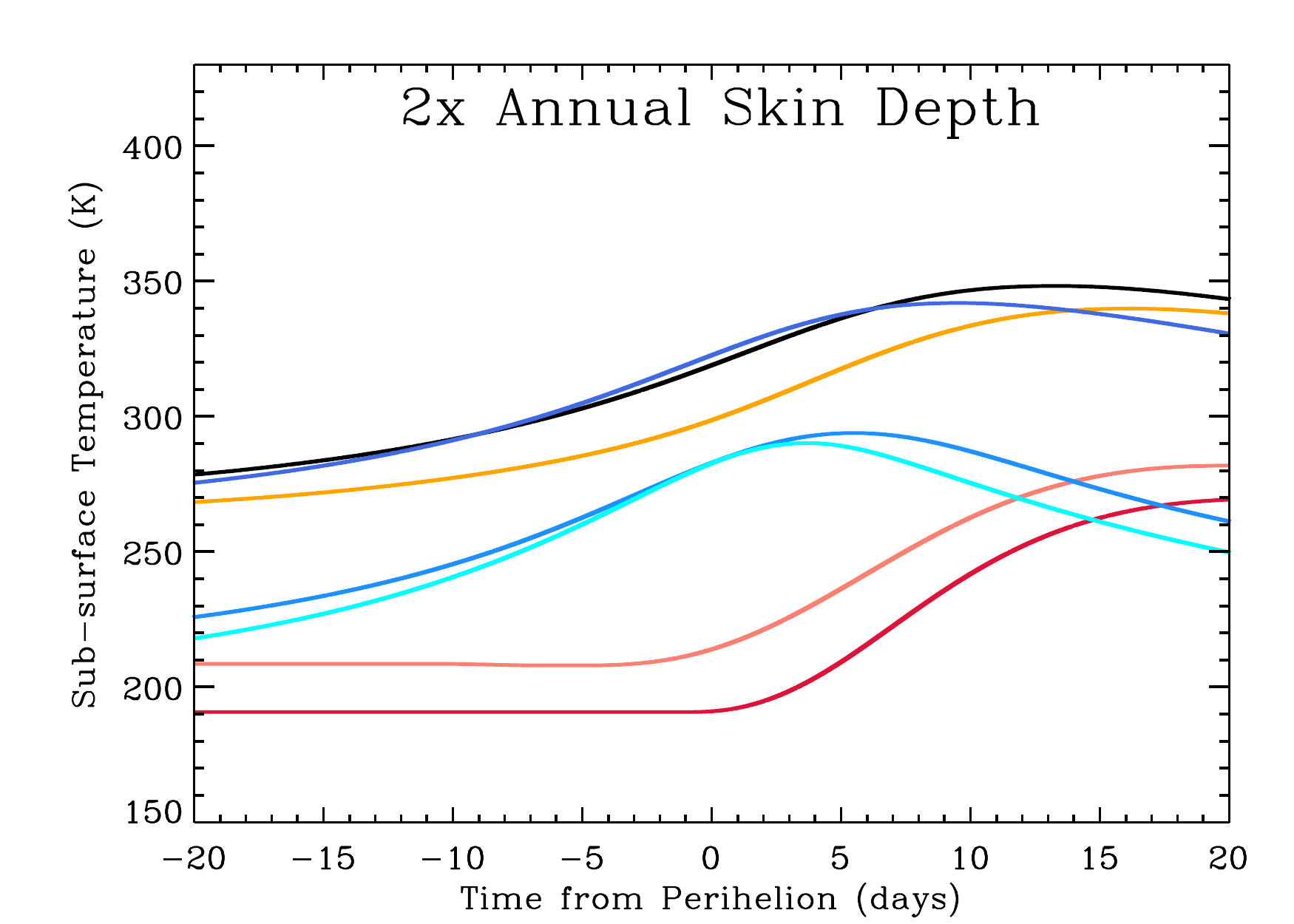}\\ \includegraphics[clip,trim=0.4cm 0.1cm 0 1.3,width=.5\linewidth]{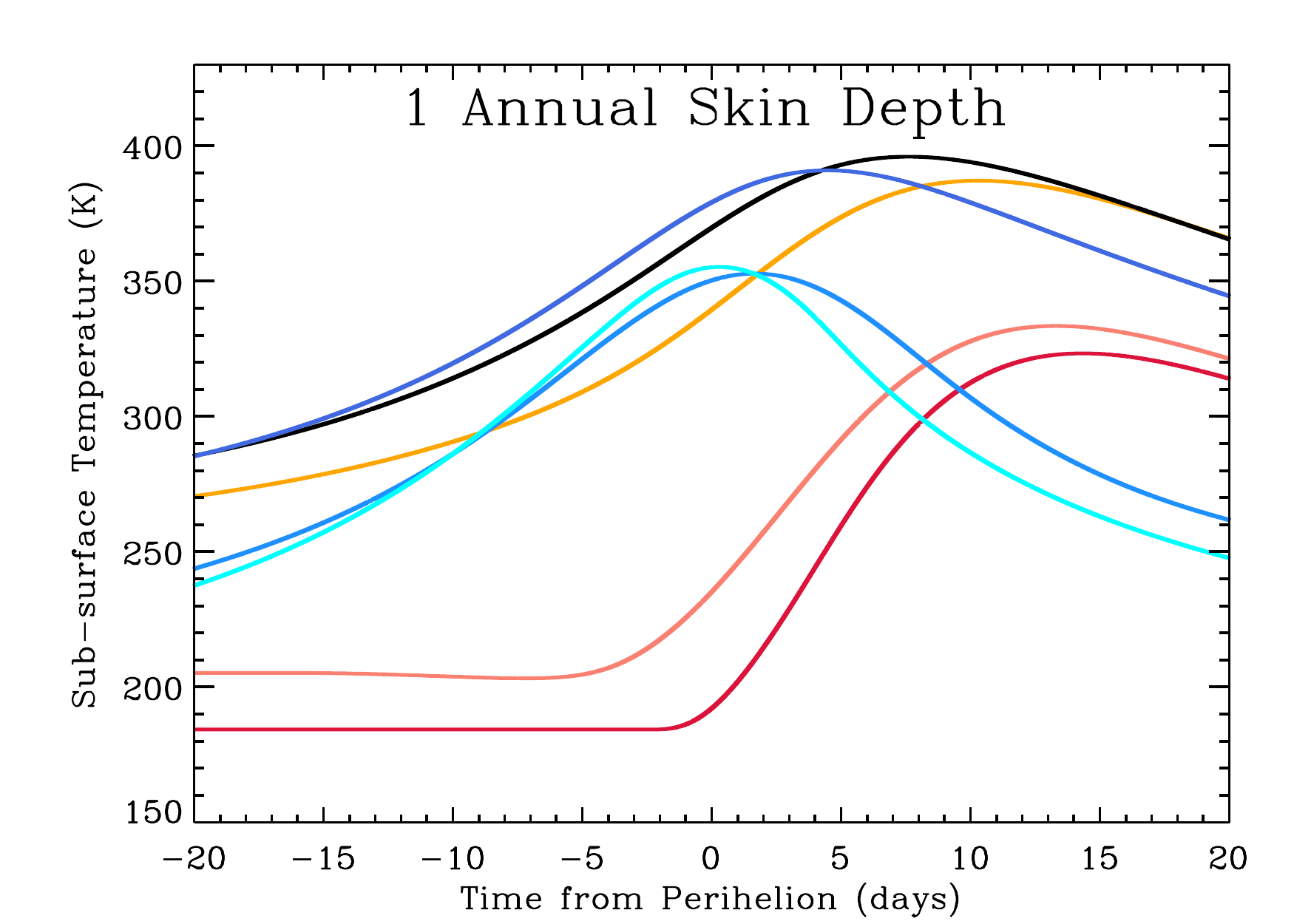}
  \end{tabular}
  \caption{The sub-surface temperatures as a function of time from perihelion at depths of 3$l^\mathit{ann}_s$, 2$l^\mathit{ann}_s$, and $l^\mathit{ann}_s$.}\label{fig:annuskinT}
\end{figure}

Lastly, we extracted the sub-surface temperature for different latitude bins at various multiples of the annual skin depth, $l^\mathit{ann}_s$. The temperatures at these depths are controlled by the changes in insolation over an entire orbit. One annual skin depth is calculated using \autoref{eq:skindepth} with $\tau = 4.522 \times 10^7 \second$ and $\tau = 4.542 \times 10^7 \second$ (the orbital period of Phaethon and UD, respectively) and the regolith thermophysical properties from \autoref{sub:tpmanalysis}. We obtain $l^\mathit{ann}_s = 1.51 \meter$ for Phaethon and $l^\mathit{ann}_s = 0.76 \meter$ for UD. \autoref{fig:annuskinT} shows this temperature dataset over the 20 days preceding and after perihelion passage. As expected, the minimum temperatures at each latitude are very close to the orbit-averaged temperature value (\autoref{fig:latorb}) and increase due to the large increase in solar energy received at smaller heliocentric distances (less than 3\% of the time is spent at $r<1\au$). In general, larger temperature changes are found closer to the surface and are more correlated in time to the changes in insolation (i.e. the time-lag is shorter). Because they are facing towards the Sun during perihelion approach, northern latitudes are warmer than southern latitudes before and until a few days after perihelion passage. In the southern hemisphere, temperatures begin to rapidly increase within a day of perihelion passage because of the drastic increase in insolation when these latitudes come out of permanent shadow. For northern latitudes, which are cast into shadow at perihelion, the situation is reversed and temperatures reach a maximum at perihelion passage in the shallow regions while temperature maxima deeper into the subsurface occur some days after perihelion passage. Temperatures at large depths are low enough to allow for volatiles to remain dormant for most of the orbit but increase during perihelion passage and can cause sublimation.

\subsubsection{Thermal History}

\begin{figure}[h!]
  \centering
  \begin{tabular}[b]{c}
	\includegraphics[clip,width=0.5\linewidth]{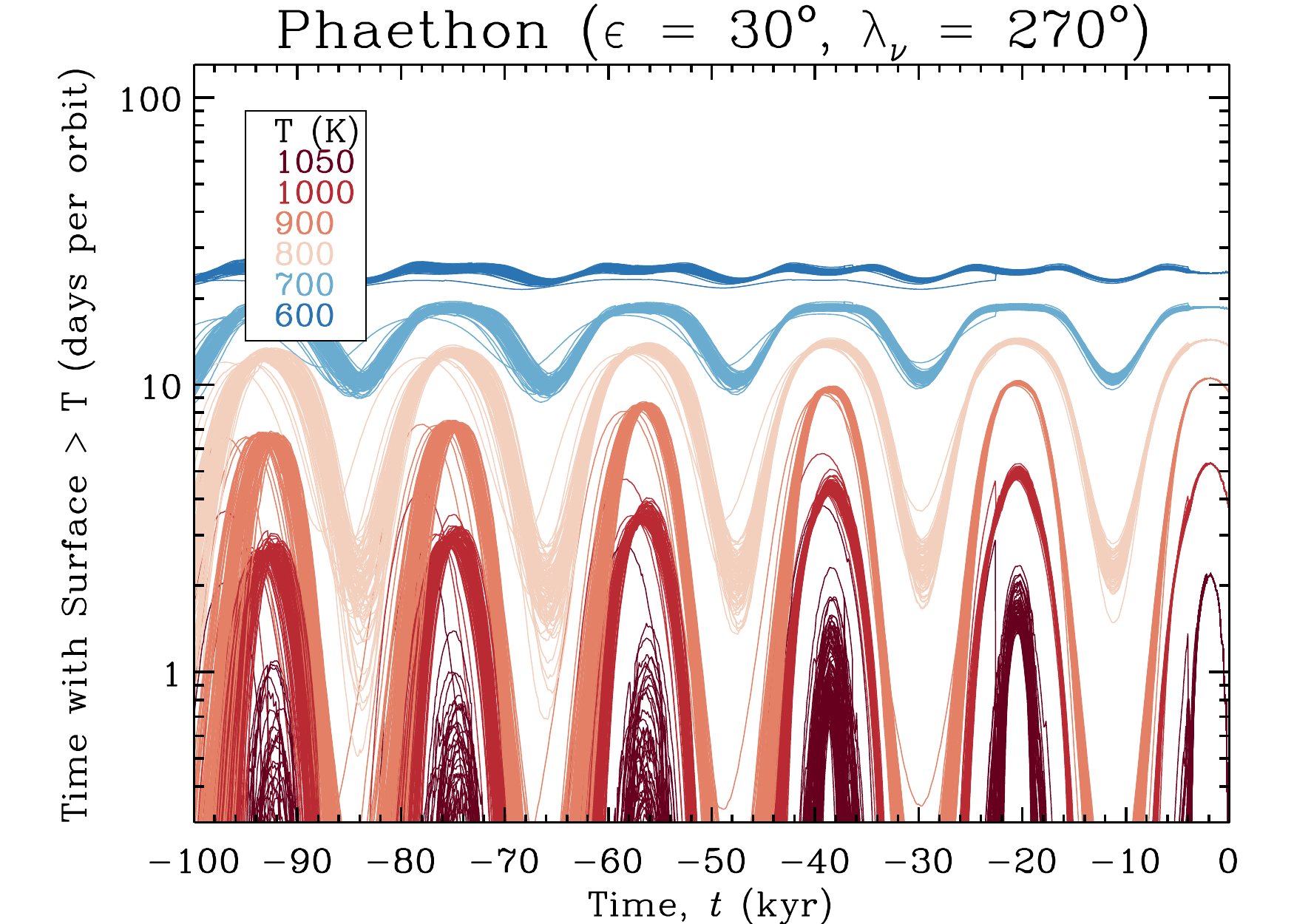}\includegraphics[clip,width=0.5\linewidth]{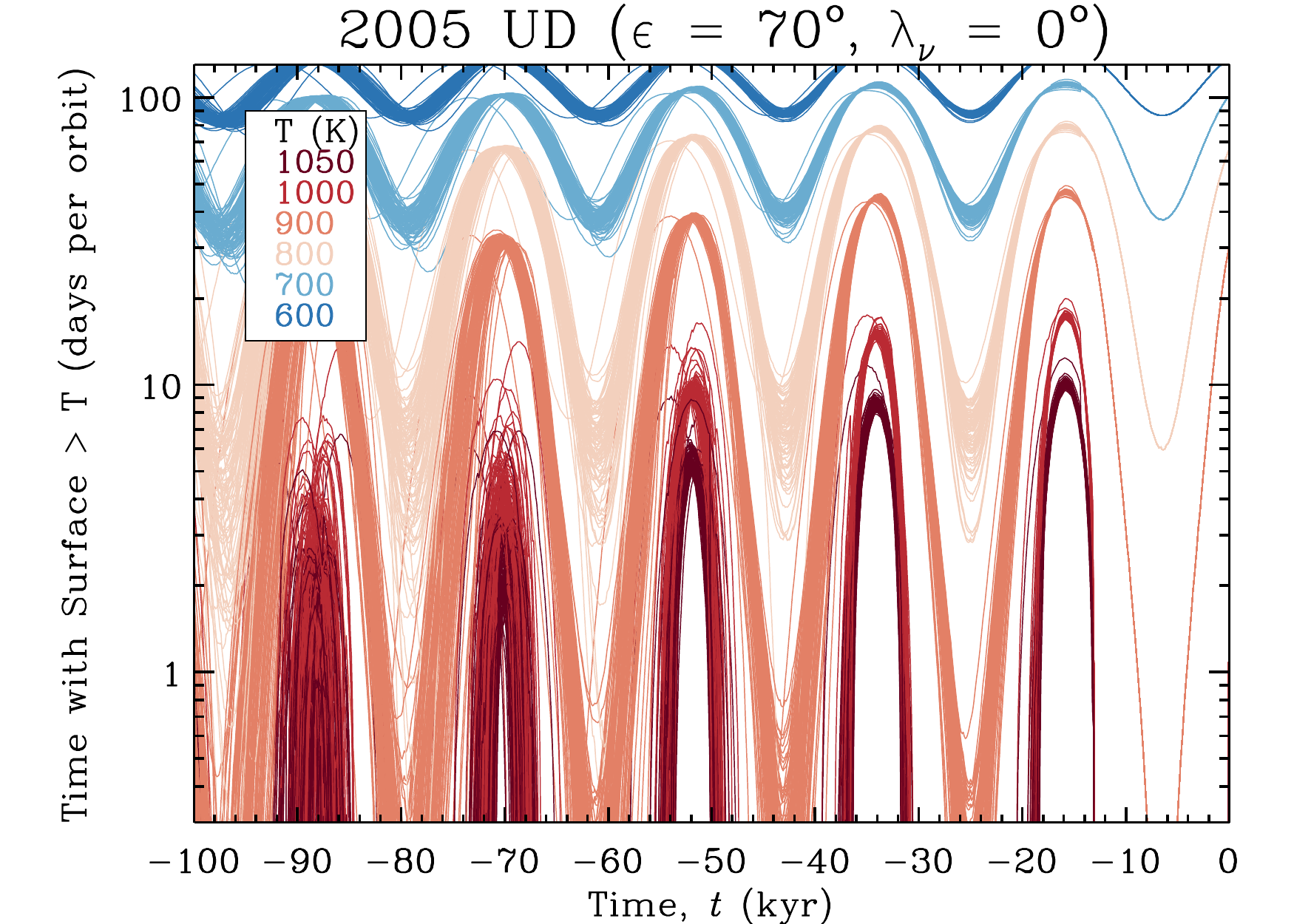} \\
	\includegraphics[clip,width=0.5\linewidth]{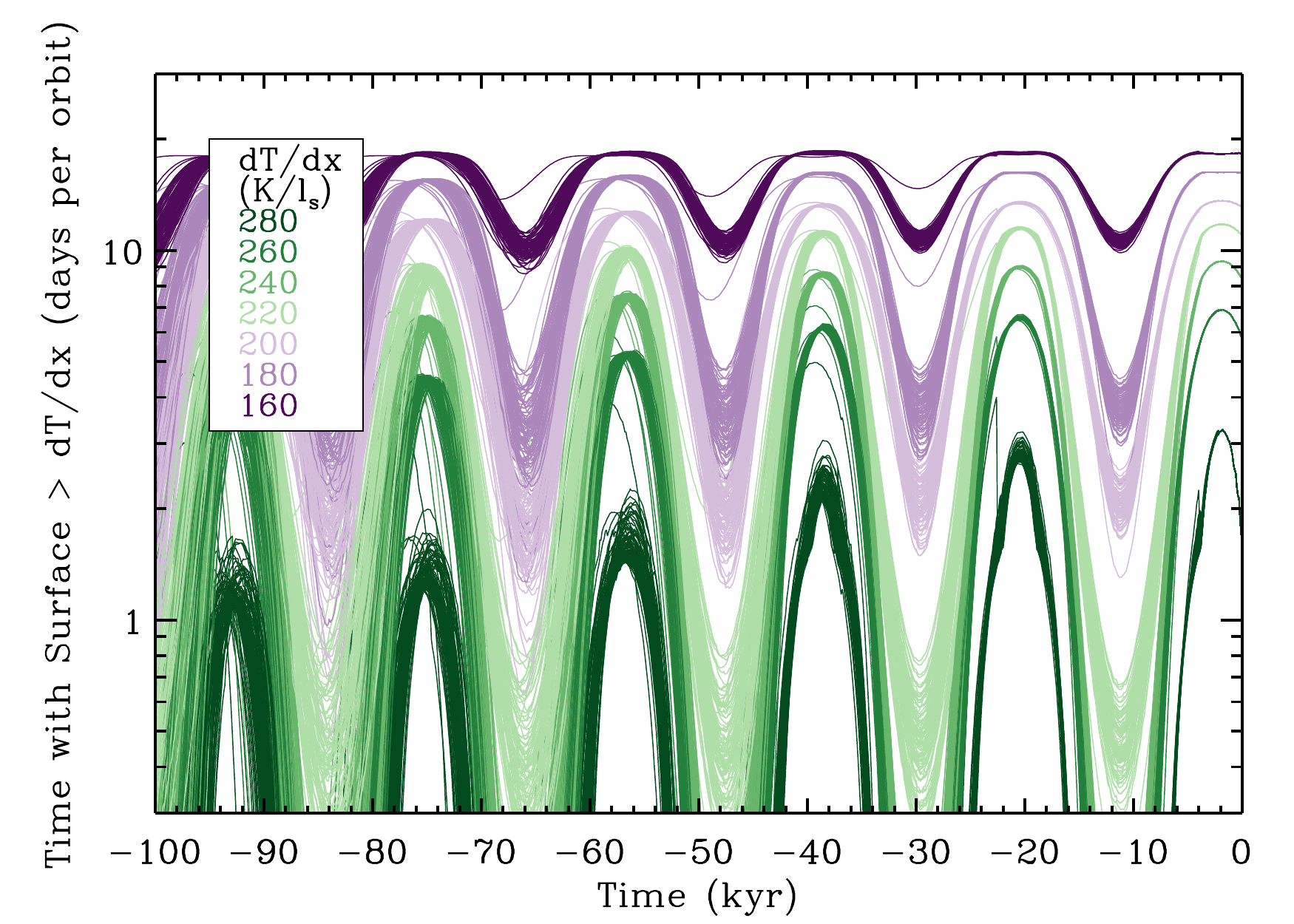}\includegraphics[clip,width=0.5\linewidth]{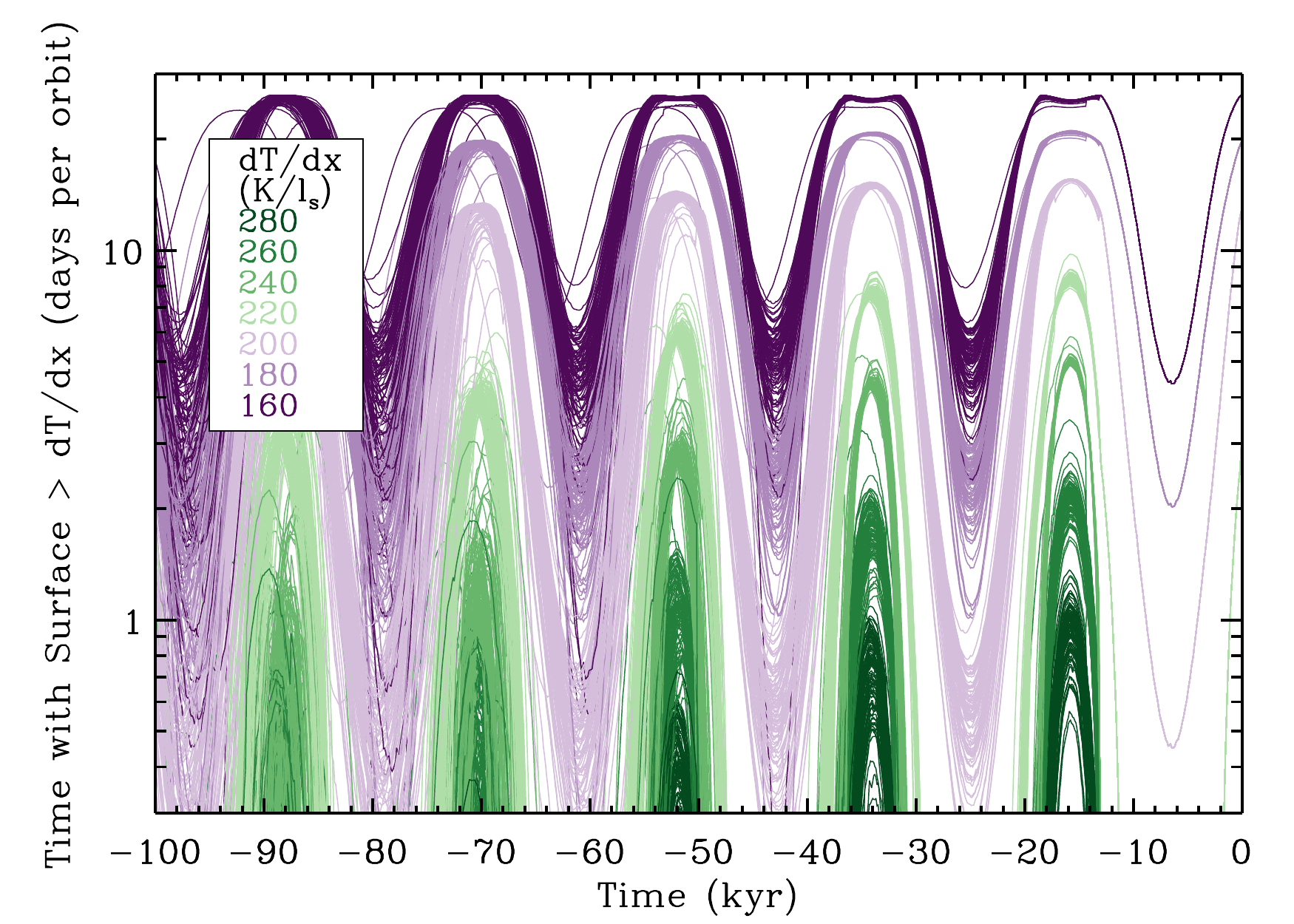} \\
  \end{tabular}
  \caption{Thermal history of Phaethon and UD shown as the number of days, per orbit, that meets the criterion of having maximum temperature (top) or having a temperature gradient (bottom) above a threshold value indicated in each panel.}\label{fig:thermhist}
\end{figure}

We now move on to characterize the change in temperature characteristics for Phaethon and UD across dynamical evolution timescales. To do so, we must distill the temperature information calculated from the orbTPM runs into a single value for each orbital configuration (i.e., for a given $a$ and $e$ pair). We do this by taking the parameters presented in \cref{sub:tpmanalysis} (surface temperature, thermal gradient, and temperature rate of change) and integrate over the time for which any part of the surface meets a particular criterion (i.e., $T_\mathrm{max}>800\K$), per orbital revolution. We perform this analysis on each of the orbTPM runs of predefined $a$ and $e$, and interpolate the values for desired orbital elements of each clone. The thermal history is only shown for the past $100\kyr$, because the orbit evolution is predictable over that time interval for both objects (\autoref{fig:thermhist}).

The history of maximum temperatures of the Phaethon and UD clones is strongly controlled by the periodic evolution of $e$. The uncertainty, shown by different clone tracks, can also be seen to gradually increase when there is a planetary encounter---in which case the semi-major axis is changed slightly ($4\kyr$ ago for Phaethon and $13\kyr$ ago for UD). The results show that both Phaethon and UD undergo periods of relatively cooler temperatures every $\sim20 \kyr$ during which surface temperatures remain under $1100\K$ and $1000\K$, respectively, most recently 9 to $13\kyr$ ago for Phaethon and 3 to $8\kyr$ ago for UD.

In the context of this work it is essential that we consider the possibility of time-dependent spin properties due to the YORP effect \citep{Rubincam00}, and relevant implications for the temperature characteristics and thermal history. In general, YORP torque can alter the spin rate and spin axis. Additionally, if Phaethon (and possibly UD) is losing mass asymmetrically (i.e., from a localized region), the result would be a change in the moment of inertia and the YORP coefficient. A change in the moment of inertia would also cause a reorientation of the spin axis that represents the spin state with the lowest possible energy. We have shown by comparing the results for Phaethon and UD that different spin rates will affect the thermal gradients. Our assumption of higher spin obliquity for UD compared to Phaethon did not appear to alter the value of the maximum surface temperatures, but rather {\it where} on the surface it occurred. The YORP rotational acceleration is proportional to $(a^2 \sqrt{1-e^2})^{-1}$ and typical YORP timescales for NEOs are on the order of $10^4\yr$ to $10^6\yr$ \citep{Rossi_etal09}. We thus do not expect large changes in the spin rate of Phaethon or UD due to the YORP effect.

\section{Discussion}

In order to place our analysis of the current and past temperature characteristics and thermal history of Phaethon and UD into context, we discuss possible temperature-dependent mass-loss mechanisms: volatile sublimation and thermal stress cycling. The possibility of volatiles existing in Phaethon and UD is related to their composition and provenance. Thus, we first discuss our investigation into the early dynamical evolution of these objects below. We mention, when possible, the likelihood of observing activity from Phaethon or UD during the DESTINY+ flyby. The location of activity and the position of the object in its orbit will provide important observational constraints on the mass-loss mechanism.

\subsection{Dynamical Origins}

The likeness of the inclinations of Pallas and Phaethon was used by \citet{deLeon_etal2010} as an argument in favor of the genetic relationship between Phaethon and Pallas. Specifically, \citet{deLeon_etal2010} demonstrated that a dynamical pathway through the 5:2 and 8:3 MMR with Jupiter could deliver an object (with 8\% efficiency) into a Phaethon-like orbit. \citet{Todorovic18} revisited this connection between Pallas and Phaethon, but specifically investigated the transfer probabilities of test particles that begin in the most unstable regions of these resonances with Jupiter. They found a much larger efficiency of 43.6\% and 46.9\% for the 5:2 and 8:3 MMR, respectively, compared to the 2\% of \citet{deLeon_etal2010}.

However, the results presented in this work suggest that a relationship between Phaethon and Pallas is improbable, given that the likelihood of an origin for the former (and possibly UD) in the inner asteroid belt is greater than an origin in the outer belt. We have also explicitly shown that using the inclination to argue for or against certain parent bodies is misleading, because the long time spent in the NEO region have allowed Phaethon and UD to significantly change the inclination they had when entering the NEO region through numerous planetary encounters. Hence the inclination-based argument for Pallas being the parent body of Phaethon is weak \citep{deLeon_etal2010}. Note also that this long evolution also makes it highly unlikely that Phaethon and UD would have separated before entering the NEO region. The similarity of the orbital elements of Phaethon and UD is unlikely to be coincidental \citep{Ohtsuka_etal2006}, and the resemblance of their visible reflectance \citep{Jewitt&Hsieh_2006} and polarimetric phase curve \citep{Devogle_etal20} implies that the two bodies were once part of the same object. This separation should have occurred {\bf more than} $100\kyr$ ago \citep[i.e.,][]{Hanus_etal2016}.

Although we cannot rule out the possibility that Pallas is the source of Phaethon and UD, our results indicate that the connection is unlikely. We therefore consider other parent bodies for the pair. Given that our results suggest that the most likely escape route for Phaethon is the $\nu_6$ resonance complex (a combination of the $\nu_6$ secular resonance and the 4:1 and 7:2 mean-motion resonances with Jupiter) followed by the 3:1 complex, we limit our search to the inner-belt region bounded by the $\nu_6$ secular resonance and the 3:1 mean-motion resonance with Jupiter. An inner-belt origin was also proposed by \citet{Ansdell_etal14} based on the fact that NEAs with retrograde sense of spin such as Phaethon are typically delivered from this region through the $\nu_6$ resonance.

\begin{figure}[b!]
    \centering
    \includegraphics[width=\textwidth]{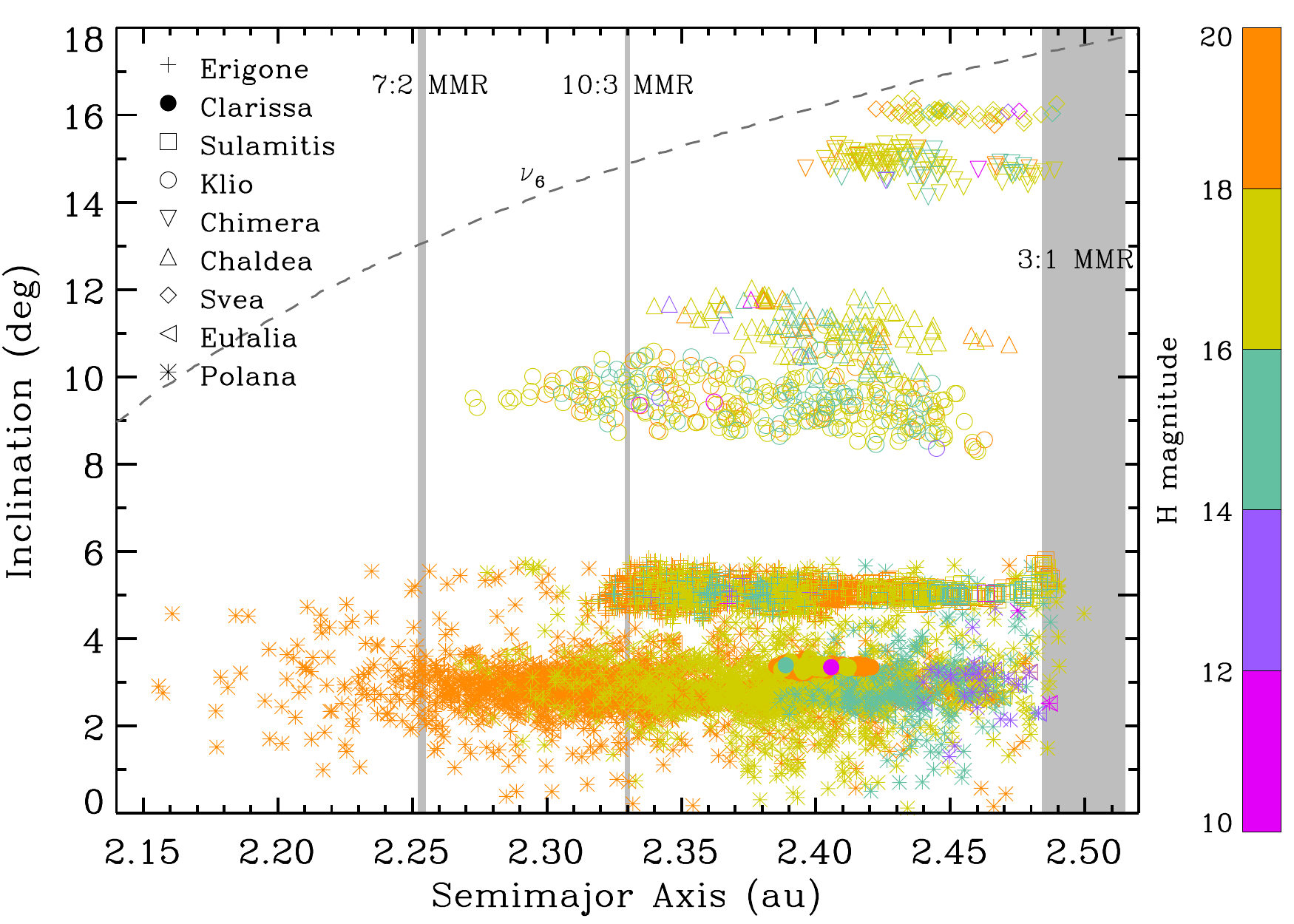}
    \caption{Primitive (C-type and B-type) asteroid families in the inner asteroid belt.}
    \label{fig:primfam}
\end{figure}

Most of the primitive (C-type and B-type) asteroid families in the inner belt have nearby secular or mean-motion resonances and could be the source family for a Phaethon-sized ($H\approx14.3\magnitude$) NEO (Fig.~\ref{fig:primfam}). Out of the currently identified families, the Svea and Polana are particularly interesting. The high-inclination ($i \approx 16\deg$) Svea family is adjacent to both the $\nu_6$ and 3:1 resonances and is capable of delivering $H\approx14.3\magnitude$ objects through both the $\nu_6$ and the 3:1 resonances. If we assume that the Yarkovsky effect has driven the dynamical evolution of Svea family members, then its more likely than not that the $\nu_6$ has delivered Phaethon and UD, as they are both likely retrograde rotators \citep{Ansdell_etal14}. Finally, we note that the inclination of the Svea family is relatively close to that of Phaethon and UD, whereas Polana family members have $i<6\deg$.

While (329) Svea is itself classified as a C-type a notable fraction of the family members are B-type \citep{Morate_etal19}. \citet{Morate_etal19} posit a slope-size correlation among primitive asteroids, with smaller bodies ($H>13\magnitude$) having a wider distribution of spectral slopes than larger asteroids ($H \approx 10\magnitude$). This proposed correlation broadens the criteria for establishing a genetic relationship among primitive asteroids. Given this line of reasoning, Phaethon's origin is consistent with the Svea family despite the C-type classification of Svea itself. On the other hand, (142) Polana is a B-type and is also a possible parent body for the active asteroid (101955) Bennu \citep{2015Icar..247..191B} and the polarimetric properties of Bennu are also consistent with members in the Polana family \citep{Cellino_etal18}. Yet, Phaethon and UD exhibit distinct polarimetric phase curves from Bennu \citep{Devogle_etal20}, which may rule-out an origin in the Polana family.

\subsection{Thermal History}

\subsubsection{Volatile Sublimation}\label{sub:volatiles}

Given Phaethon's repeated activity at perihelion, its possible that sublimation of volatiles is driving the ejection of dust. Generally speaking, supporting evidence that water ice exists at or near the surface of the Moon and Mercury \citep{Gladstone_etal10,Schultz_etal10,Lawrence_etal13} doesn't rule out the possibility that volatiles may be present inside low-$q$ asteroids, despite their large surface temperatures. Specifically, signatures of hydroxyl/water on the Moon \citep{Lawrence_etal06,Clark09,Bandfield_etal18} and on Mercury \citep{Harmon_etal01,Neumann13,Chabot_etal14,Deutsch_etal17} suggest that the conditions may exist for ice to survive buried deep beneath the cold polar regions of airless bodies at $r \leq 1\au$ \citep{Lawrence17}. Alternatively, hydrated minerals, such as those found on Bennu \citep{Hamilton_etal19}, may contain large amounts of molecular water \citep[up to 10wt\%;][]{Nuth_etal20} which may cause dust activity when sufficiently heated. We will explore both these possibilities (ice/volatiles and hydrated minerals) in this subsection.

The non-detection of a 3-micron hydration feature \citep{Takir_etal20} rules out the possibility that Phaethon's surface contains volatiles or even hydrated minerals. Yet, this evidence is applicable for the uppermost, optically-accessible region of the regolith. We thus consider the possibility that there could be volatiles and/or hydrated minerals buried at some depth below Phaethon's surface. If this is this case, the burial depth must be on the order of a few $l^\mathit{ann}_s$ (\autoref{fig:annuskinT}). As mentioned in the previous subsection, we find a high likelihood that Phaethon and UD originated in the inner part of the Main Belt. Although the colder outer-belt is thought to contain more primitive (hydrated) material, a few primitive families have been identified in the warmer, inner-belt \citep{Morate_etal19}. We note that the existence of MBC 259P/Garradd ($a=2.73\au$) may be an indicator that activity-driving volatiles are buried deep within other inner-belt asteroids \citep{Hsieh_etal21}.

Using the orbit-average surface temperature for the nominal orbits of each object, we estimate the depth to which water ice is stable at some time scale and the general lifetime of ice survival. The orbit-averaged temperatures represent the temperature at the greatest depth for each latitude value in the orbTPM. As expected, the orbit-average temperatures are correlated with $e$, but do not vary drastically ($\pm 3\K$) across kyr timescales---most likely due to the small variations in $a$. We feed the average temperatures into the idealized thermochemical model presented in a series of works by \citet{Schorghofer08}, \citet{Schorghofer16}, and \citet{Schorghofer&Hsieh18}. This series of models were used to calculate the theoretical lifetime of ice on hypothetical MBAs, but can be used for any object for which the temperatures at-depth are known.

The middle panels of \autoref{fig:latorb} show that orbit-averaged surface temperatures in Phaethon's northern polar region are $\approx195\K$. Using this temperature, a grain size of $1\cm$, and thermal conductivity of $10^{-3} \W \meter^{-1} \K^{-1}$, we calculate the desiccation time---the time for all ice to sublimate---to be $t_\mathrm{des} = 50\Myr$, assuming a $5\km$ diameter for Phaethon. This is on the same order of magnitude as the estimated delivery time of Phaethon to the NEO region, as presented in this work. \citet{Schorghofer&Hsieh18} estimates that half the volume of ice loss happens after 11\% of the desiccation time has passed, at which point the ice will have retreated to $\sim80\%$ of Phaethon's radius. This means that any ice at depths $<200\meter$ in the north polar region cannot exist after $5.5\Myr$. If $100\um$ regolith grains are assumed, then the mean-free path of the water vapour decreases and this depth cutoff changes to $<20\meter$, which is at least an order of magnitude larger than $l^\mathit{ann}_s$.

\citet{Schorghofer&Hsieh18} calculated shallow (1--$3\meter$) depths for the survival of ice over the lifetime of the Solar System within hypothetical MBAs. Sublimation-driven MBC activity is hypothesized to be triggered by small impacts that expose ice that is buried a few meters deep. Because typical $l^\mathit{ann}_s$ for these objects are on the same order as the depth at which ice can exist, increased solar heating at perihelion is thought to cause sublimation-driven activity of water ice. We have concluded above that water ice could only exist at depths $\gg l^\mathit{ann}_s$. Thus, this is inconsistent with the MBC-like activity. However, deeply buried ice in Phaethon's core could hypothetically have been activated from an impact or rotational mass shedding that removes tens of meters of regolith.

Now that we have completely ruled out the possibility of long-lived water ice in at least the topmost $20\meter$ in the northern pole of Phaethon. Another possibility is that ice is actively generated on timescales shorter than its dynamical lifetime. It has been proposed that OH is formed on surfaces via the creation of hydrogen bonds in crystal lattice defects from solar wind bombardment \citep{Farrell_etal15}. Evidence for such a process exists for the Moon \citep{Ichimura_etal12,Jones_etal18} and interplanetary dust particles \citep{Bradley_etal14}. \citet{Zhuravlev00} reviewed many experiments which have shown that hydrogen-oxygen bonds can exist for temperatures up to $463\K$. Below $l^\mathit{ann}_s$ Phaethon and UD's surface are consistently less than this temperature.

Another possibility is that a hydrated subsurface layer formed by solar wind proton implantation \citep[e.g.,][]{Starukhina01}. This process could theoretically act on igneous and metamorphic asteroids, which are thought to have lost all their volatiles in the early solar system \citep{Starukhina01}. These asteroids are rich in silicates, which have experimentally been shown to form OH bonds when bombarded with protons having energies similar to the solar-wind \citep{Schaible&Baragiola14}. Reflectance observations of NEAs have shown that spectral features in the 3-micron region due to OH/H2O are not uncommon \citep{Rivkin_etal15}, regardless of spectral type and despite their high surface temperatures relative to MBAs. This is consistent with a solar wind supply of hydrogen, but does not indicate that volatiles exist in the subsurface of these bodies. Outgassing events after meteoroid impacts on the lunar surface suggest that a hydrated layer is buried under a few meters of regolith \citep{Benna_etal19}. The temperature at these depths are in the range of 220--$255\K$, which is within the range for depths of $3l^\mathit{ann}_s$ (195--$270\K$) in the polar regions of Phaethon (\autoref{fig:annuskinT}), so the scenario solar wind supply of volatiles scenario discussed here is possible.

For primitive asteroids such as Phaethon and UD, OH/H2O may already exist within hydrated minerals such as phyllosiciates. Laboratory heating of phyllosilicates have shown that the release of adsorbed OH/H2O occurs over a time period of a few dozen hours. However, experiments that heat a sample rapidly may or may not demonstrate that energy released from within the mineralogical structure is large enough to directly eject mass from an asteroid surface. As with the MBCs, a small impact \citep{Szalay_etal19,Wiegert_etal20} could excavate material to within a few meters of buried volatiles \citep[e.g.,][]{Haghighipour_etal16}, if they exist, or hydrated minerals. This would then trigger sublimation-driven dust ejection during subsequent perihelion passages.

Observations and characterization of more low-$q$ objects (spin rate, colors/spectra, thermal inertia, and shape), and monitoring of activity or anomalous brightening will provide important insight into the possible volatile content of these objects. For example, the sungrazing object 322P/SOHO1 has not displayed a coma or tail during its past five perihelion passages, although anomalous brightening has been consistently detected after perihelion. The size and rotation period imply a bulk density of over $1000 \kg \meter^{-3}$ which is outside the range of all known comets, and its colors are most similar to a V or Q-type asteroid rather than a typical comet \citep{Knight_etal16}. The orbital elements of 322P are similar to the Jupiter-family comets, but its semi-major axis is very close to the 2:1 mean-motion resonance with Jupiter which could imply that 322P originated in the Main Belt. Could other objects from the asteroid belt be orbiting among known comet populations? The work of \citet{Hsieh_etal20} show a dynamical pathway from the Themis family into this comet population via the 2:1 resonance with Jupiter. Given the many challenges and uncertainties associated with dynamical modeling of NEOs, attempts to determine the origin of low-$q$ objects should make use of physical characterization, when possible.

The ejection of individual mm-to-cm-scale particles from Bennu throughout its orbit \citep[see][and references therein]{2020JGRE..12506549H} indicates that undetectable mass-loss can occur on B-type asteroids at temperatures $<390\K$ \citep{Rozitis_etal19}. Considering the possibility that a similar, temperature-dependent, process is responsible for Phaethon's activity, it could then explain the lack of {\it remotely-detected} activity at larger heliocentric distances \citep{Jewitt_etal2019,Ye_etal21}. Surface temperatures measured by OSIRIS-REx were used by \citet{Rozitis_etal19} to estimate the probability that volatiles were present in the subsurface. Using a TPM the authors found that small patches in Bennu's polar regions are cold enough for water ice to possibly exist. However, the occurrence of particle ejections near the equator \citep{Lauretta_etal19b} are inconsistent with sublimation-driven activity. Bennu likely originated from the inner-belt Polana or Eulalia families \citep{2015Icar..247..191B}, and the lack of evidence for sublimation may indicate that NEAs from the inner Main Belt are not likely to retain any volatiles.

\cite{Rozitis_etal19} also considered thermal cycling as the mechanism for Bennu's particle ejection. Particles originated from areas when the local time was in the late afternoon which \cite{Rozitis_etal19} noted was when a negative thermal gradient was present between a warmer subsurface and cooling surface. Exfoliation features on several boulders on Bennu's surface \citep{Molaro_etal20} are clear indicators that thermal fracturing is effective in altering NEA surfaces. We explore this mechanism for Phaethon and UD in the following subsection.

\subsubsection{Thermal Breakdown}

Rocks equal to or larger than the thermal skin depth are susceptible to the formation of large thermal gradients, which lead to internal stress fields \citep{Molaro_etal17}. Our estimate for Phaethon's skin depth of $2.55\cm$ is quite close to its characteristic grain size of 1--$2\cm$, however UD's skin depth ($1.53\cm$) is just above the upper limit of possible grain size range of 0.9--$10\mm$. It is probable that Phaethon and UD have several boulders on their surface that do not affect the overall thermal inertia of the surface. UD's thermal inertia is very close to that of Bennu's \citep[$350 \pm 20 \J \meter^{-2} \K^{-1} \second^{-1/2}$;][]{DellaGiustina_etal19}. Analysis of OSIRIS-REx images and thermal data showed an unexpectedly high fraction of boulders with high-porosity (and thus, low thermal inertia) on Bennu's surface \citep{Lauretta_etal19a}---not unlike that found for the C-type asteroid Ryugu \citep{Grott_etal19}, which has not been observed to be active. Phaethon's thermal inertia is even higher, so there is a high probability that the surface contains many large boulders.

As with the case for Phaethon, extreme thermal gradients for UD are found among the lower latitudes. It is interesting to note that while positive thermal gradients are less than that of Phaethon's, negative thermal gradients found on UD are similar in magnitude between both objects. However, our assumption of UD's large spin obliquity directly control the timing of the extreme values to occur at $\nu \pm 60\deg$ from perihelion (\autoref{fig:tpmoutput}). According to the results of \citet{ElMir_etal19}, thermal fracturing is less efficient for rocks larger or smaller than the thermal skin depth. From this fact, its possible that thermal fracturing may indeed be an efficient process on Phaethon but less likely so for UD.

\citet{ElMir_etal19} constructed a thermomechanical model to estimate the lifetime of a 10-cm-diameter rock on asteroids of various rotation periods and heliocentric distances. This model can be scaled in order to estimate lifetimes of a particular size rock on an asteroid with any orbital configuration. \citet{ElMir_etal19} showed that thermal fracturing is more efficient for rocks nearest in size to the thermal skin depth, because this is the length scale at which a significant thermal gradient exists. We note that large thermal gradients lead to large thermal stresses when they occur across particles of a size near the thermal skin depth. The breakdown timescale is $\sim10^4 \yr$ for such a rock at Phaethon and UD's perihelion distances \citep{ElMir_etal19}.

Breakdown via thermal cycling was proposed by \citet{Graves_etal19} to be the spectral freshening mechanism for asteroids with $q<1\au$. An inverse relationship between perihelion distance and spectral slope (a proxy for space weathering) was presented by \citet{Graves_etal19}, who were able to model the relationship using an approximation for the breakdown efficiency. If thermal fracturing is indeed more efficient for low-$q$ objects, then we should expect that the regolith grain sizes correlate with perihelion distance, as objects with lower $q$ are more efficiently broken down by thermal cycling. Currently, only Phaethon and UD have thermal inertia estimates, with Phaethon's value larger than UD's. This difference in thermal inertia could indicate Phaethon's loss of smaller dust particles in the past, increasing the fraction of larger rocks on the surface, which is consistent with its larger thermal inertia. In order to lend more support this thermal breakdown scenario at small heliocentric distances, further thermophysical characterization of low-$q$ objects can be used to identify a correlation between thermal inertia and $q$.

In a study by \citet{Hamm_etal19} thermal breakdown was predicted to be most efficient at lower latitudes where there are large diurnal temperature changes. The authors note that even for highly oblique spin orientations, thermal fracturing should be most efficient in a latitude band extending $\pm 30\deg$ from the equator. This thermal gradient correlation with latitude can be seen in our results for both Phaethon and UD (\autoref{fig:tpmoutput}). However, we point out that the efficiency of thermal fracturing is determined by the thermal expansion coefficient \citep{Molaro_etal17,ElMir_etal19}, $\alpha_L$, which is also temperature-dependent \citep{Opeil_etal20}. A recent experiment by \citet{Opeil_etal20} showed a strong dependence of $\alpha_L$ with temperature in the range from $210 < T < 300 \K$ for CM chondrites. Interestingly, they observed large negative values of $\alpha_L$ in some samples, which indicates contraction with increasing temperature. The authors attributed this unusual behavior to the layered nature of phyllosilicate minerals. Because the experiment only included CM chondrite meteorites and the strength of the change varied drastically from sample-to-sample, it is difficult to directly relate the results to Phaethon and UD without knowing the exact composition of their surfaces. \citet{Opeil_etal20} remarks that any phyllosiciate material should exhibit this behavior, so it is plausible that thermal expansion affects the material found on Phaethon and UD. The polar regions of Phaethon and UD are most likely to have temperatures in the range of their experiment. This extreme thermal expansion effects combined with moderate thermal gradients of $150 \K/l_s$ during perihelion passage may mean that thermal fracturing is more efficient in the polar regions of these asteroids compared to the equatorial band as predicted by \citet{Hamm_etal19}.

If the mechanical stress from thermal expansion is converted into kinetic energy with perfect efficiency, particle ejection can occur \citep{Jewitt&Li2010}. As a result of extreme diurnal insolation changes, the equatorial region of Phaethon experiences the greatest temperature change (middle and bottom rows of \autoref{fig:tpmoutput}). At perihelion, Phaethon's equatorial region has a diurnal temperature range just over $500\K$, maximum temperature rate of change of $\sim25 \K \minute^{-1}$, and spatial temperature gradient of $275\K$ over a diurnal skin depth. \citet{Jewitt&Li2010} formulate the ejection velocity as a function of temperature difference to be $v_\mathrm{ej} = \alpha_L \Delta T \sqrt{\frac{\eta Y}{\rho}}$, $\eta$ is the efficiency, and $Y$ is Young's modulus. We use $\alpha_L = 10^{-5} \K^{-1}$ \citep{Opeil_etal20}, $\Delta T = 275\K$ (\autoref{fig:tpmoutput}), $Y = 10^{11} \mathrm{N} \meter^{-2}$, and $\rho = 3110 \kg \meter^3$ to estimate an upper limit (i.e., $\eta = 1$) to the ejection velocity via thermal fracturing of $v_\mathrm{ej} \approx 15 \meter \second^{-1}$. This value is within the range of previously-estimated ejection speeds \citep{Jewitt_etal2013}, and well above the gravitational escape velocity (1.5--$2 \meter \second^{-1}$) of Phaethon \citep{Hanus_etal2018}. Yet, this velocity estimate is factors of several less than that estimated from the width of the Geminid stream \citep[$1 \km \second^{-1}$ according to][]{Ryabova16}.

To our knowledge, laboratory experiments that replicate the solar heating environment at $r\approx0.15\au$ have not been performed. We encourage experiments that mimic the large heating rates (up to $25 \K \minute^{-1}$) estimated in this work on meteorite samples, such as carbonaceous chondrites, phyllosilicate minerals or other appropriate analog materials. Furthermore, meteorite thermal expansion coefficients measured at $T>300 \K$ can be used to model the efficiency of thermal fracturing for low-$q$ asteroids and estimate dust-ejection rates.

\subsection{2005 UD: An Active Asteroid?}
No signs of evidence of activity has been observed for UD. This may be due to observing limitations related to its small size and/or low degree of activity---if it is indeed active. UD's lower surface gravity compared to Phaethon's means that the minimum escape velocity is smaller, so dust ejection is, in principle, more probable. While UD's size, albedo, and thermal inertia have been well-constrained \citep{Devogle_etal20}, UD's shape and spin characteristics remain relatively uncertain compared to Phaethon's. Furthermore, it is difficult to make accurate predictions on the temperature characteristics due to the uncertainty in the spin axis orientation of UD. If our spin obliquity assumption of $75\deg$ is accurate, we can directly apply our modeling results to speculate on possible dust activity.

If UD displays dust activity during flyby observations, the location of active region(s) will serve as crucial constraints as to the cause. We have demonstrated, for example, that thermal gradients are maximized for the equatorial region. Thus, dust activity observed at these locations would support thermal fracturing as the source of activity. This logic holds true for the assumed larger spin obliquity for UD, but is independent of the solstice position---as this would not alter the thermal characteristics of the latitudes nearest the equator where thermal cycling is thought to be most efficient \citep{Hamm_etal19}. Similarly to Phaethon, ice sublimation would theoretically be most likely to occur near the northern pole of UD---if volatiles are somehow present beneath the surface. The orbit-averaged temperature across the surface is mostly dependent on the solstice point (\autoref{sub:tpmanalysis}), which is unknown for UD.

Our dynamical modeling results show that UD has reached perihelion distances of approximately $0.14\au$ in the past. This value matches Phaethon's current perihelion and the heliocentric distance at which its dust tail is seen. If UD is subject to the same temperature-dependent mass-loss process as Phaethon, then it is possible that activity from UD was prevalent when the perihelion was smaller in the distant past. If true, then UD's original size could have been much larger than today. This has interesting implications regarding the hypothetical separation of UD from Phaethon.


\section{Conclusions}

From the results of our investigation into the dynamical and thermal history of Phaethon and UD we conclude the following.

\begin{itemize}

\item We confirm the results of previous dynamical studies of Phaethon and UD, but our interpretation is markedly different. We do not find compelling evidence that Pallas would be the most likely parent body of these objects. Rather, our results points to a parent body in the inner asteroid belt which contains families such as Svea and Polana.

\item We also confirm the finding of previous dynamical studies of Phaethon \citep[e.g.,][]{Hanus_etal2016} that the most recent minimum $q$ value occurred $\sim2000\yr$ ago and may coincide with the birth of the Geminid meteor stream (\autoref{fig:pha_orbel}).

\item Water ice cannot have survived longer than $5\Myr$ down to $200\meter$ below the surface and nowhere in Phaethon can water ice survive past $50\Myr$---our estimate for the delivery time from the Main Belt to a near-Earth orbit. Thus, if hydrogen-bearing volatiles exist anywhere beneath Phaethon's surface, they must be actively generated somehow---perhaps by similar process(es) hypothesized to supply ice on the Moon and Mercury.

\item The regoliths of Phaethon and UD are particularly susceptible to thermal fracturing, which we theoretically predict can eject particles up to 2 cm from the equatorial region. Observations from the DESTINY+ spacecraft will aid in resolving the cause of Phaethon's perihelion activity.

\end{itemize}

\section*{Acknowledgements}

This research was supported by Academy of Finland (\#316292 and \#299543) and by the Knut and Alice Wallenberg foundation (2016.0346). We also thank the two anonymous reviewers for detailed and constructive comments that greatly improved the quality of this article.


\bibliography{mybibfile}

\end{document}